\documentclass[11pt,a4paper]{article}
\usepackage{jheppub,amsmath,  amssymb,slashed,url,bm,textgreek,upgreek}
\usepackage{graphicx}
\usepackage{epstopdf}
\usepackage[normalem]{ulem}
\def\t{{ \sf t}} 

\def\tt{{\mathfrak t}}

\def\cl{{\mathrm{cl}}}

\def\MM{{\mathrm{MM}}}
\def\uU{{\mathcal U}}

\def\X{{\mathcal X}}

\def\uU{{\mathcal U}}

\def\S{{\sf S}}

\def\Sym{{\mathrm{Sym}}}
\def\g{\text{{\teneurm g}}}

\def\be{\begin{equation}}
\def\ee{\end{equation}}

\def\X{{\sf X}}

\def\tilde{\widetilde}

\def\h{\widehat}

\def\S{{\sf S}}
\def\SIgma{\Sigma}

\def\V{{\sf V}}
\def\O{{\mathcal O}}
\def\o{{\mathrm {op}}}

\def\A{{\mathcal A}}
\def\d{{\mathrm d}}

\def\bb{\sf b}
\def\R{{\mathbb R}}
\def\C{{\mathbb C}}
\def\U{{\sf U}}

\def\[{\bigl [}

\def\]{\bigr ]}

\def\N{{\mathcal N}}
\def\T{{\sf T}}

\def\Tr{{\mathrm {tr}}}

\def\t{\widetilde }
\def\h{\widehat}

\def\1{{\mathbf 1}}
\def\B{{\mathcal B}}

\def\W{{\sf W}}
\def\wW{{\mathcal W}}
\def\vV{{\mathcal V}}

\def\H{{\mathcal H}}

\def\tphi{\text{\textphi}}

\def\tilde{\widetilde}

\def\bdry{{\mathrm{bdry}}}
\def\la{\langle}
\def\ra{\rangle}
\def\a{{\sf a}}

\def\op{{\mathrm{op}}}
\def\bulk{{\mathrm{bulk}}}
\def\AdS{{\mathrm{AdS}}}

\def\SL{{{SL}}}
\def\i{{\mathrm i}}
\def\hSL{\widetilde{{SL}}}
\def\tSL{{\hSL}}
\def\x{{\sf x}}
\font\teneurm=eurm10 \font\seveneurm=eurm7  \font\fiveeurm=eurm5
\newfam\eurmfam
\textfont\eurmfam=\teneurm \scriptfont\eurmfam=\seveneurm
\scriptscriptfont\eurmfam=\fiveeurm

\font\teneusm=eusm10 \font\seveneusm=eusm7 \font\fiveeusm=eusm5
\newfam\eusmfam
\textfont\eusmfam=\teneusm \scriptfont\eusmfam=\seveneusm
\scriptscriptfont\eusmfam=\fiveeusm

\font\tencmmib=cmmib10 \skewchar\tencmmib='177
\font\sevencmmib=cmmib7 \skewchar\sevencmmib='177
\font\fivecmmib=cmmib5 \skewchar\fivecmmib='177
\newfam\cmmibfam
\textfont\cmmibfam=\tencmmib \scriptfont\cmmibfam=\sevencmmib
\scriptscriptfont\cmmibfam=\fivecmmib

\def\matt{{\mathrm{matt}}}
\def\Tr{{\mathrm{Tr}}}
\def\TFD{{\mathrm{TFD}}}
\title{Algebras and States in JT Gravity}

\author{Geoff Penington$^{1,2}$ and Edward Witten$^1$}
\affiliation{$^1$School of Natural Sciences, Institute for Advanced Study,\\ 1 Einstein Drive, Princeton, NJ 08540 USA}
\affiliation{$^2$Center for Theoretical Physics and Department of Physics, University of California,\\  Berkeley, CA 94720 USA}

\abstract{We analyze the algebra of boundary observables in canonically quantised JT gravity with or without matter. In the absence of matter,
this algebra is commutative, generated by the ADM Hamiltonian. After coupling to a bulk quantum field theory, it becomes a highly noncommutative algebra of
Type II$_\infty$ with a trivial center. As a result, density matrices and entropies on the boundary algebra are uniquely defined up to, respectively, a rescaling or shift. We show that this algebraic definition of entropy agrees with the usual replica trick definition computed using Euclidean path integrals. Unlike in previous arguments that focused on $\O(1)$ fluctuations to a black hole of specified mass, this Type II$_\infty$ algebra describes states at all temperatures or energies. We also consider the role of spacetime wormholes. 
One can try to define operators associated with wormholes that commute with the boundary algebra, but this fails in an instructive way.
In a regulated version of the theory, wormholes and topology change can be incorporated perturbatively.   
The bulk Hilbert space $\H_\bulk$ that includes baby universe states is then much bigger than the space of states $\H_\bdry$
accessible to a boundary observer. However, to a boundary observer,
every pure or mixed state on $\H_\bulk$ is equivalent to some pure state in $\H_\bdry$. 
}

\begin{document}\maketitle

\section{Introduction}\label{intro} 
JT gravity in two dimensions \cite{Jackiw,Teitelboim} with negative cosmological constant provides a simple and much-studied model of a two-sided
black hole (for example, see \cite{AP,Malda,MoreOnJT,MT,KS,ZY,HJ}).   
 JT gravity coupled to  additional matter fields, described by a quantum field theory, has also been much studied, especially in the case
that the matter theory is conformally invariant \cite{JT-CFT}.
The essential simplicity of the model is retained as long
as there is no direct coupling of the dilaton of JT gravity to other matter fields. 

In the present article, we will study JT gravity  from the point of view of understanding the algebra of observables accessible to a boundary observer
living on one side of the system.   It is believed that  in JT gravity with or without additional matter fields, it is not possible to define a one-sided black hole
Hilbert space, but it is certainly  possible to define a two-sided Hilbert space $\H$, as studied for example in \cite{HJ,ZY,KS,MLZ,JK,Lin}.    We will analyze the algebra 
$\A$ of operators acting on $\H$ that can be defined on, say,
the left boundary of the system.   

The analogous problem for more complicated systems in higher dimensions has been studied recently.   Those analyses have involved a large $N$ limit and
quantum fields propagating
in a definite spacetime,  a black hole of prescribed mass.    The starting point has been  a Type III$_1$ algebra of bulk quantum fields outside the black hole horizon,
which can be interpreted as an algebra of single-trace boundary operators \cite{LL,LL2}.   Upon including in the algebra the generator of time translations, either by including
certain corrections of order $1/N$ or by going to a microcanonical description, the Type III$_1$
algebra becomes a Type II$_\infty$ algebra \cite{GCP,CPW}.  This Type III$_1$ or Type II$_\infty$ algebra describes fluctuations about a definite spacetime, namely
the black hole spacetime that served as input.

The simplicity of JT gravity is such that it is possible to describe an algebra of boundary observables for JT gravity coupled to a definite
QFT, without taking any sort of large $N$ limit.   One obtains an algebra $\A$ of Type II$_\infty$ that is equally valid for any value of the black hole
temperature or mass.   The trace in this algebra is the expectation value in the high temperature limit of the thermofield double 
state.\footnote{This role of the high temperature limit  has also
been noted in the context of a double-scaled version of the SYK model \cite{Lin}; in that context, the algebra is of Type II$_1$.  
A description of double-scaled SYK in which the high temperature limit is conveniently accessible had been developed in \cite{Naro}.}
At high temperatures, the fluctuations in the bulk spacetime are small, and as in \cite{LL,LL2,GCP,CPW}, the operators in $\A$ can be given
a bulk interpretation.   At low temperatures, the fluctuations in the bulk spacetime are large and $\A$ cannot be usefully approximated as an algebra of bulk operators;
it has to be understood as an algebra of boundary operators.    

The fact that the algebra $\A$ can be defined in the case of JT gravity without choosing a reference temperature means that it is ``background independent.''
That is not the case for existing constructions of an algebra of observables outside a black hole horizon in more complicated models in higher dimensions.   In those
constructions, background independence is lost when one subtracts the thermal expectation value of an operator so as to get an algebra of operators that have a large 
$N$ limit.   In JT gravity coupled to matter, since we define the algebra without considering any large $N$ limit for the matter system,  background independence is retained.

The Type II$_\infty$ algebra $\A$  that describes JT gravity coupled to matter is a ``factor,'' meaning that its center consists only of $c$-numbers.  Accordingly, 
$\A$ has  a trace that is uniquely determined up to an overall multiplicative constant.  A multiplicative constant in the trace leads to an 
additive constant in the entropy, so a state of
the algebra $\A$ has an entropy that is uniquely defined up to a state-independent additive constant.    By contrast, in  JT gravity without matter, 
the algebra of boundary observables is commutative -- generated by the ADM Hamiltonian.    Therefore, in the absence of matter, 
the algebraic structure alone does not determine a unique trace or an appropriate definition of entropy.    
We will see that when  matter is present so that the algebra is of Type II$_\infty$,  the entropy of a state of the Type II$_\infty$ algebra agrees with the entropy computed
via Euclidean path integrals  \cite{GH,LM,M2} up to an overall additive constant.   Similar results were obtained previously in analyses based on large $N$ limits
 \cite{CPW}.   In contrast, previous attempts at 
understanding entanglement entropy in canonically quantised JT gravity focused on JT gravity without matter. As a result, they relied on 
the introduction of additional ingredients into the theory, such as the defect operators considered in \cite{KS,JK}, that were ``fine-tuned'' to match the Euclidean path integral results.

The bulk Hilbert space $\H_\bulk$ and the algebra $\A$ can be naturally-defined in a ``no-wormhole'' version of the theory in which the spacetime topology is assumed to
be a Lorentzian strip (or equivalently a disc in Euclidean signature), and this is quite natural for everything that we have said up to this point.  
However, it is also interesting to ask what happens if we incorporate wormholes and baby universes.  
 In pure JT gravity, there is no difficulty in studying wormhole contributions order by order in the genus of spacetime or equivalently in 
 $e^{-S}$, where $S$ is the entropy.  (The expansion in powers of 
$e^{-S}$ will break down at low temperatures.) One can even understand the theory nonperturbatively via  a dual matrix model \cite{SSS}. When the theory is coupled to matter, however, the perturbative wormhole contributions diverge because the negative matter 
Casimir energy in a closed universe leads to a divergent contribution from small wormholes. In fact this divergence plays a crucial and illustrative role in ensuring the 
consistency of the ``no-wormhole'' story described above. It does so by avoiding the presence of ``baby universe operators,'' similar to those in \cite{MM}, whose eigenvalues
would be classical $\alpha$-parameters \cite{Coleman,Giddings}.

In a more complete theory, we might expect that the Casimir divergence should be regulated. In the SYK model, for example, the divergence is suspected to be regulated by something similar 
to the 
Hawking-Page-like phase transition described in \cite{MQ}; see Section 6.1 of \cite{SSS} for discussion on this point.
As a result, we proceed somewhat formally and attempt to understand what happens to the boundary algebras and the Hilbert space in such a regulated theory. The analysis of the algebra 
$\A$ gives no major surprises: 
it  is corrected order by order in the wormhole expansion, but remains an algebra of Type II$_\infty$.  
The analysis of the Hilbert space  is more subtle and involves an interesting difference between the no-wormhole theory
and the theory with wormholes included.   With or without wormholes, a Hilbert space $\H_\bdry$ can be defined from a boundary point of view by 
first introducing states that have a reasonable
Euclidean construction, then using the bulk path integral to compute inner products among these states, and finally  dividing out null vectors and
taking a completion to get a Hilbert space.
The inner products that enter this construction have wormhole corrections, but wormholes do not affect the ``size'' of $\H_\bdry$.   On the other
hand, from a bulk point of view, once we include wormholes, to define a Hilbert space we have to include closed ``baby universes.''   The resulting Hilbert
space $\H_\bulk$ is then much ``bigger'' than it would be in the absence of wormholes.  There is a fairly simple natural definition of $\H_\bdry$ and a fairly simple
natural definition of $\H_\bulk$, but it is less obvious how to relate  them.   We make a sort of gauge choice that enables us to
define a map $\wW:\H_\bdry\to \H_\bulk$ that preserves inner products, embedding $\H_\bdry$ as a rather ``small'' subspace of $\H_\bulk$.   
States in $\H_\bulk$ that are orthogonal to $\wW(\H_\bdry)$ are inaccessible to a boundary observer.
The map  $\wW:\H=\H_\bdry\to \H_\bulk$ is
awkward to describe explicitly even for states that have a simple Euclidean construction. This map is likely far more difficult to describe for
 states that do not have such a simple construction -- for example,
states that arise from Lorentz signature time evolution starting from states with a simple Euclidean construction.

One of the main results of our study of wormholes is to learn that, from the point of view of the boundary observer, at least to all orders in $e^{-S}$ (since our analysis
is based on an expansion in this parameter), any pure or mixed state on the bulk Hilbert space
$\H_\bulk$ is equivalent to a pure state in the much smaller Hilbert space $\H_\bdry$.   Classically, one might describe this by saying that  although $\H_\bulk$
is much bigger than $\H_\bdry$, the extra degrees
of freedom in $\H_\bulk$ are beyond the observer's horizon.

Another  generalization is as follows.   Instead  of a world with a single open universe component and possible
closed baby universes, we can consider a world with two open universe components or in general
 any  number of them, plus baby universes.   In the absence of wormholes, this adds nothing essentially new: a Hilbert space for two open universes would 
 be trivially constructed from single-universe Hilbert spaces.      With wormholes included,
distinct open universes can interact with each other via wormhole exchange. However,
we can ask the following question: can an observer with access to only one asymptotic boundary of spacetime know how many other boundaries there are?
We show that the answer to this question is ``no,'' at least to all orders in $e^{-S}$, in the following sense. 
Let $\H_\bdry$ be the boundary Hilbert space for the case of a single open universe component (and any number of closed universes), and let $\A$ be the algebra of boundary operators
acting on $\H_\bdry$.
The same algebra $\A$ also acts on the bulk Hilbert space $\H_{\bulk,[n]}$ with any number $n$ of open universe components (and, again, any number of closed universes),
and every pure or mixed state on  $ \H_{\bulk,[n]}$ is equivalent, for a boundary observer, to some pure state in $\H_\bdry$.  
 Classically, one would interpret this by saying that an observer at one asymptotic end has no way to 
 know how many other asymptotic ends there are because they are all beyond a horizon.
Quantum mechanically, that language does not apply in any obvious way but the conclusion is valid.

In section \ref{bulk}, we review aspects of JT gravity and discuss from a bulk 
point of view the Hilbert space of JT gravity coupled to a quantum field theory.     In section \ref{algebra}, we construct the algebra 
$\A$ of operators accessible to an observer outside the horizon.  We define this algebra both directly within the canonically quantised theory, and via a natural alternative 
definition using Euclidean path integrals, which we argue is equivalent. This equivalence justifies the use of Euclidean path integrals to compute 
entropies in the context of JT gravity with matter.
In section \ref{operators}, we attempt to define ``baby universe operators'' that would commute with the boundary algebras, and show that this fails in an instructive fashion.
In section \ref{wormhole}, we consider wormhole corrections both to the Hilbert space constructed in section \ref{bulk} and to the algebra constructed in section
\ref{algebra}.  As already noted, once wormholes are included, the Hilbert space that is natural from a boundary point of view is a ``small'' and difficult to characterize
subspace of the Hilbert space that is natural from a bulk point of view.      In section \ref{multiple}, we consider a further
generalization to a spacetime with multiple asymptotic boundaries.      As already explained, a primary conclusion of studying  these generalizations is to learn that they are undetectable
by an observer at infinity in one asymptotic region.

The algebras discussed in the present article have been analyzed from a different point of view in \cite{DK}.  This paper contains, in particular, a precise argument for the important claim
that  the centers of the left and 
right algebras $\A_L$ and $\A_R$ consist only of $c$-numbers.   Among other things, this paper also contains illuminating and useful 
explicit formulas and a further analysis of the exotic traces
that we discuss in section \ref{operators}.   The author of \cite{DK} pointed out an error in section \ref{operators} of the original version of this article.  The error has been
corrected in the present version.   The conclusions are largely unchanged, but one important point is not very clear, as explained in section \ref{operators}.   

\section{The Bulk Hilbert Space}\label{bulk}

In this section, we first review some aspects of JT gravity -- focussing on the bare minimum needed for the present article -- and then we discuss from a bulk point of
view the Hilbert space of pure JT gravity and of JT gravity  coupled to a quantum field theory. 

\subsection{The Boundary Hamiltonian}\label{review}

The action of JT gravity with negative cosmological constant on a spacetime $M$ can be written, in the notation of \cite{HJ,JK}, as
\be\label{sjt}I_{JT}=\int_M \d^2x\sqrt{-g} \upphi (R+2) +2  \int_{\partial M} \d t \sqrt{|\gamma|}\upphi(K-1)+\cdots,\ee
where $g$ is the bulk metric with curvature scalar $R$, $\gamma$ is the induced metric on the boundary,  $K$ is the extrinsic curvature of the boundary, and 
we have omitted a topological invariant related to the classical entropy $S_0$.
Upon integrating first over $\upphi$ to impose the equation of motion $R+2=0$, with a boundary condition that fixes the boundary value of $\upphi$, the action reduces to 
\be\label{ibdry}I_{\partial M}=2  \int_{\partial M} \d t \sqrt{|\gamma|}\upphi(K-1). \ee

The condition $R+2=0$ implies  that $M$ is locally isomorphic to a portion of $\AdS_2$, a homogeneous manifold of constant curvature $-2$.
$\AdS_2$ is the universal cover of what we will call $\AdS_2^{(0)}$, namely the quadric $X^2+Y^2-Z^2=1$ with metric $\d s^2=-\d X^2-\d Y^2+\d Z^2$.
$\AdS_2^{(0)}$ has an action of $\SL(2,\R)$ generated by vector fields
\begin{align}\label{vectfields}    j_1 & = X\partial_Y-Y\partial_X \cr j_2&= Y\partial_Z+Z\partial_Y \cr
       j_3&=-X\partial_Z-Z\partial_X,\end{align}
       satisfying $[j_a,j_b]= \epsilon_{ab}^c j_c$, where the metric on the Lie algebra is $\eta_{ab}=\mathrm{diag}(-1,1,1)$.
Coordinates $T,\sigma$ with 
 \begin{align}\label{covercoord} X & = \cos T \cosh \sigma \cr
                       Y & = \sin T\cosh \sigma \cr
                         Z&=\sinh\sigma \end{align}
give a useful parametrization of the universal cover $\AdS_2$.   In these coordinates, the metric is
\be\label{adstwo} \d s^2=\d\sigma^2-\cosh^2\sigma \,\d T^2,~~~~-\infty<\sigma,T<\infty.\ee
The vector fields (\ref{vectfields}) on $\AdS_2^{(0)}$ lift to vector fields on $\AdS_2$ that generate an action of $\hSL(2,\R)$, the universal cover of
$\SL(2,\R)$.

$\AdS_2$ 
has  a ``right'' conformal boundary at $\sigma\to +\infty$ and a ``left'' conformal boundary at $\sigma\to -\infty$.    
In studies of JT gravity, $M$ is usually taken to be ``almost all'' of $\AdS_2$ \cite{AP,Malda}.   
This is achieved as follows.   First of all, the left and right boundaries could be defined simply by functions $\sigma_L(T)$, $\sigma_R(T)$.
However, in ``nearly $\AdS_2$ holography,'' one assumes that the boundary is parameterised by a distinguished parameter $t$, the time of the boundary
quantum mechanics, and one parametrizes the right boundary curve by functions $\sigma_R(t)$, $T_R(t)$, and similarly for the left boundary.
To get nearly $\AdS_2$ spacetime, one further imposes the boundary conditions
\be\label{bc}\gamma_{tt}=-\frac{1}{\epsilon^2},~~~\upphi|_{\partial M}=\frac{\tphi_b}{\epsilon}, \ee
with constant $\tphi_b$ and 
with very small $\epsilon$.  For small $\epsilon$, the condition $\gamma_{tt}=-1/\epsilon^2$ reduces to
\be\label{relno} e^{\sigma_R}=\frac{2}{\epsilon}\frac{1}{\dot T_R}, ~~e^{-\sigma_L}=\frac{2}{\epsilon}\frac{1}{\dot T_L},\ee where dots 
represent derivatives with respect to $t$.  Thus, the left and right boundary curves
lie, for small $\epsilon$, at very large negative or positive $\sigma$, and each of them is determined by a single function $T_L(t)$ or $T_R(t)$.  

It is useful to define \be\label{usefuldef} e^{\sigma_R}=\frac{2\tphi_b}{\epsilon} e^{\t\sigma_R},~~~e^{-\sigma_L}=\frac{2\tphi_b }{\epsilon} e^{-\t\sigma_L},\ee
 where (in view of eqn. (\ref{relno})) $\t\sigma_L$, $\t\sigma_R$ remain finite for
$\epsilon\to 0$. Here 
  $\t\sigma_L$, $\t\sigma_R$ are renormalized length parameters, in the sense that, for $\epsilon\to 0$,  the length of a geodesic between the left and right
boundaries is
\be\label{length}\ell =\t\sigma_R-\t\sigma_L+{\mathrm{constant}}. \ee
The constant depends only on $T_L, \,T_R$ and not on $\t\sigma_L,\,\t\sigma_R$.

With a small calculation, one finds that for  $\epsilon\to 0$, the boundary action (\ref{ibdry}) becomes 
\be\label{baction} I_{\partial M}= \tphi_b\int\d t\left( -{\dot T_R}^2+\left(\frac {\ddot T_R}{\dot T_R}\right)^2\right) +\tphi_b\int\d t\left( -{\dot T_L}^2+
\left(\frac {\ddot T_L}{\dot T_L}\right)^2 \right),\ee which is known as the Schwarzian action because it is a linear combination of the Schwarzian derivatives $\{T_R,t\}$ 
and $\{T_L,t\}$.

However, there is another convenient way to describe the problem \cite{KS,ZY}.   One term in the boundary action (\ref{ibdry}) is just $-2\upphi L$, where $L$ is the length
of $\partial M$; using the boundary condition on $\upphi$, this is $-\frac{2\tphi_b}{\epsilon}L$.      The other term involving the integral of $K$ can be expressed,
using the Gauss-Bonnet theorem, in terms of $\int_M\d^2x \sqrt g R$ (together with a topological invariant, the Euler characteristic of $M$); 
since  $R=-2$, this is just $-2A$, with $A$ the area of $M$.   Thus the boundary action is
\be\label{newbaction} I_{\partial M}=\frac{2\tphi_b}{\epsilon} \left(A-L\right)+{\mathrm{constant}}. \ee
The area form of $\AdS_2$ is $\sqrt{-g}\d\sigma\d T=\cosh\sigma \d\sigma \d T=\d(\sinh\sigma\d T)$, so 
\be\label{aform} A=\int \d t \left(\sinh\sigma_R \frac{\d T_R}{\d t}-\sinh\sigma_L \frac{\d T_L}{\d t}\right).\ee
As for the length term, the sum over all paths of length $L$ is a random walk of that length.   A random walk on a manifold describes a process of diffusion
which can also be described by the heat kernel $e^{- t \Delta}$, where $\Delta$ is the Laplacian (or its Lorentz signature analog).   On a general Riemannian
manifold, if we take for the action the usual kinetic energy of a nonrelativistic particle,  $I_{\mathrm{kin}}= \frac{1}{2}\int \d t g_{ij} \dot x^i \dot x^j$, then the corresponding
Hamiltonian is $\Delta/2$, appropriate to describe diffusion.  
The upshot is that
the $L$ term in the action can be replaced by an action of the form $I_{\mathrm{kin}}$.
    The resulting action for the right boundary is then
\be\label{rbaction}I_R=\tphi_b\int \d t \left( \dot \sigma_R^2-\cosh^2 \sigma_R \dot T_R^2\right) +\frac{2\tphi_b}{\epsilon}\int\d t \sinh \sigma_R \dot T_R,\ee
with a similar action for the left boundary.    Proofs of the relationship\footnote{This relationship involves some renormalization, leading to a divergent additive
constant in the Hamiltonian that will be dropped in the next paragraph.}  between (\ref{newbaction}) and (\ref{rbaction}) can be found in \cite{KS,ZY}, in part following
chapter 9 of \cite{Polyakov}.  

For our purposes, we will just verify that\footnote{This derivation was explained to us by Z. Yang; a similar calculation in a different coordinate system can be found in
\cite{ZY}.}   (\ref{rbaction}) is equivalent to (\ref{baction}) in the limit $\epsilon\to 0$ (apart from an additive constant that has to be dropped from the  Hamiltonian).
The canonical momenta deduced from $I_R$ are $p_{\sigma_R}=2\tphi_b \dot \sigma_R$, $p_{T_R}=2\tphi_b(-\cosh^2 \sigma_R\dot T_R+\frac{1}{\epsilon}\sinh \sigma_R)$.
The Hamiltonian is then 
\be\label{firstham}H_R=\frac{p_{\sigma_R}^2}{4\tphi_b} -\frac{1}{4\tphi_b \cosh^2\sigma_R} \left(p_{T_R}-\frac{2\tphi_b}{\epsilon}\sinh \sigma_R\right)^2. \ee
Making the change of variables (\ref{usefuldef}), where $p_{\tilde\sigma_R}=p_{\sigma_R}$, the $\epsilon\to 0$ limit of the Hamiltonian comes out to be
(after discarding an additive constant)
\be\label{smalleps} H_R=\frac{1}{2\tphi_b}\left(\frac{1}{2} p_{\t\sigma_R}^2 + p_{T_R} e^{-\t\sigma_R} +\frac{1}{2}e^{-2\t\sigma_R}\right). \ee
The action of any Hamiltonian system has a canonical form   $I_{\mathrm{can}} = \int\d t ( \sum_i p_i \dot q^i - H) $.  In the present case, this is
\be\label{canform} I_{\mathrm{can}}=\int\d t\left( p_{\t \sigma_R}\dot {\t\sigma}_R +p_{T_R}\dot T_R - \frac{1}{2\tphi_b}\left(\frac{1}{2} p_{\t\sigma_R}^2 + 
p_{T_R} e^{-\t\sigma_R} +\frac{1}{2}e^{-2\t\sigma_R}\right)\right). \ee   Here, $I_{\mathrm{can}}$  is linear in $p_{T_R}$, so $p_{T_R}$
behaves as a Lagrange multiplier
setting $e^{-\t \sigma_R}=2\tphi_b \dot T_R$.   After also integrating out $p_{\t\sigma_R}$, which appears quadratically, by its equation of motion, we see that
the action (\ref{canform}) is equivalent to the Schwarzian action\footnote{Introducing and simplifying the Hamiltonian has given an efficient way to do
this calculation; however, one can reach the same conclusion by analyzing how the solutions of the equations of motion behave for small $\epsilon$. We will
in any case need the formula for the Hamiltonian.}  ({\ref{baction}).

In terms of the variables \be\label{chidef}\chi_R=-\t\sigma_R,  ~~~~~~\chi_L=\t\sigma_L\ee used in \cite{JK}, the Hamiltonian on the right boundary is
\be\label{hamright} H_R=\frac{1}{2\tphi_b}\left(\frac{1}{2}p_{\chi_R}^2+ p_{T_R} e^{\chi_R} +\frac{1}{2} e^{2\chi_R}\right). \ee
By a similar derivation, the Hamiltonian on the left boundary is\footnote{In this derivation, a minus sign in the formula (\ref{aform}) for the area is compensated by a relative minus sign in the definitions of $\chi_R, ~\chi_L$.}
\be\label{hamleft} H_L =\frac{1}{2\tphi_b} \left( \frac{1}{2} p_{\chi_L}^2+p_{T_L} e^{\chi_L}+\frac{1}{2} e^{2\chi_L}\right). \ee
The renormalized geodesic length between the left and right boundaries is
\be\label{geolength}\ell =-\chi_R-\chi_L+\log\left(\frac{1 + \cos (T_L - T_R)}{2}\right).\ee

\subsection{The Hilbert Space of Pure JT Gravity}\label{constraints}

The left and right boundaries of $M$ are thus described by variables $T_L,\chi_L, T_R, \chi_R$ and their canonical conjugates.    Quantum mechanically,
we can describe these boundaries by a Hilbert space $\H_0$ consisting of $L^2$ functions $\Psi(T_L,\chi_L,T_R,\chi_R)$.

However  \cite{HJ,KS,ZY,MLZ,JK}, $\H_0$ is not the appropriate bulk Hilbert space for JT gravity, for two reasons.   One reason involves causality, and the second reason involves the gauge constraints.   We will discuss causality first.
Classically, one can describe a solution of JT gravity
by specifying a pair of functions $T_L(t)$, $T_R(t)$ that satisfy the equations of motion derived from the Schwarzian action (\ref{baction}).   Not all pairs of solutions
are allowed, however; one wants the two pairs of boundaries to be spacelike separated.  For the metric (\ref{adstwo}), the condition for this is that
\be\label{spacelike}|T_L(t)-T_R(t')|<\pi\ee
for all  real $t,t'$.    Quantum mechanically, the observables $T_R(t)$  at different times are noncommuting operators that cannot be simultaneously
specified; the same applies for $T_L(t)$.   So we cannot directly impose the condition (\ref{spacelike}) at all times.  Fortunately, one can check that in the classical theory it is sufficient to impose the condition (\ref{spacelike}) at one pair of times $t, t'$. As we discuss briefly below, the classical dynamics then ensure that (\ref{spacelike}) holds at all times so long as the the two trajectories have vanishing total $\tSL(2,\R)$ charge -- i.e. the solution satisfies the gauge constraints. We will define the quantum theory in the same way: we impose the condition $|T_L(t)-T_R(t')|<\pi$ at some chosen times, say $t=t'=0$, and then hope that after imposing the gauge constraints the quantum dynamics
lead to a causal answer.   We impose this initial condition by refining the definition of the Hilbert space $\H_0$ to say that it consists of $L^2$ functions
$\Psi(T_L,\chi_L,T_R,\chi_R)$ whose support is at $|T_L-T_R|<\pi$.    

Having made this definition, we then have to ask whether it leads to quantum dynamics that are consistent with causality.
For JT gravity with matter, we will eventually get a fairly satisfactory answer, along the following lines.   We will define algebras $\A_L$, $\A_R$ of observables
on the left and right boundaries, respectively.  $\A_L$ and $\A_R$ will contain, respectively, all quantum fields inserted on the left or right boundary at arbitrary
values of the quantum mechanical time.    The two algebras will commute with each other, and this will be a reasonable criterion for saying at the quantum level
that the two boundaries are out of causal contact.   For JT gravity without matter, an explanation along those lines is unfortunately not available, since there are not
enough boundary observables.    However the fact we end up with sensible boundary Hamiltonians on a Hilbert space constructed from wavefunctions with $|T_L-T_R|<\pi$ is itself evidence that the boundaries remain out of causal contact at all times.

Even after imposing the condition $|T_L-T_R|<\pi$, $\H_0$ is not the physical Hilbert space of JT gravity, because we have to impose the constraints.
Since we are interested in the intrinsic
geometry of $M$, not in how it is identified with a portion of $\AdS_2$, we have to regard two sets of variables $T_L,\chi_L,T_R,\chi_R$ that
differ by the action on $\AdS_2$ of $\hSL(2,\R)$ to be equivalent.   In other words, we have to treat $\tSL(2,\R)$ as a group of constraints.  

The constraint operators are
\be\label{constop} J_a=J_a^L+J_a^R,\ee
where $J_a^L$ and $J_a^R$ are the generators of $\tSL(2,\R)$ acting on the right and left boundaries, namely
\begin{align}\label{generators} \notag   J_1^R & = p_{T_R} \\ J_2^R&=(\cos T_R ) p_{T_R}-(\sin T_R)p_{\chi_R} +e^{\chi_R}\cos T_R +\frac{\i}{2} \sin T_R \\ \notag
J_3^R&= (\sin T_R) p_{T_R} +(\cos T_R)p_{\chi_R} +e^{\chi_R}\sin T_R -\frac{\i}{2} \cos T_R .   \end{align}
and\footnote{The formulas for $J_a^L$  used in \cite{JK} differ from these by $T_L\to T_L\pm \pi$, reversing the signs of $J_2^L$ and $J_3^L$.
We will not make this change of variables as that would make the discussion of causality less transparent.}
\begin{align}\label{generators2}    J_1^L & = p_{T_L} \cr J_2^L&=-(\cos T_L ) p_{T_L}+(\sin T_L)p_{\chi_L} -e^{\chi_L}\cos T_L -\frac{\i}{2} \sin T_L \cr
J_3^L&= -(\sin T_L) p_{T_L} -(\cos T_L)p_{\chi_L} -e^{\chi_L}\sin T_L +\frac{\i}{2} \cos T_L .   \end{align}
These operators are self-adjoint and obey $[J_a^R,J_b^R]=\i\epsilon_{ab}{}^c J_c^R$, $[J_a^L,J_b^L]=\i\epsilon_{ab}{}^c J_c^L$.    Here $\epsilon_{abc}$ is completely
antisymmetric with $\epsilon_{123}=1$; Lie algebra
indices are raised and lowered with the metric $\eta_{ab}=\mathrm{diag}(-1,1,1)$.

The derivation  of the formulas (\ref{generators}), (\ref{generators2}) can be understood as follows.   The terms in $J_a^L,J_b^R$ that are 
linear in the momenta give the $\sigma\to\pm\infty$ limit of the group action on the AdS$_2$ coordinates $(\sigma, T)$ generated by the vector
fields  (\ref{vectfields}). The imaginary terms in $J_a^L$, $J_b^R$  are there simply to make those operators self-adjoint. Finally, the  terms proportional to $e^{\chi_R}$ and
$e^{\chi_L}$ are neccessary to give the correct action on the conjugate momenta $p_{\chi_{R}}, p_{T_{R}}$ and $p_{\chi_{L}}, p_{T_{L}}$. This action can be computed from the $\tSL(2,\R)$-invariant action (\ref{rbaction}) by taking the $\epsilon\to 0$ limit. However, it is somewhat easier to instead derive the $\tSL(2,\R)$ charges in the Hamiltonian description. In this description, symmetry group generators must commute with $H_L$ and $H_R$, 
which we have already determined. This forces the inclusion of the terms proportional to $e^{\chi_R}$, $e^{\chi_L}$.   Actually, $H_R$ and $H_L$  
are essentially the quadratic Casimir operators for the action of $\tSL(2,\R)$ on the right and left boundary degrees of freedom:
\be\label{hamcas}2\tphi_b H_R=\frac{1}{2}\left(\eta^{ab}J_a^R J_b^R-\frac{1}{4}\right), ~~~~~2\tphi_b H_L =\frac{1}{2}\left(\eta^{ab}J_a^L J_b^L-\frac{1}{4}\right) . \ee

Before discussing how to impose these constraints at the quantum level, we first describe how they are implemented in classical JT gravity. 
The classical phase space procedure for dealing with a gauge symmetry is known as a symplectic quotient, and involves a two-step procedure. 
The starting point is a $\g^*$-valued function called a ``moment map''  where $\g$ is the Lie algebra of the gauge group and $\g^*$ is its dual. This moment map should generate 
the gauge group action via Poisson brackets. In our case, the moment map is just the total $\tSL(2,\R)$ charge $J_a=J_a^L+J_a^R$, where the conserved 
charges $J_a^L$ and $J_a^R$ are given by the formulas (\ref{generators}) and (\ref{generators2}) above, except that the imaginary terms can be dropped 
because we are in the classical limit. To take a symplectic quotient, we first consider the subspace of phase space on which the moment map is zero. 
To recover a symplectic manifold (i.e. a sensible phase space), we then also identify points on this constrained space that are related by the action of 
the gauge group. Each of these two steps reduces the phase space dimension by the dimension of the gauge group. In our case, the unconstrained phase 
space is eight dimensional, and the group $\tSL(2,\R)$ is three dimensional, so the physical phase space will be two dimensional.

The qualitative properties of a classical orbit depend on whether the Casimir  $\eta^{ab} J^R_a J^{R}_b$ is positive, negative, or zero.  
  If $\eta^{ab} J^R_a J^R_b<0$,
then up to an $SL(2,\R)$ rotation, we can assume that $J^R_1\not=0$, $J^R_2=J^R_3=0$.    The conditions $J^R_2=J^R_3=0$ imply via eqn. (\ref{generators})  that $p_{\chi_R}=0$
and $e^{\chi_R}=-p_{T_R}$, so that we must have $J_1^R=p_{T_R}<0$.   The $\tSL(2,\R)$ constraint implies that the left boundary particle has $J^L_a=-J^R_a$, and now the conditions $J^L_2=J^L_3=0$ lead to $J_1^L=p_{T_L}=-e^{\chi_L}<0$.   But as $J_1^R,J_1^L$ are then both negative, it is impossible to satisfy the constraint $J_1^R+J_1^L=0$.
So orbits with $\eta^{ab} J^R_a J^{R}_b<0$ cannot satisfy the constraints.   A similar analysis shows that the same is true of orbits with $\eta^{ab}J^R_a J^R_b=0$.

Thus, we have to consider orbits with $\eta^{ab} J_a^R J_b^R>0$.    Any such orbit is related by $\tSL(2,\R)$ to one with 
 $J_2^R  > 0$ and $J_1^R=J_3^R=0$; again, the $\tSL(2,\R)$ constraint requires $J_a^L=- J_a^R$.   The conditions $J_1^R=J_1^L=0$ give $p_{T_R}=p_{T_L}=0$ and the other
 conditions can be solved to give 
 \begin{align}\notag e^{\chi_R}& = J_2^R \cos T_R \\
      e^{\chi_L}&= - J_2^L\cos T_L =J_2^R\cos T_L .\end{align}
  Any orbit of this type therefore has \be 2\pi n_R -\pi/2< T_R< 2\pi n_R +\pi/2,~~~~2\pi n_L-\pi/2<T_L<2\pi n_L+\pi/2\ee for some integers $n_L,n_R$.   An element of the center
  of $\tSL(2,\R)$ will shift $n_L,n_R$ by a common integer, so only the difference $n_R-n_L$ is invariant.
  If this difference vanishes, then $|T_L(t)-T_R(t')|<\pi$ for all $t,t'$ and  two boundaries are
 spacelike separated at all times. If the difference is nonzero,  then $|T_L(t)-T_R(t')|>\pi$ always, and the two boundaries are timelike separated at  all times.
  Thus it is necessary to impose a condition that the two boundaries are spacelike separated, and if this condition is imposed at one time, it remains valid for all times.

 At this stage, we have reduced the phase space to a three-dimensional space parameterised by the value of $J_2^R$, or equivalently of the Hamiltonians $H_L = H_R = \frac{1}{4\tphi_b}(J_2^R)^2-\frac{1}{16\tphi_b}$,  along with the locations of the two boundary particles along their trajectories. To complete our analysis, we note that the gauge symmetry generator $J_2 = J_2^L + J_2^R$ preserves the gauge charges and hence preserves the two boundary trajectories. In fact (up to an energy-dependent rescaling), it generates forwards time-translation of the right boundary and backwards time-translation of the left boundary. After quotienting by this action, we obtain the final two-dimensional phase space \cite{HJ} 
  parameterised by the boundary energy $H_L = H_R$ along with the ``timeshift'' between the two boundary trajectories.

Let us now discuss what happens in the quantum theory. Because the constraint group $\tSL(2,\R)$ is non compact, imposing the constraints on quantum states is somewhat subtle. Suppose that a group $G$ acts on a Hilbert space $\H_0$, 
 with inner product $(~,~)$, and one wishes to impose $G$ as a group of constraints.   
 In our case, $G=\tSL(2,\R)$ and $\H_0$ was defined earlier.
Naively, one imposes the constraints by restricting to the $G$-invariant subspace of $\H_0$.
 This is satisfactory if $G$ is compact, but if $G$ is not compact, this procedure can be problematical because
 $G$-invariant states are typically not normalizable, so there may be few or no $G$-invariant states in $\H_0$.   
 A procedure that often works better for a 
 noncompact group and that has been extensively discussed in the context of gravity (see for example \cite{MarolfReview,marolf}) 
 is to define a Hilbert space of coinvariants of the $G$ action, rather than invariants.   This means that one considers 
 any state $\Psi\in\H_0$ 
to be physical, but one imposes an equivalence relation $\Psi\cong g\Psi$  for any $g\in G$.    The equivalence 
classes are called the coinvariants of $G$.   
$G$ acts trivially on the space of coinvariants, since by definition $\Psi$  and $g\Psi$ are in the same equivalence 
class for any $\Psi\in\H_0$, $g\in G$.   
Thus, the coinvariants are  annihilated by $G$, even if they cannot be 
 represented by invariant vectors in the original Hilbert space $\H_0$.
  If (as in the case of $\tSL(2,\R)$) the group $G$ has a left and right invariant measure $\d\mu$, one can try to define an inner product 
  on the space of coinvariants by integration over $G$:
  \be\label{coinv} \la\Psi'|\Psi\ra =\int_G \d \mu ~(\Psi',R(g)\Psi). \ee
  Here $R(g)$ is the operator by which $g\in G$ acts on $\H_0$.   If the integral  in eqn. (\ref{coinv}) is 
  convergent (as is the case for the states that will be introduced 
  presently in 
  eqn. (\ref{somewave})), then $\la\Psi'|\Psi\ra$ depends only on the equivalence
  classes of $\Psi$ and $\Psi'$, so the formula defines an inner product on the space of coinvariants 
  and enables us to define the Hilbert space $\H$ of coinvariants.
  
  The general procedure to impose constraints is really BRST quantization, or its BV generalization.   
  Both the space of invariants and the space of coinvariants are special
  cases of what is natural in BRST-BV quantization.   See \cite{shvedov} or Appendix B of \cite{CLPW} for background.
    BRST-BV quantization in general (see \cite{Henneaux} for an introduction) permits one to define something
   intermediate between the space of invariants and the space of coinvariants.   For example, 
   in perturbative string theory, where one wants to impose
  the Virasoro generators $L_n$ as contraints, one usually imposes a condition $L_n\Psi=0$, $n\geq 0$, on physical states, and also
   an equivalence relation $\Psi\cong \Psi+L_n\chi$,
  $n<0$.    This means that one takes invariants of the subalgebra generated by $L_n$ for 
  $n\geq 0$ and coinvariants of the subalgebra generated by $L_n$, $n<0$.   
 BRST quantization generates this mixture in a natural way.   Such a mixture is also natural, in general, in gauge theory and gravity. 
 
 In the case of JT gravity, such refinements are not necessary.  We can just define the Hilbert space $\H$ of JT gravity to be  the 
 space of coinvariants of the action of $\tSL(2,\R)$ on $\H_0$.   We will see that this 
 definition leads to efficient derivations of useful results,
 some of which have been deduced previously by other methods.
In fact, JT gravity is simple enough
 that  it is possible, as shown in the literature,
  to get equivalent results, sometimes with slightly longer derivations, by working with unnormalizable $\tSL(2,\R)$ invariant
 states and correcting the inner product by formally dividing by the infinite volume of $\tSL(2,\R)$. 
 
 To minimize clutter, we henceforth write just $T,T', \chi,\chi'$ for $T_R,T_L,\chi_R,\chi_L$.    For any 
 $T,T',\chi,\chi'$ satisfying the causality 
 constraint $|T-T'|<\pi$, there is always a unique element of $\tSL(2,\R)$ that sets $T=T'=0$, $\chi=\chi'$.     
 This means that the space of coinvariants is generated
 by wavefunctions of the form
 \be\label{somewave} \Psi=\delta(T)\delta(T')\delta(\chi-\chi')\psi(\chi).   \ee
 Such wavefunctions are  highly unnormalizable in the inner product of $\H_0$, but in the 
 natural inner product (\ref{coinv}) of the space $\H$ of coinvariants,
 we have simply
  \be\label{innerprod}\la \Psi,\Psi\ra =\int_{-\infty}^\infty \d \chi\,\overline\psi\psi. \ee     
  The form (\ref{somewave}) of the wavefunction is preserved by the operator $\chi$, acting by multiplication, along with
  $\t p_\chi=p_\chi+p_{\chi'}=-\i(\partial_\chi+\partial_{\chi'})$. Of course, $[\t p_\chi,\chi]=-\i$.
    In short, the physical Hilbert space $\H$ can be viewed as the space of 
    square-integrable
  functions of $\chi$ (or $\chi'$), and the algebra of operators acting on $\H$ is generated 
  by the conjugate operators $\chi$ and $\t p_\chi$.
  
  Now we can evaluate the left and right Hamiltonians $H_L$ and $H_R$ as operators on $\H$.   In doing so, we note that by definition
  any $\Psi\in\H$ is annihilated by the constraint operators $J_a=J_a^L+J_a^R$.   This statement 
  is just the derivative at $g=1$ of the equivalence
  relation $\Psi\cong g\Psi$, $g\in \tSL(2,\R)$.     Acting on a state of the form given in eqn. (\ref{somewave}), we have 
  \begin{align}\label{effform}\notag J_1 \Psi & = (p_T+p_{T'})\Psi \\ 
                                            \notag J_2\Psi & =(p_T-p_{T'})\Psi\\ 
                                              J_3\Psi& =(p_\chi-p_{\chi'})\Psi. \end{align}
   So as operators on $\H$, $p_T$  is equivalent to  $(J_1+J_2)/2$  and hence can be set to zero,
    and $p_\chi$ is equivalent to $ \frac{1}{2}\t p_\chi+\frac{1}{2}J_3$ and so can be replaced by $\frac{1}{2}\t p_\chi$.
   Likewise $p_{T'}$ can be replaced by 0 and $p_{\chi'}$ by $-\frac{1}{2}\t p_\chi$.     With these substitutions, we get
   \be\label{subham} 2\tphi_b H_L=2\tphi_b H_R= \frac{\t p_\chi^2}{8}+\frac{e^{2\chi}}{2}.\ee       
  From eqn. (\ref{geolength}) (with $\chi_L=\chi_R=\chi$, and after absorbing a constant shift in $\ell$), the renormalized length $\ell$ of the geodesic between
  the two boundaries is $\ell=-2\chi$, so  alternatively
  \be\label{ubham} 2\tphi_b H_L=2\tphi_b H_R=\frac{p_\ell^2}{2}+\frac{1}{2}  e^{-\ell}.\ee
   
   As noted in \cite{JK}, before imposing the $\tSL(2,\R)$ constraints, the operators $H_L$, $H_R$ are not positive-definite.   On the other hand, after imposing the
   constraints, we have arrived at  manifestly positive formulas for $H_L$ and $H_R$; the negative energy states have all been removed by the constraints.   This is the quantum analogue of our observation that, in classical JT gravity, orbits with $\eta^{ab} J^R_a J^{R}_b<0$ cannot satisfy the constraints.   
   
   The fact that $H_L=H_R$ after imposing constraints is analogous to the fact that in higher
dimensions, the ADM mass of an unperturbed Schwarzschild spacetime is the same at either end. It can be deduced directly from the relation    (\ref{hamcas}) between the Hamiltonians and the Casimir
   operators.   We have
   \be\label{relham}2\tphi_b(H_R-H_L)=\frac{1}{2}\left(\eta^{ab}J_a^R J_b^R
-\eta^{ab}J_a^L J_a^R\right)=\frac{1}{2} \eta^{ab}(J_a^R+J_a^L)(J_b^R-J_b^L). \ee
The operator on the right hand side annihilates physical states, since any operator of the general form $\sum_a J_a X^a$, where $J_a$ are the constraint
operators and $X^a$ are any operators, annihilates $\H$.      Hence
$H_R-H_L=0$ as an operator on $\H$.      Once we know this, it follows easily that $H_R$ and $H_L$ are positive after imposing the constraints.
Since $H_R=H_L$ as operators on $\H$, if one of them is negative, so is the other.   From eqns. (\ref{hamright}) and (\ref{hamleft}), we see that for this
to happen, $p_{T}$ and $p_{T'}$ must be negative, but in this case $J_1=p_{T}+p_{T'}$ is negative, contradicting the fact that $J_1$ annihilates
physical states.
      
 \subsection{Including Matter Fields}\label{incmatter}
 
It is pleasantly straightforward to include matter fields in this construction.    As we will see, $H_L$ and $H_R$ remain positive. 

As in many recent papers, we add to JT gravity a ``matter'' quantum field theory that does not couple directly to the dilaton field $\upphi$ of JT gravity.  Quantized
in $\AdS_2$, such a theory has a Hilbert space $\H^\matt$.      Since $\t\SL(2,\R)$ acts on $\AdS_2$ as a group of isometries, any relativistic field theory on
$\AdS_2$, whether conformally invariant or not, is $\t\SL(2,\R)$-invariant.    Hence the group $\t\SL(2,\R)$ acts naturally on $\H^\matt$, say with generators
$J_a^\matt$, obeying the $\tSL(2,\R)$ commutation relations.

In the context of coupling to JT gravity, the matter system should be formulated on a large piece $M$ of $\AdS_2$, not on all of $\AdS_2$.    However,
in the limit $\epsilon\to 0$ that was reviewed in section \ref{review}, this distinction is unimportant because the boundary of $M$ is, in the relevant sense, near
the conformal boundary of $\AdS_2$.   Hence we can think of the matter theory as ``living'' on all of
$\AdS_2$.     Therefore, prior to imposing constraints, we can take the Hilbert space of the combined system to be
 $\H_0\otimes \H^\matt$, where $\H_0$ is defined as in 
section \ref{constraints}.  

On this we have to impose the $\tSL(2,\R)$ constraints.    The relevant constraint operators are now the sum of the constraint operators of the gravitational
sector and the matter system:
\be\label{constsum} J_a=J_a^R+J_a^L+J_a^\matt. \ee

Now it is straightforward to impose the constraints and construct the physical Hilbert space $\H$.    We define $\H$ to be the space of coinvariants of the action
of $\tSL(2,\R)$ on $\H_0\otimes \H^\matt$.    As before, because $\tSL(2,\R)$ can be used to fix $T=T'=0$, $\chi=\chi'$ in a unique fashion, 
the coinvariants are generated by states of the form
\be\label{stateform}\Psi =\delta(T)\delta(T')\delta(\chi-\chi') \psi(\chi). \ee
The only difference is that $\psi(\chi)$, instead of being complex-valued, is now valued in the matter Hilbert space $\H^\matt$.      Evaluation of the inner
product (\ref{coinv}) now gives 
\be\label{statenorm}\la \Psi,\Psi\ra =\int_{-\infty}^\infty \d \chi \,\,(\psi(\chi),\psi(\chi)), \ee
where here $(~,~)$ is the inner product on $\H^\matt$.  So  the Hilbert space of coinvariants is $\H=L^2(\R)\otimes \H^\matt$, where $L^2(\R)$ is the space of
$L^2$ functions of $\chi$.   The algebra of operators on $\H$ is generated by $\chi$, $\t p_\chi,$ and the operators on $\H^\matt$.   

Now we want to identify the boundary Hamiltonians $H_R$ and $H_L$ as operators on $\H$.   To do this, we just have to generalize
 eqn. (\ref{effform}) to  include $J_a^\matt$. 
On a state of the form (\ref{stateform}), the constraint operators $J_a$ act by
  \begin{align}\label{zeffform}\notag J_1 \Psi & = (p_T+p_{T'} +J_1^\matt)\Psi \\ 
                                            \notag J_2\Psi & =(p_T-p_{T'}+J_2^\matt)\Psi\\ 
                                              J_3\Psi& =(p_\chi-p_{\chi'}+J_3^\matt)\Psi. \end{align}
With the aid of these formulas, one finds that as operators on $\H$,
\begin{align}\label{asops}\notag 2\tphi_b H_R& = \frac{1}{8}(\t p_\chi-J_3^\matt)^2-\frac{1}{2}(J_1^\matt+J_2^\matt)  e^\chi +\frac{1}{2} e^{2\chi}     \\
         2\tphi_b  H_L&     = \frac{1}{8}(\t p_\chi+J_3^\matt)^2-\frac{1}{2}(J_1^\matt-J_2^\matt)  e^\chi +\frac{1}{2} e^{2\chi}.   \end{align}     
The operators $(\t p_\chi\pm J_3^\matt)^2$, $e^{2\chi}$,  and $e^\chi$  are manifestly positive, and in a moment, we will show that the operators
$-(J_1^\matt\pm J_2^\matt)$ are non-negative.  So $H_L$ and $H_R$ are positive as operators on the physical Hilbert space $\H$.   
One can also verify using eqn. (\ref{asops}) that $[H_L,H_R]=0$, as expected since this is true even before imposing the constraints.

To understand the statement that the operators $-(J_1^\matt\pm J_2^\matt)$ are non-negative, we need to discuss in more detail the meaning of the constraints.
Let $\Phi$ be one of the matter fields that can be inserted on the boundary of $\AdS_2$, say on the right side. The constraints are supposed to commute with boundary
insertions such as $\Phi(T(t))$, while reparameterising $T$.    Since $J_1^R=p_T=-\i 
\partial_T$,
we have $[J_1^R,T(t)]=-\i$.    To get $[J_1^R+J_1^\matt,\Phi(T(t))]=0$, we then need $[J_1^\matt,\Phi(T)]=+\i \partial_T\Phi(T)$.   Comparing to the standard
quantum mechanical formula
$[H,\Phi(T)]=-\i\partial_T\Phi$, where $H$ is the Hamiltonian, we conclude that actually $J_1^\matt=-H$.
In quantum field theory in $\AdS_2$,  $H$ is non-negative and annihilates only the $\tSL(2,\R)$-invariant ground state.   So therefore $J_1^\matt$ is non-positive.
For $-1<a<1$, the operator $J_1^\matt+ a J_2^\matt$ is conjugate in $\tSL(2,\R)$ to a positive multiple of $J_1^\matt$, so it is again non-positive.  Taking
the limit $|a|\to 1$, the operators $J_1^\matt\pm J_2^\matt$ are likewise 
non-positive, and therefore $-(J_1^\matt\pm J_2^\matt)$ is non-negative, as claimed in the last paragraph.   

More generally, the operators $J_a^R$ act on $T(t)$ by
\be\label{motto} [J_a^R,T]=-\i f_a(T), \ee
where $f_a(T)=(1,\cos T,\sin T)$, and the same logic implies that
\be\label{zotto} [J_a^\matt, \Phi(T)]=+\i f_a(T) \partial_T \Phi(T). \ee
One might worry that the relative sign between eqn. (\ref{zotto}) and eqn. 
(\ref{motto}) would spoil the $\tSL(2,\R)$ commutation relations, but actually
this sign is needed for the commutation relations to work out 
correctly.\footnote{Concretely, we have $[J_a^\matt,[J_b^\matt,\Phi(T) ]]=-f_b \partial_T (f_a\partial_T \Phi)$,
leading to $$[J_a^\matt,[J_b^\matt,\Phi(T)]]-[J_b^\matt, [J_a^\matt,\Phi(T) ]] =
+ (f_a \partial_T f_b -f_b\partial_T f_a) \partial_T \Phi(T).$$  
 By contrast, $$[J_a^R,[J_b^R,T]]-[J_b^R,[J_a^R,T]]=- (f_a\partial_T f_b-f_b\partial_T f_a). $$  The commutation relations are 
 satisfied, since the signs on the right hand sides of 
those two formulas are opposite, like the  signs on the right hand sides of (\ref{motto}) and (\ref{zotto}). }

We will describe in a little more detail the relation of boundary operators of the matter system to bulk quantum fields.  Typically
in the AdS/CFT correspondence, with a metric along the boundary
of the local form $\frac{1}{r^2}(-\d T^2+\d r^2)$, if a bulk field $\phi(r,T)$ vanishes for $r\to 0$ as $r^\Delta$,
then a corresponding boundary operator $\Phi_\Delta$ of dimension $\Delta$ is defined by 
\be\label{adscftrel} \Phi_\Delta(T)=\lim_{r\to 0} r^{-\Delta}\phi(r,T).\ee 
In the context of JT gravity coupled to matter, we want to view both $r$ and $T$ as functions of the time $t$ of the boundary quantum mechanics.
Moreover, since the $\tSL(2,\R)$ symmetry is spontaneously broken along the boundary by the cutoff field $\chi$,  it is possible to define
the boundary operator to have dimension 0, not dimension $\Delta$.   
The starting point in our present derivation was
the $\AdS_2$ 
metric $\d\sigma^2-\cosh^2\sigma\, \d T^2$, which for $\sigma\to\infty$ can be approximated as $\frac{1}{r^2}(-\d T^2+\d r^2)$ with 
$r= 2e^{-\sigma}= \frac{\epsilon}{\tphi_b}  e^{\chi}$.   So $r^{-\Delta}\phi(r,T)= \left(\frac{\epsilon}{\tphi_b}\right)^{-\Delta} e^{-\Delta \chi(t)} \phi(\chi(t),T(t)).$
Since $e^{-\Delta \chi(t)}$ is already one of the observables in the boundary description (before imposing constraints),  we can omit this factor and define
\be\label{opdef} \Phi(t) =  \left(\frac{\epsilon}{\tphi_b}\right)^{-\Delta}  \phi(\chi(t),T(t))\ee
as a boundary observable.    The advantage is that $\Phi(t)$ defined this way is $\tSL(2,\R)$-invariant.   

Before imposing constraints, it is manifest that the left Hamiltonian $H_L$ commutes with operators inserted on the right boundary, and vice-versa.
The same is therefore also true after imposing constraints. Explicitly, at $T_R = 0$, 
\begin{align}
[J_1^\matt,\phi(\chi(t),T(t))] = [J_2^\matt,\phi(\chi(t),T(t))] = +i \partial_T \phi(\chi(t),T(t)), 
\end{align}
while 
\begin{align}
[\t p_\chi, \phi(\chi(t),T(t))] = - [J_3^\matt,\phi(\chi(t),T(t))] = - i \Delta\, \phi(\chi(t),T(t)). 
\end{align} $H_L$ is constructed from $\t p_\chi+J_3^\matt$, $J_1^\matt-J_2^\matt$, and $e^\chi$, all of which commute with $\phi(\chi(t),T(t))$.
So $[H_L, \phi(\chi(t),T(t))] = 0$.

\section{The Algebra}\label{algebra}

In the rest of this paper, we will study the algebra of observables in JT gravity, in general coupled to a matter theory.

In quantum field theory in a fixed spacetime $M$, one can associate an algebra $\A_\uU$ of 
observables to any open set $\uU$ in spacetime.   In a theory of gravity, one has to be more careful,
since spacetime is fluctuating and in general it is difficult to specify a particular region in spacetime.  To the extent that 
fluctuations in the spacetime are small, one has an approximate notion of a spacetime region and a corresponding
algebra.   In JT gravity, however,
at low temperatures or energies, the spacetime fluctuations are not small, so we cannot usefully define an algebra associated to a general bulk spacetime region.

Instead, as in the AdS/CFT correspondence, we can define 
an algebra of boundary observables.   In the AdS/CFT correspondence,
this would be an algebra of observables of the conformal 
field theory (CFT) on the boundary, possibly restricted to a region of the boundary.    In favorable cases,
one has some independent knowledge of the boundary CFT.    In JT gravity coupled to a two-dimensional quantum field theory, 
there is not really a full-fledged boundary quantum mechanics, since there is no one-sided Hilbert space.     
But one can nevertheless define an algebra of boundary observables.   More precisely, one can define algebras $\A_R$ and
$\A_L$ of observables on the right and left boundaries.   These will be the main objects of study in the rest of this article.

\subsection{Warm up: Pure JT Gravity}
 Before considering theories with matter, it is helpful to first study the simpler case of pure JT gravity. As we saw in section \ref{constraints}, 
 even in pure JT gravity, imposing the $\tSL(2,\R)$ constraints on the Hilbert space required working with coinvariants. At the level of operators, 
 however, imposing the constraints simply means restricting to operators that commute with the group of constraints.    

We would like to associate subalgebras $\mathcal{A}_R$ and $\mathcal{A}_L$ of gauge-invariant operators to the right and left boundaries. Classically,  in JT gravity without matter, 
an observable on the right boundary is an  $\tSL(2,\R)$-invariant function on the unconstrained phase space $\varPhi_R$ of the right boundary.     Here $\varPhi_R$ is four-dimensional, and the constraint group is 
three-dimensional, so the quotient $\Lambda_R =\varPhi_R/\tSL(2,\R)$ is 
one-dimensional.   So classically, the algebra of $\SL(2,\R)$-invariant functions on $\varPhi_R$ is generated
by a single function that parametrizes $\Lambda_R$.  For this function, we can choose the Hamiltonian $H_R$. 
Similarly, the algebra of invariant functions on the left boundary is generated by $H_L$.
$H_R$ and $H_L$ are equal in classical JT gravity without matter after imposing the constraints \cite{HJ,KS,ZY,MLZ,JK}.  

 All of these statements remain valid quantum mechanically. The only gauge-invariant right and left boundary operators are functions 
 of the Hamiltonians $H_R$ and  $H_L$ respectively, which are equal as operators on the
 constrained Hilbert space (as we saw in section \ref{constraints}). Thus
 in JT gravity without matter, the boundary algebras $\A_L$ and $\A_R$ are commutative and  equal and generated only by $H=H_R=H_L$. Because $H$ has a nondegenerate
  spectrum, any operator that commutes with $H$ is actually a function of $H$ and is contained in both $\A_L$ and $\A_R$.   So 
  the algebras $\A_L$ and $\A_R$ are commutants, meaning that $\A_R$ is the algebra of operators that commute with $\A_L$, and vice-versa.

Given any algebra $\A$, ``states'' on $\A$ are defined to be normalized, positive linear functionals -- linear maps from $\A$ to complex-valued 
``expectation values'' such that positive operators have real positive expectation values and the expectation value of the identity is 1. Because the algebra 
$\A_R$ is classical, these states are in fact in one-to-one correspondence with probability distributions $p(H_R)$, where the expectation value of a function $f(H_R)$ is
\begin{align}
 \langle f(H_R) \rangle_p = \int_0^\infty dH_R\, p(H_R) f(H_R).
\end{align}

It is natural to ask whether one can define a notion of entropy for such states, and indeed one can. An obvious definition is the continuous (or differential) Shannon entropy
\begin{align}\label{eq:contshannonentropy}
S(p) = - \int_0^\infty dH_R\, p(H_R) \log p(H_R).
\end{align}
There are two problems with this definition, however. The first problem is that it gives completely different answers to those given by Euclidean path integral computations. The second, related problem is that the continuous Shannon entropy is not invariant under reparameterisations where $H_R$ is replaced by $\t H_R = g(H_R)$ for some arbitrary invertible function $g$. The probability distribution $\t p(\t H_R)$ for $\t H_R$ by definition satisfies
\begin{align}
\langle f(H_R) \rangle_p = \int dH_R\, p(H_R) f(H_R) =  \int d\t H_R \,\t p(\t H_R) f(g^{-1}(\t H_R)).
\end{align}
However, this means that the continuous Shannon entropy
\begin{align}
\t S(p) = - \int d \t H_R\, \t p(H_R) \log \t p(H_R) = - \int d H_R\, p(H_R) \log \left(\left[\frac{d g}{d H_R}\right]^{-1} p(H_R)\right)
\end{align}
defined using $\t H_R$ does not agree with the entropy (\ref{eq:contshannonentropy}) defined using $H_R$. In fact this second problem mildly ameliorates the first: if we choose $g$ to be the integral of the Euclidean density of states then one obtains the ``correct'' Euclidean answer for the entropy. However there is nothing within the canonically quantised theory that picks out this choice of $g$.  Without additional input from Euclidean path integral calculations, any other choice appears equally valid. 

The origin of this ambiguity can be understood as follows. A state $p$ is a linear functional on an algebra $\A$. However to define an entropy we need to associate to this state an operator $\rho \in \A$ that is normally called the density matrix of $p$. The state $p$ and the density matrix $\rho$ are related by
\begin{align}
\langle \a \rangle_p = \Tr\,[\rho \a].
\end{align}
for any $\a \in \A$. Here the trace $\Tr$ on the algebra $\A$ is some faithful positive linear functional\footnote{Here faithful means that the trace of any nonzero positive operator is nonzero. This condition is required to ensure the existence and uniqueness of $\rho$.} on $\A$ such that
\begin{align} \label{eq:trace}
\Tr[\a\bb] = \Tr[\bb\a]
\end{align}
for all $a ,b \in \A$. The entropy is then defined by the usual formula
\begin{align}
S(p) = - \Tr[\rho \log \rho] = - \langle \log \rho\rangle_p
\end{align}
However, for a commutative algebra such as $\A_R$, the condition (\ref{eq:trace}) is trivial. As a result, any faithful positive linear functional is a valid choice of trace. The particular trace being used needs to be specified as part of the definition of the entropy $S(p)$. For example, if we define the trace by
\begin{align} \label{eq:traceH_R}
\Tr[f(H_R)] = \int d H_R f(H_R),
\end{align}
then the density matrix of a state $p$ is simply the probability distribution $\rho = p(H_R)$ viewed as an operator in $\A_R$. We find that $S(p)$ is the continuous Shannon entropy with respect to $H_R$.    If (\ref{eq:traceH_R}) is replaced by some other positive linear functional (e.g. by replacing $H_R$ by $\t H_R$), then one can obtain other definitions of entropy (one for each choice of functional), including e.g. the Euclidean definition. In the absence of a preferred choice of trace included as an independent element of the theory, all of these definitions are equally natural.\footnote{From an algebraic perspective, the defect operators of \cite{KS,JK} play exactly this role; they are additional structure added to the theory that picks out a preferred choice of trace.}

\subsection{Definition using canonical quantisation}\label{lordef}

The fact that the boundary algebras in pure JT gravity have a nontrivial center in the intersection $\A_L \cap \A_R$ is in contrast with the general expectation in AdS/CFT that each asymptotic boundary constitutes an independent set of degrees of freedom; it has therefore been dubbed the factorisation problem \cite{HJ}.\footnote{We are using a convention here suggested by Henry Maxfield where different spellings are used to contrast this problem with the (related) factorization problem, where spacetime wormholes cause partition functions not to factorize on a set of disconnected asymptotic spacetime boundaries.} As we shall now see, adding matter to the theory replaces the commutative boundary algebras by Type II$_\infty$ von Neumann factors. The center is thus rendered trivial, although,  because the algebras are Type II rather than Type I,
 the Hilbert space does not factorize into a tensor product of Hilbert spaces on each boundary, as would be expected in full AdS/CFT at finite $N$,

On the right boundary, we have the Hamiltonian $H_R$ and also the QFT  observables $\Phi(t)$ at an arbitrary
value of the quantum mechanical time $t$, inserted at the corresponding point $(\chi(t),T(t))$ on the right boundary.
These operators generate the right algebra $\A_R$.
Of course,  $H_R$ generates the evolution in $t$:
\be\label{evt} \Phi(t) =e^{\i H_R t}\Phi(0) e^{-\i H_R t}.    \ee
However, in order to make possible simple general statements, we want to define $\A_R$ as a von Neumann algebra, acting
on the Hilbert space $\H$ that was analyzed in sections \ref{constraints}, \ref{incmatter}.
For this, we should consider not literally $H_R$ and $\Phi(t)$ but bounded functions of those operators.
  Examples of bounded functions of $H_R$ are $e^{\i H_R t}$
and (since $H_R\geq 0$) $\exp(-\beta H_R)$, with $t\in\R,\,\beta>0$.  For $\Phi(t)$, matters are more subtle.
Experience with ordinary
quantum field theory (in the absence of gravity) indicates that expressions such as $\Phi(t)$ are really operator-valued
distributions, which first have to be smeared to define an operator (a densely defined unbounded operator, to be precise); 
then one can consider bounded functions of such operators.
One can smear in real time, defining
\be\label{zolt} \Phi_f =\int \d t \,f(t) \Phi(t),\ee
where $f(t)$ is a  smooth function of compact support, or one can smear by imaginary time evolution, defining
$\Phi_\epsilon(t) =\exp(-\epsilon H_R) \Phi(t) \exp(-\epsilon H_R),$ $\epsilon>0$.  

Similarly, the left boundary $\A_L$ is generated by bounded functions of $H_L$ and matter operators $\Phi_L(t)$, inserted at the position $(\chi'(t),T'(t))$ of the left boundary at quantum mechanical time $t$.   

We would like to establish a few basic facts about these algebras:

(1) They commute with each other; more specifically the commutant of $\A_L$, which is defined as
the algebra $\A_L'$ of all bounded
operators on $\H$ that
commute with $\A_L$, satisfies  $\A_L'=\A_R$, and likewise $\A_R'=\A_L$.

(2)  In the absence of matter, $\A_R$ and $\A_L$ were commutative, with the single generators $H_L=H_R$.    However,
after coupling to a matter QFT that satisfies reasonable conditions, we expect that $\A_R$ and $\A_L$ become
``factors,'' 
 meaning that their centers are trivial, and consist only of complex scalars.  
 
 (3) In the presence of matter, $\A_R$ and $\A_L$ are algebras of Type II$_\infty$.  (In the absence of matter, they are,
 as just noted, commutative, and therefore are direct integrals of Type I factors.)

Some of these assertions are most transparent in the context of a Euclidean-style construction of the algebras which we
present in section \ref{euclideanstyle}.    Here we will make some general remarks.

$\A_L$ is generated by left boundary operators at time zero, together with $H_L$. 
We do not need to include $\Phi(t)$ for $t\not=0$ as an additional generator, since it is obtained from $\Phi(0)$ by conjugation by
$e^{\i t H_L}$.
  Similarly, $\A_R$ is generated
by right boundary operators at time zero together with $H_R$.    But at time zero, the matter operators and Hamiltonian
 on the left boundary commute with the matter operators and the Hamiltonian on the right boundary, and vice-versa.
 This statement is true even before imposing constraints.   So $\A_L$ and $\A_R$ commute, a statement that is conveniently written
 $[\A_L,\A_R]=0$.
  As was already explained in section \ref{constraints},  the assertion $[\A_L,\A_R]=0$ 
  is a statement of causality, a quantum version of the statement that the left and right
boundaries are out of causal contact.

  The sharper statement  $\A_L=\A_R'$, $\A_R=\A_L'$
means that the set of operators generated by $\A_L$ and $\A_R$ together is complete, in the sense that the
algebra $ B(\H)$ of all bounded operators on $\H$ is the same as the algebra $\A_L\vee \A_R$ generated by 
$\A_L$ and $\A_R$ together.
Semiclassically, one might think that this is not the case, since JT gravity coupled to matter can describe long wormholes, and one
might think that operators acting deep in the interior of the long wormhole, far from the horizons of an observer on the left or right
side, would not be contained in $\A_L\vee \A_R$.    Entanglement wedge reconstruction, however, motivates the idea
that the algebra $\A_L\vee \A_R$ is nevertheless 
complete, with $\A_L$ accounting for operators that act to the left of the RT or HRT surface,
and $\A_R$ accounting for operators that act to the right.   But entanglement  wedge reconstruction is really only formulated
and understood in semiclassical situations, that is, under the assumption that there is a definite semiclassical spacetime, to a good
approximation.   In JT gravity coupled to matter, at low temperatures or energies, that is far from being the case.
Thus the statement $\A_L'=\A_R$, $\A_R'=\A_L$ can be viewed as being at least a partial counterpart of
 entanglement wedge reconstruction that holds
even without a semiclassical picture of spacetime.

The relation to entanglement wedge reconstruction -- which is a very subtle, nonclassical  statement in the case that
 a  long wormhole is present -- 
suggests that there will be no immediate, direct argument to show that $\A_L'=\A_R$, $\A_R'=\A_L$.   However, these facts
will  be evident in the Euclidean-style approach.

Now we discuss the question of the centers of the algebras $\A_R$, $\A_L$.   For it to be true that these algebras have
trivial center after coupling to a bulk QFT, it has to be the case that the QFT itself does not have any boundary operators
that are central. 
(A non-trivial condition is needed, because abstractly we could tensor a matter QFT on $\AdS_2$  with a topological field
theory that lives only on the conformal boundary of $\AdS_2$ and that might have central operators.)   
For example, we expect that there are no central boundary operators if all boundary operators $\Phi(t)$ of the QFT are limits of bulk operators
$\phi(r,T)$ by the limiting procedure described in eqn. (\ref{opdef}).    In that case, operator products
such as $\Phi(t)\cdot \Phi(t')$ will inherit short distance 
singularities from the singularities of bulk operator products $\phi(r_1,T_1)\cdot \phi(r_2,T_2)$, 
and so $\Phi(t)$ will be non-central.   These short distance singularities also imply that $\Phi(t)$
depends nontrivially on $t$, implying after coupling to JT gravity that $H_R$ does not commute with $\Phi(t)$ and is
non-central.   

Of course, one might ask whether $\mathcal{A}_R$ contains some other more complicated operator
that is central. We do not have a formal proof that no such operator exists (other than $c$-numbers), but we find the possibility that one does highly implausible 
on general physical grounds.  The dynamics of JT gravity are chaotic, which should mean that there are no conserved charges except for obvious ones. 
A central operator would be much more special than a new conserved quantity, since a conserved quantity only needs to commute with the 
Hamiltonian, while a central operator has to commute with every element of the algebra. A more precise argument can be made in the high-energy limit, 
where the fluctuations of the boundary particle become small. In that limit, the algebra $\A_R$ becomes the crossed product of the algebra of bulk QFT 
operators in the boundary causal wedge by its modular automorphism group \cite{GCP, CPW}. And one can prove that this crossed product algebra has 
trivial center whenever the bulk QFT algebra is a Type III$_1$ von Neumann factor. As a result, any hypothetical central operator in $\A_R$ would have to act trivially at high energies.

Finally we discuss the assertion that in the presence of matter, $\A_R$ and $\A_L$ are of Type II$_\infty$.   Once one
knows that $\A_R$ or $\A_L$ is a factor, to assert that it 
  is  of Type II$_\infty$ means that on this algebra one can define a trace which
is positive but  is not defined for all elements of the algebra.\footnote{$\A_R$ and $\A_L$ are not of Type I,
since in JT gravity coupled to matter, there is no one-sided Hilbert space.}     Here a trace on an algebra $\A$ is a complex-valued linear function
$\Tr:\A\to \C$ such that $\Tr \, \a\a'=\Tr\,\a'\a$, $\a,\a'\in\A$; the trace is called positive if $\Tr\,\a\a^\dagger>0$ for all $\a\not=0$.  

We can argue as follows that the algebras $\A_R$ and $\A_L$ do have such a trace.    For this, we consider first the thermofield
double state of the two-sided system at inverse temperature $\beta$.  
Although JT gravity coupled to matter does not have a one-sided Hilbert space, there is a natural definition
in this theory of thermal expectation values of boundary operators.   For an operator $\a\in \A_R$ (or $\A_L$),
its thermal expectation value at inverse temperature $\beta$, denoted $\la \a\ra_\beta$, is defined by evaluating
a Euclidean path integral on a disc whose boundary has a renormalized length $\beta$, with an insertion of the operator
$\a$ on the boundary.   Alternatively, there is a thermofield double state $\Psi_\TFD(\beta)$ such that thermal expectation values
are equal to expectation values in the thermofield double state:
\be\label{tfdthermal} \la\Psi_\TFD(\beta) |\a|\Psi_\TFD(\beta) \ra =\la \a\ra_\beta. \ee  
  In the case of JT gravity with or without matter, the thermofield double description is not obtained by doubling anything,
since there is no one-sided Hilbert space. However, in the two-sided Hilbert space of JT gravity,
there is a state $\Psi_\TFD(\beta)$ that satisfies eqn. (\ref{tfdthermal}) \cite{HJ,ZY,KS,Saad}.    
It can be defined by a path integral on a half-disc
with an asymptotic boundary of renormalized length $\beta/2$ (and a geodesic
boundary on which the state is defined; see section \ref{euclideanstyle}). 
Defined this way,  $\Psi_\TFD(\beta)$ is not in general normalized, but satisfies $\la\Psi_\TFD(\beta)|\Psi_\TFD(\beta)\ra = Z(\beta)$ where $Z(\beta)$ is the Euclidean partition function on a disc with renormalized boundary length $\beta$.
Because this Euclidean path integral has no matter operator insertions, any matter fields present are  in the $\tSL(2,\R)$-invariant ground state $\Psi_\mathrm{gs}$.
Therefore, the thermofield double state in the presence of matter  is simply the tensor product of the thermofield double state $\Psi_\TFD(\beta)$ for pure JT gravity with $\Psi_\mathrm{gs} \in \H^\matt$.

 As in the case of an ordinary quantum system, expectation values in the thermofield double state satisfy a KMS condition:
\be\label{welco}\la\Psi_\TFD(\beta)|\Phi(t) \Phi(t')|\Psi_\TFD(\beta)\ra =\la\Psi_\TFD(\beta)|\Phi(t') \Phi(t+\i\beta)|\Psi_\TFD
(\beta)\ra. \ee 
More generally, for any $\a,\a'\in \A_R$, with the definition $\a(t)=e^{\i H_R t} \a e^{-\i H_R t}$, we have
\be\label{telco}   \la\Psi_\TFD(\beta)|\a\a'|\Psi_\TFD(\beta)\ra =\la\Psi_\TFD|\a'  \a(\i\beta)|\Psi_\TFD
(\beta)\ra. \ee   We define $\Tr\,\a= \lim_{\beta\to 0}\la\Psi_\TFD(\beta)|\a|\Psi_\TFD(\beta)\ra$ for any $\a\in \A_R$ such
that this limit exists.   
The limit certainly does not exist for all $\a$; for example, 
if $\a=1$, then $\la\Psi_\TFD(\beta)|\a|\Psi_\TFD(\beta)\ra$ is equal to the partition function $Z(\beta)$, which diverges for $\beta\to 0$.      
But it is equally clear
that there exist $\a\in \A_R$ such that the limit does exist.    For example, for $\a=e^{-\epsilon H_R}$, $\epsilon>0$, we get
$\lim_{\beta\to 0} \la\Psi_\TFD(\beta)|\a|\Psi_\TFD(\beta)\ra=\lim_{\beta\to 0} Z(\beta+\epsilon)=Z(\epsilon)$, so $\a$
(and similarly any operator regularized by a factor such as $\exp(-\epsilon H_R)$) has a well-defined trace.    For
operators such that the limits exist, the $\beta\to 0$ limit of the KMS condition shows that the function $\Tr$ satisfies the
defining property of a trace.  As for
positivity, one has 
\be\label{posit} \la \Psi_\TFD(\beta)|\a^\dagger \a|\Psi_\TFD(\beta)\ra=\la\a \Psi_\TFD(\beta )|\a\Psi_\TFD(\beta)\ra \geq 0,\ee
with vanishing if and only if $\a\Psi_\TFD(\beta)=0$.  
 Since $H_L$ commutes with $\a,\a^\dagger \in\A_R$, and $e^{-(\beta_2-\beta_1)H_L/2}\Psi_\TFD(\beta_1)=
\Psi_\TFD(\beta_2)$, we have 
\begin{align}\notag \la \Psi_\TFD(\beta_2)|\a^\dagger \a|\Psi_\TFD(\beta_2)\ra &= \la \Psi_\TFD(\beta_1)|\a^\dagger \exp(-(\beta_2 - \beta_1)H_L) \a|\Psi_\TFD(\beta_1)\ra \\ & \leq \la \Psi_\TFD(\beta_1)|\a^\dagger \a|\Psi_\TFD(\beta_1)\ra\end{align} for $\beta_2 > \beta_1$.   Hence (\ref{posit}) is a monotonically decreasing function of $\beta$. Thus as $\beta \to 0$, (\ref{posit}) always either converges to a finite positive limit or tends to positive infinity. We conclude that $\Tr( \a^\dagger \a) \in [0,+\infty]$ is in fact well defined in the extended positive real numbers for any positive operator $\a^\dagger \a$.}  We will argue in section \ref{euclideanstyle} that the algebras $\A_R, \A_L$ are cyclic-separating for $\Psi_\TFD(\beta)$. As a result, $\a\Psi_\TFD(\beta)=0$ implies $\a = 0$ and the trace $\Tr$ is faithful.   We should add that the existence of a faithful trace will anyway be perhaps
more obvious in section \ref{euclideanstyle}.

 There is an alternative definition of the trace $\Tr$ that was used in Appendix I of \cite{susywormholes} to give an algorithm for computing Euclidean disc partition functions from canonically quantised pure JT gravity (although the interpretation as an algebraic trace on the boundary algebras was not noted there). In the high temperature limit, the wavefunction $\Psi_\TFD(\beta)$ becomes tightly peaked as a function of $\chi$ around a saddle-point value $\chi_c$ such that $\chi_c \to \infty$ as $\beta \to 0$. Equivalently, it is peaked around a semiclassical renormalized geodesic length $\ell_c = - 2 \chi_c$ such that $\ell_c \to -\infty$ as $\beta \to 0$. As a result the trace of an operator $\a$ with matrix elements $\a(\chi_1,\chi_2) \in \B(\H^\matt)$ is given by
\begin{align} \label{eq:alttracedef}
\Tr(\a) = \lim_{\chi \to \infty}  \exp(\chi + 8e^{\chi})\la \Psi_\mathrm{gs}|\a(\chi,\chi) | \Psi_\mathrm{gs}\ra.
\end{align}
The correct scaling of the prefactor in (\ref{eq:alttracedef}) may be determined from the normalization of $\Psi_\TFD(\beta)$ as a function of the saddle-point value $\chi_c$ as $\beta \to 0$. Alternatively, it may be determined by analyzing the universal decay as $\chi \to \infty$ of the matrix elements of operators that e.g. project onto finite-energy states and hence should have finite trace.

Let us use this trace to compute the entanglement entropy of the thermofield double state $\Psi_\TFD(\beta)$, or, more precisely, of the normalized thermofield double state 
\begin{align}
\widehat\Psi_\TFD(\beta) =\frac{ \Psi_\TFD(\beta)}{Z(\beta)^{1/2}}.
\end{align}
 It follows from the definition using path integrals (and can be verified explicitly using the formulas from \cite{ZY}) that
\begin{align}
 e^{-\beta_1 H_R/2} \Psi_\TFD(\beta_2) = \Psi_\TFD(\beta_1 + \beta_2).
\end{align}
As a result, for any $\a \in \A_R$, we have
\begin{align}
\la  \Psi_\TFD(\beta) | \a | \Psi_\TFD(\beta) \ra &= \lim_{\beta' \to 0} \la  \Psi_\TFD(\beta') |  e^{-\beta H_R/2} \a  e^{-\beta H_R/2} | \Psi_\TFD(\beta') \ra
\\&= \Tr[e^{-\beta H_R/2} \a  e^{-\beta H_R/2}] = \Tr[e^{-\beta H_R} \a ].
\end{align}
We therefore conclude that the density matrix of the normalized thermofield double state $\widehat\Psi_\TFD(\beta)$ on $\A_R$ is $\rho = e^{-\beta H_R}/Z(\beta)$. The entropy of this state is
\begin{align}
S(\rho) = - \la \log \rho \ra = \la \beta H_R \ra + \log Z(\beta),
\end{align} 
which matches the Euclidean answer. 

 Crucially, unlike in JT gravity without matter, we did not need to add any additional ingredients by hand in order to obtain this result: if the algebra $\A_R$ is  a von Neumann factor, that is, its center is trivial,
 then the trace (if it exists) is unique up to rescaling.\footnote{More precisely, on a Type I or II factor, the trace is unique if one requires it to be normal and semifinite; see the discussion at the end of section \ref{operators} for details.} Consequently, the entropy formula derived here is unique up to an additive constant. Even though we used Euclidean path integrals as a convenient way of discovering the trace, the definition itself was forced upon us by the structure of the algebra.

Since the algebra is Type II$_\infty$, there is no canonical choice of normalization for the trace, and hence no canonical choice for the additive constant in the definition of entropy. There is a similar additive ambiguity in Euclidean path integral entropy computations. The JT gravity action includes a topological term that evaluates to $-S_0 \chi$ where $\chi$ is the Euler characteristic of the spacetime manifold. To remove contributions from higher genus spacetimes containing wormholes, one needs to take the limit $S_0 \to \infty$. This leads to a state-independent infinite contribution $S_0$ to the entanglement entropy, which describes the universal divergent entanglement of the Type II$_\infty$ algebra. To define a finite renormalized entanglement entropy we need to subtract this piece, which leads to the same additive ambiguity that we found above from an algebraic perspective.

\subsection{Definition using Euclidean path integrals}\label{euclideanstyle}

We now offer an alternative definition of the algebras $\A_L$ and $\A_R$ based on Euclidean path integrals. Although we will eventually argue that this definition is equivalent to the one given above, it is helpful because  a) it makes certain expected properties of $\A_R$ and $\A_L$ (such as the fact that they are commutants) easier to justify, and b) it justifies the use of Euclidean replica trick computations for computing entropies on $\A_R$ or $\A_L$. 

Our starting point is a formal algebra $\A_0$, built out of strings of symbols, each of which is either $e^{-\beta H}$, with some $\beta>0$, or else 
 one of the boundary operators $\Phi_\alpha$ of the matter system.   The two types of symbol are required to alternate and  the string is required to begin and
end with a symbol of the type $e^{-\beta H}$.    Thus here are some examples of allowed strings:
\begin{align} \label{strings} \notag &e^{-\beta H} \\ \notag  &e^{-\beta H}\Phi e^{-\beta' H} \\  &e^{-\beta_1 H}\Phi_1e^{-\beta_2 H}\Phi_2 e^{-\beta_3 H} .\end{align}
Strings are multiplied in an obvious way by joining them end to end and using the relation $e^{-\beta H}e^{-\beta' H}=e^{-(\beta+\beta')H}$.
Thus for example if $\S_1= e^{-\beta_1 H}\Phi_1e^{-\beta_2 H}$ and $\S_2=e^{-\beta_3 H}\Phi_2 e^{-\beta_4 H}$, then  
$\S_1\S_2 =e^{-\beta_1 H}\Phi_1 e^{-(\beta_2+\beta_3)H} \Phi_2 e^{-\beta_4 H}$.     Eventually, we will reinterpret these strings as the Hilbert space operators that  these expressions
usually represent, but to begin with we consider them as formal symbols.

We can define an algebra $\A_0$ whose elements are complex linear combinations of strings, multiplied as just explained. 
This is an algebra without an identity element; we could add an identity element as an additional generator of $\A_0$
but this will not be convenient.

 \begin{figure}
 \begin{center}
   \includegraphics[width=5in]{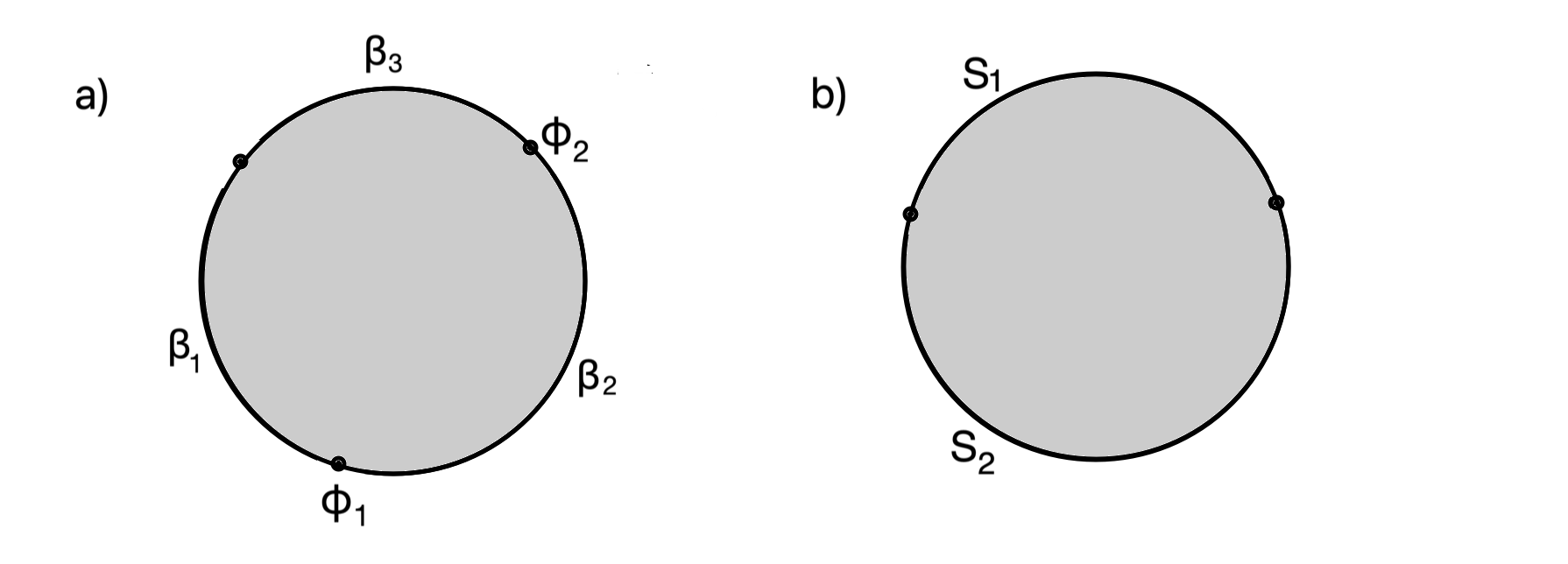}
 \end{center}
\caption{\footnotesize (a)   The path integral on a disc that computes $\Tr\,\S$ with   $\S= e^{-\beta_1 H}\Phi_1 e^{-\beta_2 H}\Phi_2 e^{-\beta_3 H}$.   The boundary of the disc is made of
three segments with renormalized lengths $\beta_1$, $\beta_2,$ and $\beta_3$.   At two junctions of segments, operators $\Phi_1$ and $\Phi_2$ are inserted.   At the
third junction, the two ends of $\S$ are joined together.   (b)   The path integral on a disc that computes $\Tr\,\S_1\S_2$.   The boundary of the disc consists of two
segments labeled respectively by $\S_1$ and by $\S_2$.   There is no intrinsic ordering of the two segments so $\Tr\,\S_1\S_2=\Tr\,\S_2\S_1$.     \label{AAA}}
\end{figure} 

The Euclidean path integral on a disc
 can be used to define a trace on the algebra $\A_0$.  
In this article, a disc path integral, when not otherwise specified, is a path integral on a disc whose boundary is an asymptotic boundary on which the boundary quantum
mechanics is defined.  Thus, in the limit that the usual cutoff is removed, the boundary of the disc is at conformal infinity in $\AdS_2$.
We do not assume time-reversal symmetry, so discs, and more general two-dimensional spacetimes considered later,
 are oriented, as are their boundaries.    In the figures,
the orientation runs counterclockwise along the boundary (thus, upwards or ``forwards in imaginary time'' on right boundaries
and downwards or ``backwards in imaginary time'' on left boundaries).   

  To define $\Tr\,\S$ for a string $\S$, we view $\S$, with its ends sewn
together, as a recipe
to define a boundary condition on the boundary of the disc.    
For example (fig. \ref{AAA}(a)), for the case $\S= e^{-\beta_1 H}\Phi_1 e^{-\beta_2 H} \Phi_2 e^{-\beta_3 H}$,
$\Tr\,S$ is computed by a path integral on a disc whose renormalized circumference is $\beta=\beta_1+\beta_2+\beta_3$, with insertions of the operators
$\Phi_1$ and $\Phi_2$ at boundary points separated by imaginary time $\beta_2$.   With this recipe, a simple rotation of the path integral picture shows that for any two strings
$\S_1$, $\S_2$, we have $\Tr\,\S_1\S_2=\Tr\,\S_2\S_1$  (fig. \ref{AAA}(b)).  Hence  $\Tr$ is indeed a trace. 

So far the elements of $\A_0$ are just symbols,  However, 
we  can extract more information from the path integral on a disc.    First, we define the ``adjoint'' $\S^\dagger$ of a string $\S$.  $\S^\dagger$ is defined
by reversing the order of the symbols in $\S$ and replacing each matter operator $\Phi$ with its adjoint $\Phi^\dagger$.  For example, the adjoint of 
$\S= e^{-\beta_1 H}\Phi e^{-\beta_2 H}$ is $\S^\dagger =e^{-\beta_2 H}\Phi^\dagger e^{-\beta_1 H}$.    So we can define a hermitian inner product on
$\A_0$ by $\la \S_1,\S_2\ra=\Tr\,S_1^\dagger S_2$.    We will see shortly that this inner product is positive semi-definite but has plenty of null vectors.  If $\N$ is the subspace of
null vectors, then $\A_0/\N$ is a vector space with a positive-definite hermitian inner product.   It can therefore be completed to a Hilbert space.

 \begin{figure}
 \begin{center}
   \includegraphics[width=4in]{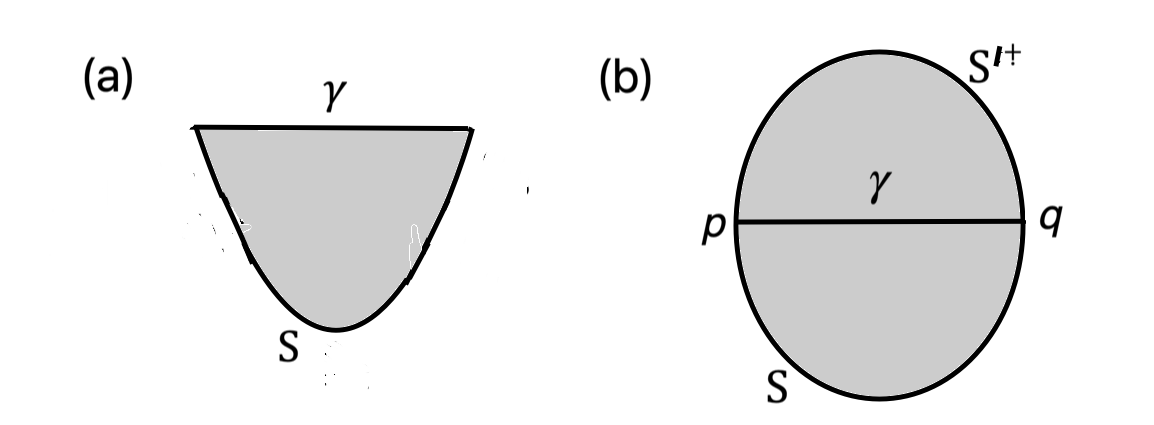}
 \end{center}
\caption{\footnotesize (a) The path integral on a half-disc that computes the map from a string $\S$ to a Hilbert space state $\Psi_\S$.  
The half-disc has an asymptotic boundary labeled by the string $\S$ and a geodesic boundary $\gamma$.
  (b) The path integral that computes $\la \S',\S\ra$ and can be used to demonstrate that the map $\S\to \Psi_\S$ from a string to a bulk state
preserves inner products.
  \label{BBB}}
\end{figure} 
But in fact, this Hilbert space  is none other than the 
Hilbert space $\H$ of JT gravity plus matter, described in section \ref{incmatter}.   We recall that an element of $\H$ is a square-integrable function $\Psi(\chi)$ that is valued
in the matter Hilbert space $\H^\matt$, where the renormalized length of a geodesic between the two boundaries is $\ell=-2\chi$; in other words, $\H=\H^\matt\otimes L^2(\R)$,
where $\chi$ acts on $L^2(\R)$ by multiplication.    A path integral on what we will call a half-disc gives a linear map $\S\to \Psi_\S\in \H$.   By a half-disc, we mean a disc
whose boundary consists of two connected components, one an asymptotic boundary on which the dual quantum mechanics is defined, 
and one an ``interior'' boundary at a finite distance. The structure of an asymptotic boundary is defined by a string.
Interior boundaries are always assumed to be geodesics. 
With this understanding, the path integral on a half-disc
 can be used to define
a linear map $\S\to \Psi_\S\in \H$ (fig \ref{BBB}(a)).  
We compute $\Psi_\S$  by a  path integral on a half-disc that has an asymptotic boundary determined by $\S$ and an interior   geodesic boundary of renormalized
length $\ell=-2\chi$.   For given $\chi$, the output of this path integral is a state in $\H^\matt$, and letting $\chi$ vary we get the desired state $\Psi_\S(\chi)\in\H$.
   
The map $\S\to \Psi_\S$ preserves inner products in the sense that
\be\label{innerrel} \la\S',\S\ra =\la \Psi_{\S'},\Psi_\S\ra,\ee
where the inner product on the left is the one on $\A_0$, and the inner product on the right is the one on $\H$.   To justify eqn. (\ref{innerrel}), we simply consider (fig. \ref{BBB}(b))
the path integral that computes $\la\S',\S\ra=\Tr\,\S'^\dagger \S$.  This is a path integral on a disc $D$ with an asymptotic boundary that consists of segments labeled respectively by
$\S$ and by $\S'^\dagger$, joined at their common endpoints  $p$, $q$.    In the standard procedure to analyze the path integral of JT gravity, possibly coupled to matter, the first step
is to integrate over the dilaton field.   This gives a delta function such that
 the metric on the disc becomes the standard $\AdS_2$ metric of constant negative curvature (cut off near the
conformal boundary, as reviewed in section \ref{review}).   In this metric, there is  a unique geodesic $\gamma$ from $p$ to $q$.   This geodesic divides $D$ into a ``lower'' part
$D_-$ and an ``upper'' part $D_+$.   The path integral on $D_-$ computes the ket $|\Psi_\S\ra$, the path integral on $D_+$ computes the bra $\la \Psi_{\S'}|$, and the integral
over degrees of freedom on $\gamma$ sews these two states together and computes their inner product $\la\Psi_{\S'},\Psi_\S\ra$.    So this establishes eqn. (\ref{innerrel}), which in 
particular confirms that the inner product $\la ~,~\ra$ on $\A_0$ is positive   semi-definite,

As an example of this construction, let $\S=e^{-\beta H/2}$.   The corresponding state $\Psi_\S=|e^{-\beta H/2}\ra$ is actually the thermofield double state of the two-sided
system, at inverse temperature $\beta$.   Indeed, for this choice of $\S$, the recipe to compute $\Psi_\S$ is just the standard recipe to construct the thermofield double
state by a path integral on a half-disc.  The thermofield double state was already discussed in section \ref{lordef}.

The map $\A_0\to \H$ is surjective, in the sense that states of the form $\Psi_\S$, $\S\in \A_0$ suffice to generate $\H$.   This is particularly clear if the matter theory is
a conformal field theory (CFT). Let $\Omega$ be the CFT ground state.
  The operator-state correspondence says that any state in $\H^\matt$ is of the form $\Phi|\Omega\ra$ for some unique local CFT operator $\Phi$.   A consequence is that states $\Psi_\S$ 
  for $\S$  of the highly
  restricted form $\S =e^{-\beta H/2} \Phi e^{-\beta H/2}$ actually suffice to generate $\H$.   Indeed, we can choose $\Phi$ to generate any desired state of the matter system,
  multiplied by a function of $\chi$ that depends on $\beta$.\footnote{One has to be slightly careful here because the operators $e^{-\beta H/2}$ act nontrivially on the matter Hilbert space. As a result, the reduced state of $\Psi_\S$ on $\H^\matt$ will not necessarily be the state dual to $\Phi$. However we do not expect this fact to alter the basic conclusion that a dense set of states in $\H$ can be prepared using strings $\S$ of the form described above.}   Taking linear combinations of the states we get for different values of $\beta$, we can approximate any desired
  function of $\chi$; consequently, states $\Psi_\S$ for $\S$ of this restricted form suffice to generate $\H$.   All of the other strings that we could have used, with more than
  one CFT operator, are therefore redundant in the sense that they do not enable us to produce any new states in $\H$.   So the map $\A_0\to \H$ has a very
  large space $\N$ of null vectors, as asserted earlier.

   \begin{figure}
 \begin{center}
   \includegraphics[width=3in]{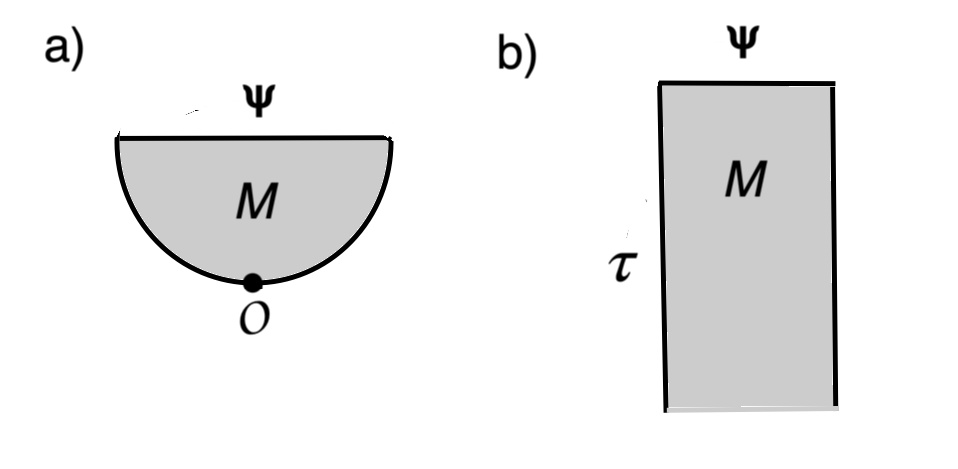}
 \end{center}
\caption{\footnotesize  Two views of a spacetime $M$ which is half of AdS$_2$ (in Euclidean signature).   Viewing AdS$_2$ as a hyperbolic disc, half of AdS$_2$ is the half-disc
shown in (a); on the other hand, the AdS$_2$ metric can be put in the static form $\d\sigma^2+\cosh^2\sigma \d\tau^2$, and in this form, half of AdS$_2$ looks like a semi-infinite 
strip with $\tau\leq 0$, as shown in (b).  
 \label{YYY}}
\end{figure} 

  Even if the matter system is not conformally invariant, the same idea applies, basically because the $\tSL(2,\R)$ symmetry of $\AdS_2$ is the conformal group of the 
  boundary.   The relevant facts are actually familiar in the AdS/CFT correspondence, where typically the bulk theory is not at all 
  conformally invariant but the boundary theory is conformally invariant, and any bulk state can be created by a local operator on the boundary.
  In our context, this reasoning applies to the matter sector, which possesses unbroken $\tSL(2,\R)$ symmetry (not to the full system including JT gravity).
  The basic setup is depicted in  fig. \ref{YYY}, which shows two views of a spacetime $M$ that is half of Euclidean AdS$_2$.   
   For any matter QFT, the path integral in  in (a) gives  a map from a local operator $O$
inserted at on the conformal boundary, as shown, to a bulk state $\Psi$ observed on the upper, geodesic boundary of $M$.   From (b), we can get a map
in the opposite direction.  Suppose that the state $\Psi$ is an energy eigenstate with energy $E_0$.   Cut off the strip by restricting to the range
$-\tau_0\leq \tau\leq 0$ and input the state $\Psi$ at the bottom of the strip. The path integral in the strip will then give back the same state $\Psi$ at the top,
multiplied by $e^{-\tau_0 E_0}$.  To compensate for this, multiply the path integral in the strip by $e^{+\tau_0 E_0}$.  Then upon taking the limit
$\tau_0\to\infty$, the picture in (b) becomes equivalent to the one in (a), with a state inserted in the far past turning into a local operator $O$ inserted on the boundary.

  Now we want to show that the quotient of $\A_0$ by its subspace of null vectors, namely $\A_1=\A_0/\N$, is an algebra in its own right and has a trace.
  To show that the linear function $\Tr:\A_0\to \C$ makes sense as a function on $\A_1$, one needs to show that for $\S\in \A_0$, $\Tr\,\S$ is invariant under $\S\to \S+\S_0$ with $\S_0\in\N$.   In other words, one has
  to show that $\Tr\,\S_0=0$.   $\S_0$ being null means $(\S_1,\S_0)=0$ for any $\S_1$.   In particular, taking $\S_1=e^{-\epsilon H}$, we have
  $0=\la e^{-\epsilon H},\S_0\ra=\Tr\,e^{-\epsilon H}\S_0$, and hence
  \be\label{vantrace}0=\lim_{\epsilon\to 0} \Tr\,e^{-\epsilon H} \S_0=\Tr\,\S_0,\ee as desired.

  What is involved in showing that  $\A_1=\A_0/\N$ is an algebra in its own right?
  Consider two equivalence classes in $\A_0/\N$ that can be represented by elements $\S_1,\S_2\in\A_0$.    To be able to consistently multiply equivalence classes,
  we need the condition that if we shift $\S_1$ or $\S_2$ in its equivalence class by $\S_1\to \S_1+\S_0$ or $\S_2\to \S_2+\S_0$ where $\S_0$ is null,
  then $\S_1\S_2$ should shift by a null vector.   In other words, the condition we need is that if $\S_0$ is null, then $\S \S_0$ and $\S_0\S$ are null, for any $\S\in \A_0$.
  
 \begin{figure}
 \begin{center}
   \includegraphics[width=4in]{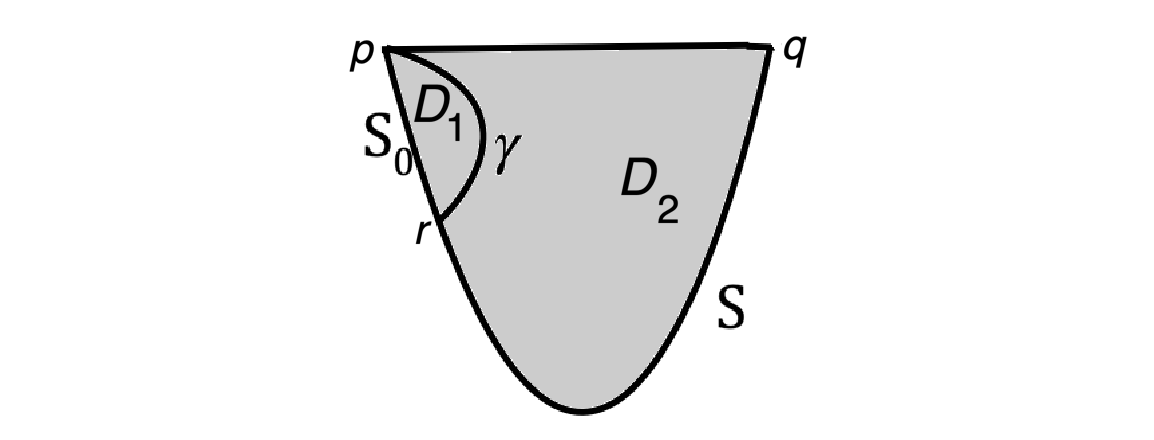}
 \end{center}
\caption{\footnotesize    Depicted here is a half-disc $D_0$ with an asymptotic boundary labeled by $\S_0\S$ and a geodesic boundary (the horizontal line at the top).  The path integral on $D_0$
computes $\Psi_{\S_0\S}$.    $\gamma$ is a geodesic
that connects the endpoints $p,r$ of the boundary segment labeled by $\S_0$.    If $\Psi_{\S_0}=0$, then the path integral in the region $D_1$ bounded
by $\S_0$ and $\gamma$ vanishes, regardless of the fields on $\gamma$, and therefore $\Psi_{\S_0\S}=0$.
 \label{CCC}}
\end{figure} 
To prove this, we consider the path integral on a half-disc $D_0$ that computes $\Psi_{\S_0\S}$.    We want
to show that if $\S_0$ is null, this path integral is identically zero, regardless of $\S$ and regardless of the renormalized length of the geodesic boundary of $D_0$.    The boundary of $D_0$
consists of a geodesic, say with endpoints $p$ and $q$, and an asymptotic boundary that is the union of two intervals labeled by $\S_0$ and by $\S$, which meet at a common
endpoint $r$ (fig. \ref{CCC}).    Let $pr$ be the segment labeled by $\S_0$.   The points $p$ and $r$ are joined in $D_0$ by a unique geodesic $\gamma$.   This geodesic divides
$D_0$ into two pieces.   One piece is a smaller half-disc $D_1$ whose asymptotic boundary is labeled by $\S_0$, and which has $\gamma$ for its geodesic boundary.  Let
$D_2$ be the rest of $D_0$.   The path integral on $D_0$ can be evaluated by first evaluating separately the path integrals on $D_1$ and on $D_2$, keeping fixed the fields   on $\gamma$
($\chi$ and
the matter fields),
and then at the end integrating over the  fields on $\gamma$.     The statement that $\S_0$ is null means that the path integral on $D_1$ vanishes, for any values of the fields on
$\gamma$.    Hence the path integral on $D_0$ vanishes, showing that $\S_0\S$ is null.   By similar reasoning, $\S\S_0$ is null if $\S_0$ is null.    Arguments similar to the one just explained
will recur at several points in this article.   

The function $\Tr:\A_1\to \C$ obeys the usual condition $\Tr\,\S_1\S_2=\Tr\,\S_2\S_1$, since this was already true on $\A_0$.   Moreover, $\Tr$ is positive as a function on $\A_1$,
in the sense that $\Tr\,\S^\dagger\S>0$ for all $\S\not=0$, since we have disposed of null vectors in passing to $\A_1$.

We can now reinterpret  strings as Hilbert space operators.   If $\S,$ $\T$ are strings, we say that $\S$ acts on the state $\Psi_\T$ by $\S\Psi_\T=\Psi_{\S\T}$.
  This definition is consistent, since if $\T$ is null (so that $\Psi_\T=0$), then $\S\T$ is also null (so $\Psi_{\S\T}=0$).   
   Since states $\Psi_\T$ are dense in $\H$ and the operators
  $\S$ defined this way are bounded,
  the rule $\S\Psi_\T=\Psi_{\S\T}$ completely defines $\S$ as an operator on $\H$.  
    Finally, since $\S\Psi_\T=\Psi_{\S\T}=0$ if $\S$ is null, the operator corresponding to $\S$ only depends on the equivalence class of $\S$ in $\A_1=\A_0/\N$.   Thus we get an action of
    $\A_1$ on the Hilbert space $\H$.
    
     \begin{figure}
 \begin{center}
   \includegraphics[width=4in]{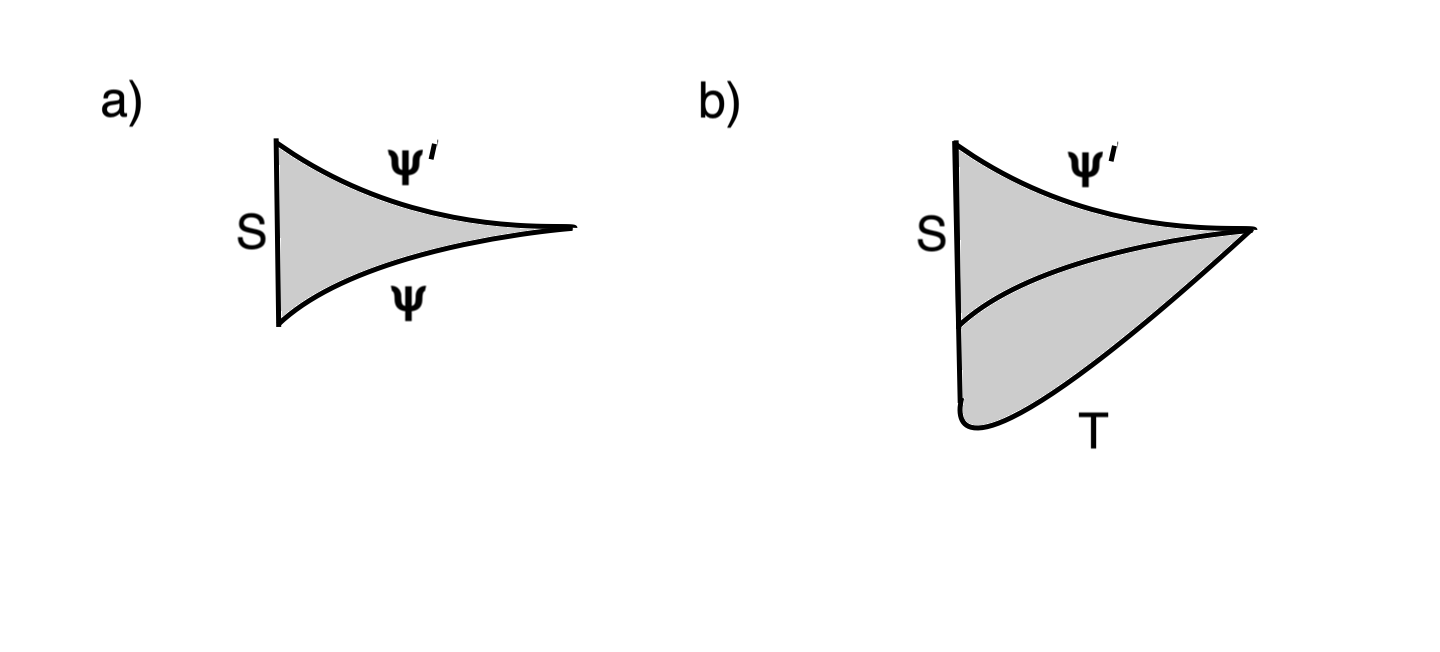}
 \end{center}
\caption{\footnotesize (a)   This figure shows the path integral that would be used to compute a matrix element $\la \Psi'|\S|\Psi\ra$ of a Hilbert space
operator corresponding to a string $\S$ between initial and final states $\Psi, \Psi'$ in the bulk Hilbert space $\H$.  $\Psi$ and $\Psi'$ are inserted on geodesic
boundaries that asymptotically meet at a point on the right boundary.  (b)   In the special case that the initial state is
$\Psi=\Psi_\T$, by gluing onto (a)   the path integral preparation of the state $\Psi_\T$, we get a representation of the matrix element
$\la\Psi'|\S|\Psi_\T\ra$.   But this  coincides with the path integral representation we would use for the inner product $\la\Psi'|\Psi_{\S\T}\ra$, showing that
the standard interpretation of $\S$ as a Hilbert space operator is consistent with $\S\Psi_\T=\Psi_{\S\T}$. Note that this picture can also be read to show
that if $\Psi_\T=0$ then $\Psi_{\S\T}=0$.  \label{DDD}}
\end{figure} 
   The operator that acts on $\H$ by $\Psi_\T\to \S\Psi_\T$ is actually the standard
  Hilbert space operator that one would associate to the string $\S$, acting on the left boundary of a two-sided spacetime.  That
  is true because the path integral rules that we have given agree with the standard recipe to interpret a string $\S$ as a Hilbert space operator.  
  To define $\S$ as an operator between states in $\H$, we would consider according to the standard logic a path integral on a hyperbolic two-manifold
  with geodesic boundaries on which initial and final states in $\H$ are inserted, and an asymptotic boundary labeled by $\S$ (fig. \ref{DDD}(a)).     This path integral will compute a matrix element of $\S$ between initial and final states in $\H$.
  Now if we want to let $\S$ act on $\Psi_\T$, we just glue onto the lower geodesic boundary in fig \ref{DDD}(a) the path integral construction of the state $\Psi_\T$, adapted
  from fig. \ref{BBB}(a).
  The resulting picture (\ref{DDD}(b)) is just the natural path integral construction of the state $\Psi_{\S\T}$.  So the rule $\S\Psi_\T=\Psi_{\S\T}$ agrees with the
  standard definition of a Hilbert space operator corresponding to $\S$, acting on the left boundary of a two-sided state.    
    To get operators acting on the right boundary, we would consider the operation $\Psi_\T\to \Psi_{\T\S}$.   This gives the commutant or opposite algebra, as we discuss presently.

At this stage, in particular we know that $\A_1=\A_0/\N$ is an algebra that acts on a Hilbert space $\H$.   We can therefore complete $\A_1$ to get a von Neumann algebra
$\A$ that acts on $\H$.   Although $\A_1$ does not contain an identity element, $\A$ does.   The reason for this is the following.   Although $\A_1$ does not contain an identity
element, it does contain the elements $e^{-\epsilon H}$ for arbitrary $\epsilon>0$.   When we complete $\A_1$ to get a von Neumann algebra, we have to include all operators
on $\H$ that occur as limits of operators in $\A_1$.  In particular, we have to include the identity operator $\1$, since it arises as $\lim_{\epsilon\to 0} e^{-\epsilon H}$.  
The reason that we did not include an identity operator in $\A_0$ at the beginning is that this would have prevented us from being able to define the map $\S\to \Psi_\S$,
since there is no Hilbert space state that corresponds to the identity operator $\1$.    Since the state that corresponds to $e^{-\beta H/2}$ is the thermofield double
state at inverse temperature $\beta$, a state $|\1\ra$ corresponding to $\1=\lim_{\beta\to 0} e^{-\beta H/2}$ would be the infinite temperature limit of the thermofield double state.
But there is no such Hilbert space state; its norm would be $\la\1,\1\ra=\lim_{\beta\to 0} \Tr\,e^{-\beta H}=\lim_{\beta\to 0} Z(\beta)=\infty$.
   Rather, one can interpret $|\1\ra$ as a ``weight'' of the von Neumann algebra $\A$, which means roughly that it is an unnormalizable
state that has well-defined inner products with a dense set of elements of $\A$.   Indeed, $\la\1,\S\ra=\Tr\,\S$ is well-defined for any $\S\in \A_1$, and by definition $\A_1$
 is dense in $\A$.   
 
 Since $\A_1$ has a trace that is positive-definite, the same is true of its completion $\A$.   However, taking the completion adds to $\A_1$ elements -- such as the identity
 element $\1$ -- with trace  $+\infty$.   Since the trace in $\A$ is accordingly not defined for all elements of $\A$, it follows  that $\A$ is of Type II$_\infty$, not Type II$_1$.
 $\A$ is not of Type I because there is no one-sided Hilbert space for it to act on. It is not of Type III because it has a trace.
 
 Now we can analyze the commutant $\A'$ of the algebra $\A$. What makes this straightforward is the close relation between $\A$ and $\H$: they were both obtained by completing
 $\A_1$, albeit in slightly different ways.  Let $\T$ be a linear operator on $\H$ that commutes with $\A$.  Consider any $\S,\U\in \A_1\subset \A$.
For $\T$ to commute with $\S$ as operators on $\H$ implies in particular that $\S \T \Psi_\U=\T\S\Psi_\U=\T\Psi_{\S\U}$.   Now set $\U=e^{-\epsilon H}$ and take the limit $\epsilon\to 0$.
In this limit, $\S\U\to\S$ and $\Psi_\U\to |\1\ra$, so we
 get $\T\Psi_\S= \S \T|\1\ra$.   We can approximate $\T|\1\ra$ arbitrarily well by $\Psi_\W$ for some $\W\in \A_1$, since states $\Psi_\W$ are dense in $\H$.   Hence
we learn that a dense set of operators in $\A'$ are operators that act by $\T\Psi_\S =\S \Psi_\W=\Psi_{\S\W}$ for some $\W\in\A_1$.   This means that right multiplication in $\A_1$ by
$\S\to \S\W$ gives a dense set of operators in $\A'$.   $\A'$ is the closure of this set.

 What is happening here is that there are always two commuting algebras that act on an algebra $\A$.   $\A$ can act on itself by left multiplication, and $\A$ acting on
 itself in this way commutes with another algebra $\A'$ that acts on $\A$ by right multiplication.   $\A'$ is isomorphic to what
  is called the opposite algebra of $\A$, sometimes denoted $\A^\op$.   
 Elements of $\A^\op$ are in one-to-one correspondence with elements of $\A$, but they are multiplied in the opposite order.  For $\S\in \A$, write $\S^\op$ for the corresponding element
 of $\A^\op$.    Multiplication in $\A^\op$ is defined by $\S^\op \T^\op =(\T\S)^\op$, which agrees with right multiplication of $\A$ on itself, showing that $\A'\cong \A^\op$. 
 The mathematical statement here is called the commutation theorem for semifinite traces.   It says that a von Neumann algebra $\A$ with semifinite trace $\Tr$ and the opposite algebra $\A^\op$ acting on it from the right are commutants on the Hilbert space $\H = \{ \a \in \A: \Tr\,\a^\dagger \a <\infty\}$.  

If a string $\S$ corresponds to an invertible operator (even if the inverse is an unbounded operator affiliated to $\A$ rather than an element of $\A$), the state $\Psi_\S$ is cyclic-separating for $\A$ and $\A'$; an example is
 $\S = e^{-\beta H/2}$ with $\Psi_\S$ the thermofield double state.
 
 The intersection $\A\cap\A'$ consists of operators that commute with $\A$ (since they are in $\A'$) and with $\A'$ (since they are in $\A$).  So the intersection
  is the common center of $\A$ and $\A'$.   Under hypotheses discussed in section \ref{lordef}, this common center is trivial, $\A\cap\A'=\C$.
 Since $\A$ and $\A'$ are von Neumann algebras that are commutants, a general theorem of von Neumann asserts that the algebra $\A\vee \A'$ generated by $\A$ and $\A'$
 together is the whole algebra $B(\H)$ of bounded operators on $\H$.   We will challenge this claim in section \ref{operators} by using baby universes to define what will
 appear to be operators on $\H$ that commute with both $\A$ and $\A'$.   This claim will turn out to fail in an instructive fashion.
 
 To complete the story, we would like to show that the algebras $\A$ and $\A'$ coincide with the algebras $\A_L$ and $\A_R$ that were defined in the Lorentz
 signature picture in section \ref{lordef}.   In one direction, this is clear.   
 $\A$ was defined as the smallest von Neumann algebra containing operators that correspond to the strings in eqn. (\ref{strings}), acting on the left side of a two-sided system.
 All these strings correspond to bounded operators built from $H$ and the matter operators $\Phi$. $\A_L$ was defined as the algebra of all bounded operators built from $H_L$
 and matter operators $\Phi_L$, acting on the left boundary.    So $\A\subset \A_L$.   Similarly $\A'\subset \A_R$.
 Since $\A$ and $\A'$ are commutants (meaning that they are each as large as they can be while commuting with the other),
 and $[\A_L,\A_R]=0$, it is impossible for $\A_L$ to be bigger than $\A$ or for $\A_R$ to be bigger than $\A'$.
 Thus $\A_L=\A$, $\A_R=\A'$.    
 
In this discussion, we started with an algebra $\A_0$ of strings and then we formally defined a state  $\Psi_\S$ for every $\S\in\A_0$.  At this level, then, there is trivially
a state for every element $\S\in \A_0$.
 Then we took a completion
of the space generated by the states $\Psi_\S$ to get a Hilbert space $\H$, and a completion of $\A_0$ 
to get the algebra $\A$.   One can ask whether after taking completions there is still a Hilbert space state for every element of the algebra.  The answer to this question is ``no,''
because the state formally associated to an algebra element $\x$ might not be normalizable.  For example, as we have already discussed, the state $|\1\ra$ that would be formally
associated to the identity element $\1\in \A$ is not normalizable and so is not an element of $\H$.    But this is the only obstruction.  Since the norm squared of a state
$|\x\ra$ corresponding to an algebra element $\x$ is supposed to satisfy $\la\x|\x\ra = \Tr\,\x^\dagger \x$, the necessary condition for the existence of a state $|\x\ra\in\H$ that corresponds
to an algebra element $\x$ is simply
\be\label{algcond}\Tr\,\x^\dagger\x<\infty. \ee

If such a state $|\x\ra$ does exist, then for every $\a\in\A$,
\be\label{migcond} \la\x|\a|\x\ra=\Tr\,\a \x\x^\dagger. \ee
This formula says that the density matrix of the state $|\x\ra$ on $\A_R$ is $\rho = \x \x^\dagger$. 
If $\x \in \A_0$ is a string, then the string describing $\x \x^\dagger$ is formed by concatenating $\x$ with a reversed-ordered copy 
of itself. Similarly, $\Tr\,\rho^n =  \Tr\,(\x \x^\dagger)^n$ is computed by evaluating a Euclidean path integral on a disc with boundary formed by gluing together 
$n$ copies of $\x \x^\dagger$.  It should be clear that the rule we have just described for computing $\Tr\,\rho^n$ using a 
Euclidean gravitational path integral is exactly the usual rule used in replica trick  entropy computations in Euclidean gravity. This rule is usually justified either by appealing to the AdS/CFT dictionary to relate the gravitational path integral to microscopic CFT entropy computations \cite{LM,M2}  or, in settings where no explicit microscopic theory is known, simply by its success in giving sensible answers \cite{GH}.   In contrast, we started with an explicit asymptotic boundary algebra $\A_R$ in a canonically quantised gravity theory. We argued that this algebra has (up to an additive constant) a unique definition of entropy. Finally, we showed that, given a state $\Psi$ prepared by some Euclidean path integral, 
we can compute the entropy of $\Psi$ on the algebra $\A_R$ -- in the canonically quantised theory -- using the usual rules for replica trick Euclidean gravity computations.

\section{Baby Universe ``Operators''}\label{operators}

Up to this point we have assumed the spacetime topology to be a disc (in Euclidean signature) or 
a strip (in Lorentz signature).   But in a theory of gravity, it is natural to consider
more general topologies.   An obvious direction, which we explore starting in section \ref{wormhole}, is to include wormholes and topology change in the dynamics.  First, 
however, we will consider wormholes and closed baby universes as external probes.    Via such probes, we can define what 
will appear at first sight to be operators with paradoxical properties.   
The paradox will be resolved in an instructive fashion.

     \begin{figure}
 \begin{center}
   \includegraphics[width=4in]{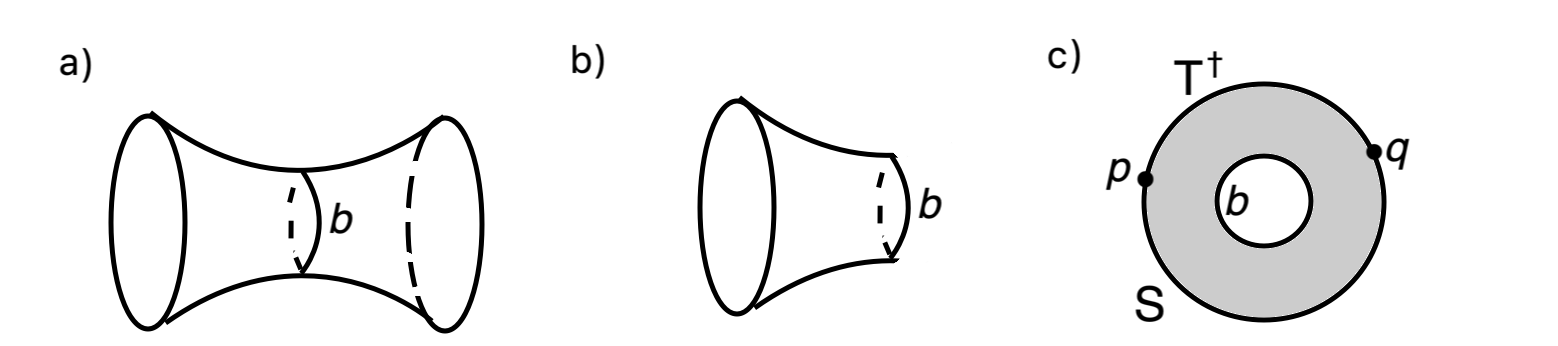}
 \end{center}
\caption{\footnotesize{ (a)  A double trumpet.  The boundary quantum mechanics is defined on the left boundary, and the boundary condition on the right  boundary
is chosen so the  internal geodesic has circumference $b$ and the matter fields are in state $\Lambda$.   
(b)  Such pictures can be abbreviated   by omitting a ``trumpet'' that connects to the external boundary (in this example, the omitted region is the portion to the right of the closed geodesic).
The omitted region can always be glued back in a unique way, and this is always assumed.
 (c) A face-on view of the ``trumpet'' in (b), which topologically is an annulus.   The inner boundary is a geodesic of circumference $b$ and the quantum mechanics is defined on the 
 outer boundary.   The outer boundary has been labeled by strings $\S$, $\T^\dagger$, separated by marked points $p,q$.
 The path integral on this Euclidean spacetime computes $\la \Psi_\T|\O_{b,\Lambda}|\Psi_\S\ra$. }   \label{A1}}
\end{figure}

   \begin{figure}
 \begin{center}
   \includegraphics[width=2.5in]{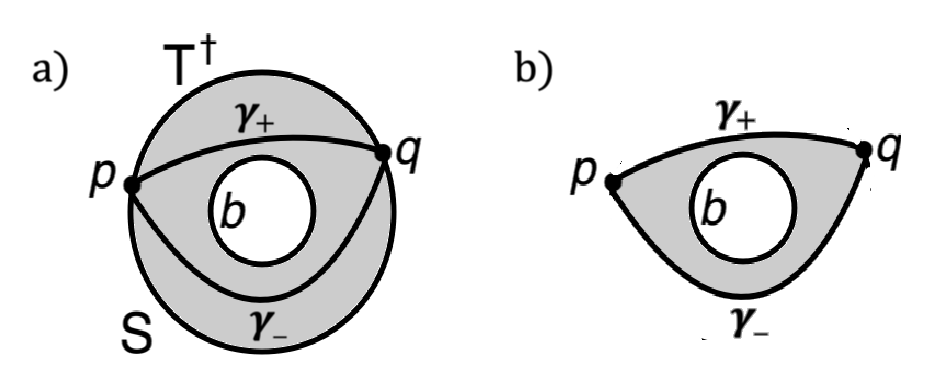}
 \end{center}
\caption{\footnotesize    (a)  To interpret the path integral in fig. \ref{A1}(b) as a matrix element $\la\Psi_\T|O_{b,\Lambda}|\Psi_\S\ra$, we introduce the indicated geodesics $\gamma_+$ and $\gamma_-$ 
 that go ``below'' and ``above'' the hole.   (b) The operator $\O_{b,\Lambda}$ is represented by the path integral in the ``middle region''
$\Sigma_0$ between $\gamma_+$ and $\gamma_-$.  The path integral in the region below $\gamma_-$ prepares the ket $|\Psi_\S\ra$, and that in the region above $\gamma_+$
prepares the bra $\la\Psi_\T|$.  \label{A2}}
\end{figure} 

To use  wormholes as  probes, we adapt to the present context a construction made  in \cite{MM}.    In Euclidean signature, 
instead of assuming spacetime to have just one asymptotic boundary on which the dual quantum mechanics
is defined, we add a second asymptotic boundary that creates a closed baby universe.   So the Euclidean spacetime becomes a ``double trumpet'' (fig. \ref{A1}(a)).
A hyperbolic metric on the double trumpet has a single real modulus, namely the circumference $b$ of the simple closed geodesic in its ``core.'' 
We assume that the boundary quantum mechanics is defined at the ``left'' end of the double trumpet ($\tau\to-\infty$); thus, at this end a cutoff of the usual type is imposed
near the conformal boundary.  Along the ``right'' boundary ($\tau\to+\infty$), which we will call an ``external'' boundary, we do not place such a cutoff, but instead impose a condition on 
the asymptotic behavior of the metric which ensures that the circumference
of the closed geodesic will be $b$.  Concretely, to do this, we observe that the hyperbolic metric of the double trumpet has a standard form
\be\label{stanform}\d s^2=\d \tau^2+\cosh^2\tau \d\phi^2,\ee
with $\phi\cong \phi+b$, $-\infty< \tau<\infty$.   The closed geodesic that is homologous to the 
boundary is at $\tau=0$ and its circumference is $b$.   Setting $y=\frac{b}{4\pi}e^{-\tau}$, $\sigma=\frac{2\pi}{b}\phi$,
the metric takes the form
\be\label{anform} \d s^2= \frac{\d y^2+ \d\sigma^2+ y^2\frac{b^2}{8\pi^2} \d\sigma^2+\O(y^4)}{y^2},\ee
and we see that $b$ can indeed be encoded in the coefficient of a subleading term of the metric near the conformal boundary at $y=0$. 
In pure JT gravity, the boundary condition that we want on the external boundary can be defined  just by fixing a value of $b$.    In JT gravity coupled to matter,
we additionally need a boundary condition on the matter fields.   Such a boundary condition can be determined by any rotation-invariant state $\Lambda$ in the closed universe
matter Hilbert space $\H^\matt_\cl$.
 By rotation invariance, we mean invariance under  $\phi\to \phi+{\mathrm {constant}}$.\footnote{\label{notinv} If $\Lambda$ is not invariant under shifts of $\phi$, then in doing the path integral on the double
 trumpet, we will have to integrate over a twist of the left of the double trumpet relative to the right, and this will  effectively replace $\Lambda$ by its rotation-invariant
 projection. In section \ref{closed}, we describe the closed universe Hilbert space in a more leisurely fashion.} 
  
We can slightly simplify the following  by ``cutting'' along the closed geodesic at $\tau=0$ and discarding the ``exterior'' piece ($\tau>0$).  Upon doing so,
$\Sigma$ becomes an ordinary trumpet (rather than a double trumpet) with a quantum boundary at big distances and a closed geodesic boundary of circumference $b$
on which the matter state $\Lambda$ is inserted (fig. \ref{A1}(b)).   Such truncations also exist  and are convenient in cases discussed later with multiple external boundaries.   In what follows,
we always draw the truncated version of the spacetime.

In section \ref{euclideanstyle}, given a pair of strings $\S,\T$, we defined inner products $\la \Psi_\T,\Psi_\S\ra$ via a path integral on a disc with its boundary labeled by $\T^\dagger\S$.
An obvious idea now is to consider a similar path integral on a Riemann surface $\Sigma$ that is a disc with a hole labeled by some $b,\Lambda$ (fig. \ref{A1}(c)).
$\Sigma$ has two marked points on its outer boundary, namely the endpoints $p,q$ of the seqment labeled by $\S$.   

 A natural expectation is that this path integral  can be interpreted
as $\la\Psi_\T|\O_{b,\Lambda}|\Psi_\S\ra$ for some operator $\O_{b,\Lambda}$.     To justify this expectation, we note (fig. \ref{A2}(a))   that the points $p,q$ are connected by a unique embedded geodesic $\gamma_-$  that goes
``below'' the hole and also by a unique embedded geodesic $\gamma_+$ that goes ``above'' the hole.  These geodesics exist because on a hyperbolic two-manifold,
there is always a unique geodesic in each homotopy class of paths; this is a fact that we will use repeatedly.  Correspondingly, $\Sigma$ is the union of a portion $\Sigma_-$ below $\gamma_-$,
a portion $\SIgma_+$ above $\gamma_+$, and a portion $\Sigma_0$ in between.   The path integral on $\Sigma_-$ computes $|\Psi_\S\ra$
and the path integral on $\Sigma_+$ computes $\la\Psi_\T|$, so to interpret the path integral on $\Sigma$ as a matrix element $\la\Psi_\T|\O_{b,\Lambda}|\Psi_\S\ra$,
the operator $\O_{b,\Lambda}$ has to be represented by the path integral on $\Sigma_0$.   
In fact, let $\ell_-$ and $\ell_+$ be the renormalized lengths of the geodesics $\gamma_-$ and $\gamma_+$.   The path integral on $\Sigma_0$ with specified
values of $\ell_-$ and $\ell_+$ computes the kernel $\O_{b,\Lambda}(\ell_+,\ell_-)$ in the length basis.    This kernel is an operator acting on the matter Hilbert space
$\H^\matt_\cl$, though we do not indicate that explicitly in the notation.   

The renormalized lengths  $\ell_+$ and $\ell_-$ are not uniquely determined by the complex structure of $\Sigma_0$; they depend also on the positions of the boundary
particles or in other words on the cutoffs near the boundary points $p,q$ (the cutoff variables were called $\sigma_L,\sigma_R$ or $\chi_L,\chi_R$ in section \ref{review}).   
However, the difference $\Delta\ell=\ell_+-\ell_-$ does not depend on the cutoffs and is a modulus
of the Riemann surface $\Sigma_0$.   The positions of the boundary particles, together with one modulus that $\Sigma_0$ has even if we neglect the boundary particles, determine the separate values of $\ell_+$ and $\ell_-$, and one additional length, which
one can take to be the renormalized distance from the left boundary to the interior geodesic.
To compute the kernel $\O_{b,\Lambda}(\ell_+,\ell_-)$, we keep $\ell_+$ and $\ell_-$ fixed and integrate only over the one additional modulus.

 However, consideration of the $\O_{b,\Lambda}$ leads to  an apparent contradiction.    As we explain momentarily, one can argue formally that the $\O_{b,\Lambda}$ commute
 with the boundary algebras $\A_L$, $\A_R$ that were introduced in section \ref{algebra}.   
 In the absence of matter, there is no problem with this. 
 In pure JT gravity without matter, the only relevant choice of $\Lambda$ is $\Lambda=1$, so we drop $\Lambda$ from the notation and consider the baby universe operators $\O_b$.
 $\A_L$ and $\A_R$ in the absence of matter are commutative algebras, generated by the Hamiltonian, so $\O_b$ commutes with $\A_L$ and $\A_R$ if and only if it
 is diagonal in the energy basis.   Indeed, a computation in \cite{Saad} (see eqn. (3.30) of that paper) shows that this is true.\footnote{{See also Appendix C of \cite{StanfordYang} for a computation of $\O_{b,\Lambda}(\ell_+,\ell_-)$ that more directly matches our discussion above.}}  More specifically, in pure JT gravity, if  $\Psi_E$ is a state of energy $E$, then
 \be\label{tmore} \O_b \Psi_E= \frac{{2\pi} \cos(b\sqrt{2E})}{\sqrt{2E} \,{\sinh(2\pi\sqrt{2E})}}\Psi_E. \ee

 For JT gravity coupled to matter, things are completely different.   In that case, 
the description of $\A_L$ and $\A_R$ in terms of
 left and right multiplication of an algebra $\A$ on itself makes it fairly clear that $\A_L$ and $\A_R$ are commutants -- apart from $c$-numbers, there are no operators on the
 Hilbert space $\H$ of the theory that commute with both of them.
 
  \begin{figure}
 \begin{center}
   \includegraphics[width=4in]{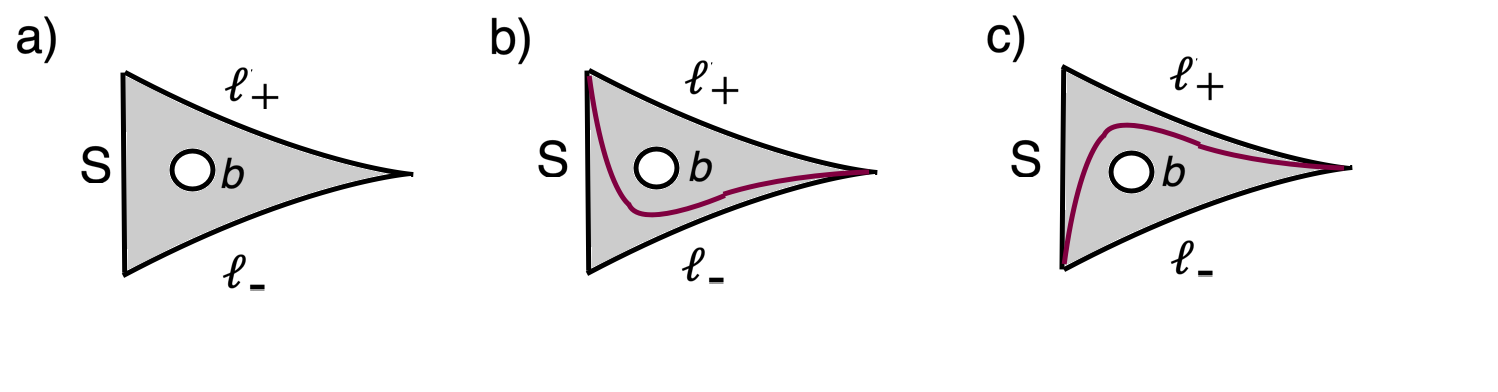}
 \end{center}
\caption{\footnotesize (a) Adding a geodesic hole of circumference  $b$ to the configuration of fig. \ref{DDD}(a), we get a Riemann surface that is a candidate
for describing the kernel of the operator product $\S \O_{b,\Lambda}$ or $\O_{b,\Lambda}\S$ in the length basis.   
(b) and (c)   By considering a geodesic that starts at the upper or lower corner of the
diagram and goes below or above the hole, we decompose the picture of (a) in two pieces that respectively describe $\S$ and $\O_{b,\Lambda}$.   This confirms that
the picture in (a) does compute this operator product and that $\S\O_{b,\Lambda}=\O_{b,\Lambda}\S$.
\label{A4}}
\end{figure} 

However, we can formally argue as follows that operators $\O_{b,\Lambda}$ commute with $\A_L$ and $\A_R$.  Since operators associated to strings are dense in $\A_L$,
 to  show that  $\O_{b,\Lambda}$ commutes with
$\A_L$, it is enough to show that it commutes with the element of $\A_L$ associated to any string $\S$ acting on the left.  That $\O_{b,\Lambda}$ commutes with $\A_R$ is 
proved in the same way.   

Given a string $\S$, how would we compute a product of operators $\S \O_{b,\Lambda}$ or $\O_{b,\Lambda}\S$?    The obvious way is to add a hole to the Riemann surface that we would
use to compute a matrix element of $\S$ between states of prescribed length (fig. \ref{DDD}(a)).   The candidate spacetime to compute $\S \O_{b,\Lambda}$ or $\O_{b,\Lambda}\S$ in
the length basis is shown in fig. \ref{A4}(a).
We note immediately that there is no natural ordering between $\S$ and $\O_{b,\Lambda}$ in this spacetime, so if  $\S\O_{b,\Lambda}$ and $\O_{b,\Lambda}\S$
can be computed in this fashion,  these operators  must commute.
That is in fact the case, as we see by drawing an appropriate geodesic above or below the hole (figs. \ref{A4}(b,c)).  With one choice of geodesic, one learns that the
path integral on this surface computes $\O_{b,\Lambda}\S$; with the other choice, one learns that the same path integral computes $\S\O_{b,\Lambda}$. 
The only modulus in either picture is the length of the intermediate geodesic.    

Though some puzzles remain, as will be clear in what follows, 
 it appears that what is wrong in the claim 
that the $\O_{b,\Lambda}$ are operators that commute with $\A_L$ and $\A_R$ is simply that, in the presence of matter,
the $\O_{b,\Lambda}$ do not make sense as Hilbert space operators.   
They do make sense as quadratic forms,
meaning that they have well-defined matrix elements between a dense set of Hilbert space states.   For example, for strings $\S,\T$, there is no problem in defining the 
matrix element $\la\Psi_\T|\O_{b,\Lambda}|\Psi_S\ra$.   However, to define a Hilbert space operator, one needs more.   If an object $\O$ is supposed to act as an operator
on a Hilbert space $\H$, there should be at a minimum a dense set of states $\Psi\in\H$ such that $\O\Psi$ can be defined as a vector in $\H$.   This condition requires
$|\O\Psi|^2<\infty$ or $\la\Psi|\O^\dagger\O|\Psi\ra<\infty$.    So in order for $\O_{b,\Lambda}$ to be defined as a Hilbert space operator, the
product $\O^\dagger_{b,\Lambda} \O_{b,\Lambda}$ should have a finite expectation value in a dense set of states.   In fact, the adjoint of $\O_{b,\Lambda}$ is
$\O_{b,\bar\Lambda}$, where $\bar\Lambda$ is the $\sf{CPT}$ conjugate of $\Lambda$.  So we need the product $\O_{b,\bar\Lambda}\O_{b,\Lambda}$ to have finite matrix
elements, in a dense set of states.   

In pure JT gravity, the formula (\ref{tmore}) shows that the $\O_{b,\Lambda}$ are well-defined as Hilbert space operators.   They are unbounded, because of the $1/E$ singularity in
eqn. (\ref{tmore}).   But they make sense as unbounded, self-adjoint operators whose domain includes any square-integrable state whose wavefunction in the energy basis  vanishes sufficiently
rapidly for $E\to 0$.  

   \begin{figure}
 \begin{center}
   \includegraphics[width=4.5in]{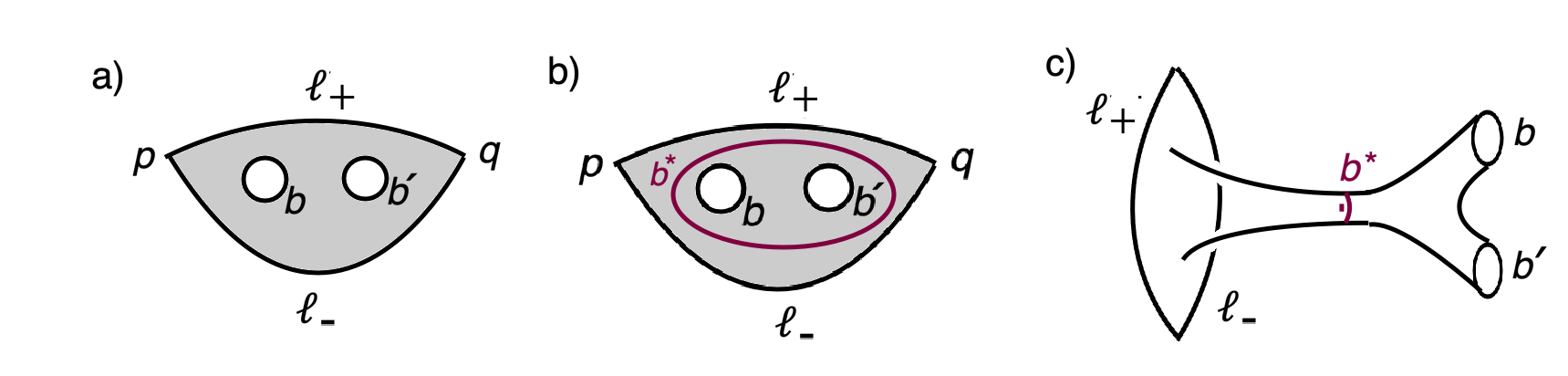}
 \end{center}
\caption{\footnotesize (a)  
The configuration that intuitively is appropriate to compute, in the length basis, the matrix element of a product of baby universe operators.     The upper and lower boundaries are geodesics that have been labeled
by their renormalized lengths $\ell_+$ and $\ell_-$.  (b) There  is a unique closed geodesic $\gamma^*$, as shown, that circles around the two holes.  We denote its length as $b^*$.
(c) In JT gravity coupled to matter, but not in pure JT gravity, the path integral that computes a matrix element of $\O_{b',\bar\Lambda}\O_{b,\Lambda}$ diverges for $b^*\to 0$, because of what in string theory would be called a closed string tachyon. \label{AA}}
\end{figure} 

\begin{figure}
 \begin{center}
   \includegraphics[width=4.5in]{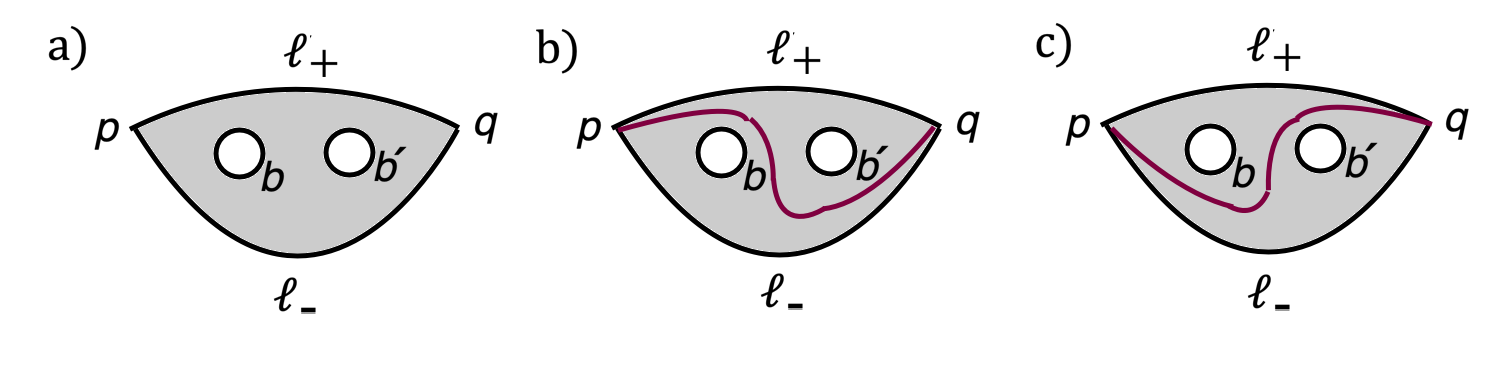}
 \end{center}
\caption{\footnotesize An embedded curve from $p$ to $q$ in this spacetime can always be deformed to an embedded geodesic in the same homotopy class.
This implies the existence of geodesics $\gamma$ and $\gamma'$  that go above one hole and below the other, or vice-versa, as sketched in (a) and (b).    \label{A3}}
\end{figure}

What happens when matter is included?   In fig. \ref{AA}, we have sketched an argument showing that matrix elements of products of wormhole 
operators such as  $\O_{b',\bar \Lambda} \O_{b,\Lambda}$
are always divergent in JT gravity coupled to matter.      First of all, an obvious guess is that to compute a matrix element of a product of baby universe operators, we should just
add another hole in the spacetime of fig. \ref{A2}(b), arriving at fig. \ref{AA}(a).  This is analogous to claims that we have made in other cases.  Now we note a mathematical
fact that will be used many times in the rest of this article:   
in a hyperbolic two-manifold $\Sigma$, any embedded closed curve is homotopic to a  unique embedded geodesic.\footnote{\label{generalization} An analogous statement holds for embedded  curves
with specified endpoints at infinity, that is, on the conformal boundary  of $\Sigma$: any such curve is homotopic  (keeping its endpoints fixed) to 
a unique embedded geodesic. This will be important at many points in this
article, for example in the discussion of fig. \ref{A3}.}
So in particular
there is a unique closed geodesic $\gamma^*$  that
wraps once around the two holes (fig. \ref{AA}(b)).    The length  $b^*$ of $\gamma^*$ is a modulus of the spacetime of fig. \ref{AA}, and to compute the amplitude associated to this
spacetime, one must integrate over all values of $b^*$.   In pure JT gravity, there is no difficulty with this integral, which can be computed explicitly using the arguments
of  \cite{SSS} and \cite{Saad}.   
But in any conventional two-dimensional QFT, the path integral on this surface will
 diverge  for $b^*\to 0$ because the ground state energy of a two-dimensional field theory  on a small circle is always negative.    For example, a CFT with central charge $c$ on a circle of circumference $b^*$ has a ground state energy $-\pi c/6 b^*$.   Any QFT that is conformally
invariant at short distances similarly has a negative ground state energy on a small circle.    For small $b^*$, the hyperbolic metric on $\Sigma_1$ has a very long tube separating
the part of $\Sigma_1$ where the dual quantum mechanics is defined from the two holes (fig. \ref{AA}(c)).    Propagation of a negative energy ground state down that long tube will give a contribution
that grows exponentially for $b^*\to 0$, leading to a divergence in the path integral that describes  the product of two baby universe ``operators.''     This is
a divergence that occurs in the limit that the intermediate geodesic of  fig. \ref{A3} goes to infinity, t because a geodesic that goes above one of the two holes and below the other in fig.
\ref{A3} must in fig. \ref{AA}(c)
go all the way up the tube and back again, so its length diverges for $b^* \to 0$.   By analogy with  familiar terminology in string theory, one might call this a divergence in the
``closed universe channel.''

The conclusion, then, is that the objects $\O_{b,\Lambda}$ make sense as quadratic forms, with well-defined matrix elements between suitable states, but they do not make sense
as Hilbert space operators.   In particular, the $\O_{b,\Lambda}$ do not have eigenvectors and eigenvalues.  If $\Psi$ were an eigenvector of $\O_{b,\Lambda}$, say with eigenvalue
$w$, then we would have $\la\Psi|\O_{b,\bar\Lambda}\O_{b,\Lambda}|\Psi\ra=|\O_{b,\Lambda}\Psi|^2=|w|^2 |\Psi|^2<\infty$, contradicting the universal nature of the divergence
in the $\O_{b,\bar\Lambda}\O_{b,\Lambda}$ product.\footnote{It does not help to assume that $\O_{b,\Lambda}$ has a continuous spectrum.   Let $\Pi$ be the projection operator
onto states with $|\O_{b,\Lambda}|\leq w$, and let $\Psi$ be in the image of $\Pi$.   Then we would have $|\la\Psi|\O_{b,\bar\Lambda}\O_{b,\Lambda}|\Psi\ra|\leq |w|^2|\Psi|^2$,
again contradicting the universal nature of the divergence in $\O_{b,\bar\Lambda}\O_{b,\Lambda}$.}
Thus, in JT gravity coupled to matter, one cannot define $\alpha$ parameters
as eigenvalues of the $\O_{b,\Lambda}$.

Since this phenomenon may seem unfamiliar, we will mention an elementary situation in which something similar occurs.   Let $\phi$ be a local
operator in some quantum field theory  in Minkowski spacetime $M$ of any dimension $D\geq 2$.
Consider two complementary Rindler wedges in $M$, with respective operator algebras $\A_L$, $\A_R$.  According
to the Bisognano-Wichman theorem
\cite{BisWic}, $\A_L$ and $\A_R$ have trivial center and are commutants.   We can reach an apparent contradiction as follows.
  Let $p$ be a point in the bifurcation surface where the two wedges meet and
consider the ``operator'' $\phi(p)$.   One can formally argue that $\phi(p)$ commutes with\footnote{  $\A_L$ or $\A_R$ is generated by functions of smeared quantum fields.
The smearing functions are 
 smooth functions   supported  in the Rindler wedge. 
The interior of the Rindler wedge is spacelike separated from the point $p$, so a  nonzero commutator of such a smeared field with $\phi(p)$ must arise
from a contribution on the boundary of the Rindler wedge.    There is no such contribution, since a smooth function with support in the Rindler wedge
vanishes to all orders near the boundary, killing any singularity that commutators of quantum fields may have along the diagonal or at null separation.}
 $\A_L$ and $\A_R$, seemingly contradicting the theorem.   The resolution is that
$\phi(p)$ 
 makes sense as a quadratic form, since it has well-defined matrix elements between a suitable dense set of states,
but does not make sense as an operator.     For example, in free field
theory, $\phi(p)$ has well-defined matrix elements between Fock space states.   However, $\phi(p)$ does not make sense as an operator and does not have eigenvalues and eigenvectors,
because if $\Psi$ is any Hilbert space state, $\phi(p)\Psi$ is unnormalizable.    The norm squared of $\phi(p)\Psi$ would equal $\lim_{q\to p}\la \Psi|\phi^\dagger(q)\phi(p)|\Psi\ra$, and this
is divergent because the product $\phi^\dagger(q)\phi(p)$ is singular for $q\to p$.   

The $\O_{b,\Lambda}$  would  presumably be far more  significant  if they could be defined as operators, for then their eigenvalues (``$\alpha$-parameters'')  could
be used to decompose the Hilbert space.  Note that  ``operators'' $\phi(p)$ on the bifurcation surface are not very useful in studying physics in the Rindler wedge.

While we believe that it is true that in the presence of matter, the $\O_{b,\Lambda}$ do not make sense as Hilbert space operators, there is a puzzle that we have not been able
to resolve.   Let us return to the spacetime of fig. \ref{AA}(a), which heuristically describes a matrix element of a product of baby universe operators $\O_{b,\Lambda}\cdot \O_{b',\Lambda'}$.
To try to prove that the interpretation is correct, we observe that in the figure, there exist embedded geodesics between the two ``corners'' $p$ and $q$ that go above
one hole and below the other, or vice-versa (fig. \ref{A3}).   These geodesics exist and are unique in their homotopy classes by virtue of the general 
statement in footnote \ref{generalization}.   The path integral with a given embedded geodesic that separates the two holes has a clear interpretation as a matrix
element of a product of baby universe operators: the path integral in fig. \ref{A3}(a) computes a matrix element of the product 
$\O_{b',\Lambda'}\cdot \O_{b,\Lambda}$ between states $\ell_-, \ell_+$, and the path integral in fig. \ref{A3}(b) similarly computes a matrix element of the
product $\O_{b,\Lambda}\cdot \O_{b',\Lambda'}$.   If, therefore, the separating geodesics $\gamma$ and $\gamma'$ that are drawn in fig. \ref{A3}(a,b) were unique, it would
follow that the products $\O_{b',\Lambda'}\cdot \O_{b,\Lambda}$ and $\O_{b,\Lambda}\cdot \O_{b',\Lambda'}$ are equal, in other words these operators
commute, and moreover that the path integral in fig. \ref{AA} does indeed compute the matrix element of either  of these products.   

However, $\gamma$ and $\gamma'$  are only unique in their homotopy classes, and these homotopy classes are actually far from being unique.   The reason for this
is that in, say, fig. \ref{A3}(a), for  any integer $n$, one can vary the spacetime $\Sigma$ by letting one hole move $n$ times around the other, returning to its original starting point.  During this process,
$\gamma$ can be varied in such a way that it is always embedded.   When the two holes return to their original positions, the original curve $\gamma$ is replaced by a new embedded
curve whose homotopy class depends on $n$; in this homotopy class there is an embedded geodesic $\gamma_{[n]}$ that can play the same role as $\gamma$.  

The path integrals with a particular choice of $\gamma$ or $\gamma'$ really do compute the matrix elements of products $\O_{b',\Lambda'}\cdot \O_{b,\Lambda}$ or $\O_{b,\Lambda}\cdot \O_{b',\Lambda'}$.   Because there are infinitely many possible choices of $\gamma$ or $\gamma'$, the path integrals in fig. \ref{A3}(a,b) should be divergent; these path integrals
should be equivalent to that of fig. \ref{AA}(a), multiplied by an infinite factor that counts the number of possible choices of $\gamma$ or $\gamma'$.

It may seem that we have reached a happy outcome, another way to show that operator products  $\O_{b',\Lambda'}\cdot \O_{b,\Lambda}$ and $\O_{b,\Lambda}\cdot \O_{b',\Lambda'}$ 
are divergent.   However, this argument seems to prove too much, because it does not depend on the presence of matter.   The argument seems to show that
products of baby universe operators $\O_{b'}\cdot \O_b$ are divergent even in pure JT gravity without matter.  But that conclusion seems to contradict the explicit formula of eqn. (\ref{tmore}),
which implies that on states of nonzero energy, we can act with baby universe operators any number of times.

There is not really a contradiction, because eqn. (\ref{tmore}) only shows that the baby universe operators, and their products, are well-defined when acting on states that vanish sufficiently
rapidly at zero energy.    The length eigenstates that are considered as initial and final states in figs. \ref{AA} and \ref{A3} do not have that property.    One can hope that
given a linear combination of length eigenstates that vanishes sufficiently rapidly at zero energy,  the corresponding linear combination of the path integrals of fig. \ref{A3}(a,b) 
is actually finite, despite the infinite over-counting that seems to come from the choice of separating geodesic.   We do not know how to demonstrate this.   If it is true, demonstrating
it probably depends on a careful analysis of conditionally convergent integrals.

Even if it is true that the $\O_{b,\Lambda}$ make sense only as quadratic forms and not as operators, this is not the end of the story.
One might still still worry about an apparent conflict with our claim in section \ref{algebra} that the trace on a von Neumann factor is unique up 
to rescaling. Specifically, for each $b,\Lambda$, we can define a new ``trace'' $\Tr_{b,\Lambda}(\a)$ by evaluating a Euclidean path integral 
on the annulus shown in fig. \ref{A1} with boundary conditions at infinity defined using $\a$ as in section \ref{euclideanstyle}. By construction, 
this satisfies  $\Tr_{b,\Lambda}(\a \a') =  \Tr_{b,\Lambda}(\a' \a)$ for all $\a,\a' \in \A$.       However, 
in contrast to the usual trace, 
there is no formal argument based
on reflection positivity of the bulk path integral that $\Tr_{b,\Lambda}$ is positive on positive operators, and that is actually not true.
If  $\Lambda$ is the CFT ground state, then from  eqn. (\ref{tmore}) (eqn. (3.30) of \cite{Saad}), one can deduce that for any function $f(H)$ of the Hamiltonian 
$H$,
\be\label{notpos} \Tr_{b,\Lambda}\,f(H)=\int_0^\infty \d E f(E)\frac{ \cos(b\sqrt{2E})}{\pi\sqrt{2E}}\ee
 (up to a constant factor that depends on the regularization of the matter 
path integral), showing the lack of positivity.
The same formula actually applies for any $\Lambda$, since there is actually no coupling between the gravitational sector and the matter sector in the path integral that
computes $\Tr_{b,\Lambda}\,f(H)$.   

For any operator $\a \in\A$, the operator $e^{-\varepsilon H} \a$ has a finite trace $\Tr_{b,\Lambda}(e^{-\varepsilon H} \a)$. Since $e^{-\varepsilon H} \a$ 
converges to $\a$ as $\epsilon\to 0$, the trace $\Tr_{b,\Lambda}$ (just like the disc trace $\Tr$ defined in section \ref{algebra}) is finite on a dense set of 
operators. And it is not related to $\Tr$ by a rescaling.

To understand what is  going on here, we need to be a bit more precise.
A Type I or II von Neumann factor has a  trace that is unique if one requires it to be normal and semifinite; more technically, one says that such a factor
has a unique semifinite normal tracial  weight. A weight is a linear map $\phi$ from the positive elements\footnote{A weight is defined only for positive elements to 
avoid difficulties that one would encounter with $\infty-\infty$ if one attempts in an infinite von Neumann algebra
to extend the definition of a typical weight to indefinite elements.} of the algebra $\A$ to $[0,\infty]$. It is tracial if $\phi(\a\bb) = \phi(\bb\a)$ for all 
$\a,\bb \in \A$.   We will briefly discuss semifiniteness  at the end of this section. 
The important qualification for our purposes is that uniqueness depends on  the trace being  normal. 
Being ``normal'' is roughly a condition of continuity, but in an infinite von Neumann algebra, this has to be stated with care.   A precise definition is that a weight $\phi$ is normal
if given an increasing sequence of positive operators $\a_n\in \A$ that converge to $\a$, we have  $\lim_{n\to\infty} \phi(\a_n) = \phi(\a)$.   The reason for requiring the sequence $\a_n$ to be 
increasing is that in an infinite von Neumann algebra -- Type I$_\infty$ or Type II$_\infty$ -- one can have, for example, projection operators $p_n$ of arbitrarily large trace.
So   for a normal weight $\phi$, one could have a sequence of positive operators, say $\a_n=\a+p_n/n$, with $\lim_{n\to\infty}\,\a_n=\a$ but $\lim_{n\to\infty}\,\phi(\a_n)>\phi(\a)$.   
This is described by saying that $\phi$ is lower semicontinuous; it can jump downward but not upward in a limit.  In the case of an increasing sequence, lower semicontinuity becomes
ordinary continuity.

An obvious example of a normal weight is the functional $\phi(\a)=\la \Psi|\a|\Psi\ra$, where $\Psi$ is any vector in a Hilbert space $\H$ on which the algebra $\A$ acts. A positive functional of this kind (for any choices of $\Psi$ and $\H$) is said to be ``ultraweakly continuous.'' A function $f: \R \to \R$ is lower semicontinuous if and only if it can be written as the limit of a monotonically increasing sequence of continuous functions $f_n$. Similarly, a weight $\phi$ is normal if and only if it can be written as the limit of a sequence of monotonically increasing ultraweakly continuous weights $\phi_n(\a)=\la \Psi_n|\a|\Psi_n\ra$.

A Type II$_\infty$ factor  does have additional densely defined traces -- such as we are finding with $\Tr_{b,\Lambda}$ -- if one drops the conditions of normality and semifiniteness.   
For example the tensor product of a Dixmier trace on a Type I$_\infty$ factor
with the standard trace on a Type II$_1$ factor gives a trace on a Type II$_\infty$ factor that is densely defined and positive, but not normal.   This example does not seem very similar to
our $\Tr_{b,\Lambda}$, however.

The uniqueness statement about traces is simpler  in the case of an algebra of Type II$_1$, because then there is an upper bound on the trace of a projector.   In our context, we can transfer
the discussion to an algebra of Type II$_1$ by introducing the projection operator $P_0$ onto states with energy less than some large cutoff energy $E_0$.   
Such a projection arose naturally in \cite{CLPW} in the analysis of de Sitter space.  As in that case, the projected algebra $\t\A=P_0\A P_0$ is of Type II$_1$.  To prove this,
one just observes that $P_0$ is the identity in $\t\A$, and $\Tr\,P_0<\infty$, showing that $\t\A$ is of Type II$_1$, not II$_\infty$.

A theorem about Type II$_1$ factors\footnote{For example, see Corollary 6.1.19 in \cite{VJ}.}  asserts
 that the usual trace is the unique tracial, ultraweakly continuous linear functional on the algebra. Note that there is no assumption 
here that the functional must be positive.   There is also no analog of semifiniteness; instead the trace is assumed to be defined for all elements of the algebra.
A general (not necessarily positive) linear functional $\phi$ on an algebra $\A$ is ultraweakly continuous if we can write it as
 $\phi(\a)=\la \chi|\a|\Psi\ra$, where $\chi$, $\Psi$ are   vectors in some
Hilbert space on which the algebra $\A$ acts.   In our problem, for the usual trace, we can take $\H$ to be the usual Hilbert space and assume $\chi=\Psi$.  The state
\begin{align}
\Psi_0 = \lim_{\beta \to 0}\, P_0 \Psi_\TFD(\beta)
\end{align}
is normalizable.   If $\a_0$ is an element of $\t\A$, then, as $\a_0=P_0\a_0 P_0$,
   \begin{align}\notag \Tr\,\a_0&= \lim_{\beta\to 0}\langle \Psi_\TFD(\beta) | \a_0|\Psi_\TFD(\beta)\rangle \\
&= \lim_{\beta\to 0}\la \Psi_\TFD(\beta)|P_0\a_0 P_0|\Psi_\TFD(\beta)\ra=\la \Psi_0|\a_0|\Psi_0\ra.\end{align} So the usual trace $\Tr$ is indeed ultraweakly continuous on $\t\A$. 

 \begin{figure}
 \begin{center}
   \includegraphics[width=2in]{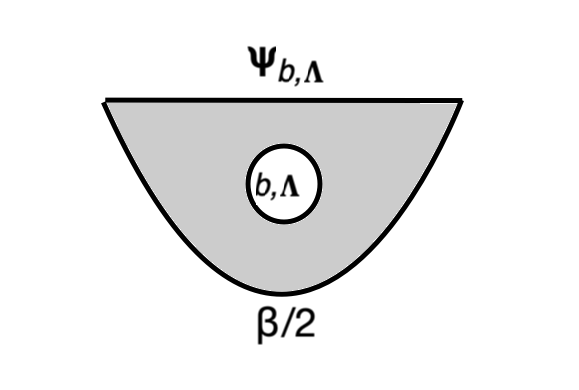}
 \end{center}
\caption{\footnotesize The path integral that computes $\Psi_{b,\Lambda}$.   \label{BB}}
\end{figure}

What about $\Tr_{b,\Lambda}$?   One can try to write $\Tr_{b,\Lambda}(\a_0) =  \langle \Psi_{b, \Lambda} | \a_0 |\Psi_0\rangle$ where
\begin{align}
\Psi_{b, \Lambda} = \lim_{\beta \to 0}\, P_0 \Psi_{b,\Lambda}(\beta)
\end{align}
and $\Psi_{b,\Lambda}(\beta)$ is prepared using a path integral (fig. \ref{BB})
on an annulus with an asymptotic boundary of renormalized length $\beta/2$, a geodesic 
boundary on which the state is defined, and an additional closed geodesic boundary labeled by  $(b, \Lambda)$. However, $\langle \Psi_{b, \Lambda} | \Psi_{b, \Lambda} \rangle$ 
diverges for exactly the same reason that $\O_{b,\Lambda}$ does not make sense as an operator. So $\Psi_{b,\Lambda}$ is not a Hilbert space state 
and we do not succeed in proving that $\Tr_{b,\Lambda}$ is 
 ultraweakly continuous.   Hence there is no contradiction with $\t\A$ being a Type II$_1$ factor.

We conclude by demonstrating that the trace $\Tr$ that we defined originally does indeed satisfy all the expected properties of the  standard trace on the full Type II$_\infty$ factor $\A$. 
We already know that it is tracial. To see that it is  normal, we note that the functional $F_\beta(\a)=
 \la\Psi_\TFD(\beta)|\a|\Psi_\TFD(\beta)\ra$
increases as $\beta$ tends to zero (since $\partial_\beta e^{-\beta E}\leq 0$ for $E\geq 0$), so normality of the trace follows from 
$\Tr\,\a=\lim_{\beta\to 0} \la\Psi_\TFD(\beta)|\a|\Psi_\TFD(\beta)\ra$.  
 Finally, a weight is semifinite if for every nonzero positive operator $\a \in \A$ there exists a positive operator $\a' \leq \a$ such that 
 $\Tr[\a']$ is finite. For any positive $\a \in \A$, the operator $\a^{1/2} P_0 \a^{1/2}\leq \a$ converges to $\a$ in the strong operator topology as $E_0 \to \infty$; consequently $\a^{1/2} P_0 \a^{1/2}$ 
 is nonzero for sufficiently large $E_0$. Since 
\begin{align}
\Tr[\a^{1/2} P_0 \a^{1/2}] = \Tr[P_0 \a P_0] =  \la \Psi_0|\a|\Psi_0\ra
\end{align}
is finite, this gives us the desired result.

\section{Wormhole Corrections}\label{wormhole}

\subsection{Overview}\label{overview}

We will now explore what happens when one includes wormhole corrections to the analysis of 
 section \ref{euclideanstyle}.  In other words, we will  allow for the possibility that
spacetime is a (connected) oriented two-manifold $M$ with a specified boundary but otherwise with any topology.\footnote{\label{reflection}We do not assume time-reversal symmetry or equivalently (by the
two-dimensional version of the 
CPT theorem) spatial reflection symmetry.   If one does assume such symmetry, one should allow the possibility that $M$ is
unorientable. On an unorientable two-manifold, the contribution of very small cross-caps is such that the path integral of JT
gravity is divergent, even in the absence of matter \cite{SW}.  This divergence 
is somewhat analogous to the small $b$ divergence that occurs in the presence of matter in the orientable case.}  This means that the open universe Hilbert space that we have
considered so far will be extended by including a Fock space of closed baby universes, as considered by a number of previous authors \cite{Giddings,MM}.
 
In JT gravity coupled to matter, we do not have a framework to
discuss the wormhole contributions nonperturbatively.
In JT gravity without matter, the matrix model gives such a framework, which has been exploited very
successfully for some purposes \cite{JohnsonA,JohnsonB}, though it is not clear whether it is useful for 
the sort of questions that we consider in the present paper.    

Still, whether matter is present or not, we can certainly study wormholes order by
order in an expansion in the genus, which is a non-negative integer $g$.    In JT gravity with or without matter, a 
genus $g$ contribution is suppressed relative to the $g=0$ contribution by a factor $e^{-2 g S}$, where $S$ is the 
 the black hole entropy.   Assuming the classical contribution $S_0$ to the black hole
entropy is large, $S$ is large except at extremely low temperatures or energies, where the theory becomes strongly
coupled and the genus expansion will break down.   At moderate or high temperatures, the 
genus expansion is a reasonable framework for studying wormhole contributions, and we will work in that
framework.   

The wormhole expansion in JT gravity 
coupled to matter has a well-known technical problem, which we already encountered in section \ref{operators}.    
In a path integral with a dynamical wormhole, one will have to integrate over the circumference $b$ of the wormhole,
and the integral will diverge for $b\to 0$, because the ground state energy of a quantum field theory on a circle is negative and of order $-1/b$
(it is $-\pi c/6b$ for a CFT with central charge $c$).   We will proceed formally, ignoring this issue.
 Since the divergence at small $b$ is an ultraviolet issue, one can
reasonably  hope that JT gravity with matter is an approximation to a 
better theory in which our general considerations are applicable and 
the wormhole contributions are convergent.   Of course, in section \ref{operators}, the divergence at $b\to 0$ was crucial to the left and right boundary operator
algebras being commutants with trivial center.   How that story is modified (or not) in a regulated theory with wormholes  potentially depends on the details of the regulated theory. We discuss some possibilities at the end of  section \ref{remarks}.

It is also true that the divergence in the wormhole amplitudes does not arise for JT gravity without matter.  But we do not
want to be limited to that case, since in that case the algebra of boundary observables is commutative and not
so interesting.    

In section \ref{closed}, we discuss from a bulk point of view the Hilbert space of JT gravity coupled
to matter in the presence of closed universes.   In section \ref{boundary}, we analyze the natural
Hilbert space from a boundary point of view.   Purely from a boundary point of view, it is straightforward
to include wormhole contributions to path integrals and thereby to generalize the definitions of a trace,
a Hilbert space, and an algebra of observables that were given in section \ref{euclideanstyle}.
In sections \ref{comparison}-\ref{furthersteps}, we make contact between the boundary analysis and the bulk analysis.
This is not nearly as straightforward as it was in section \ref{euclideanstyle} in the absence of wormholes.

What we learn can be summarized as follows.  With wormholes included, the algebra of boundary observables is modified but is still of Type II$_\infty$.
 In the theory with wormholes, the natural boundary Hilbert space
$\H_{\bdry}$ is a small and hard to characterize subspace of a much bigger bulk Hilbert space $\H_\bulk$.
However, the difference is undetectable by a boundary observer, in the sense that every pure or mixed state on $\H_\bulk$
is equivalent, for a boundary observer, to some  state on $\H_\bdry$.   In fact, the state on $\H_\bdry$ can be assumed to be
pure.   Roughly, not being able to see beyond the horizon, a boundary observer cannot detect the extra degrees of freedom described by $\H_\bulk$.

\subsection{The Hilbert Space From A Bulk Point Of View}\label{closed}

In Lorentz signature, a (connected) closed universe with constant scalar curvature $R=-2$ 
can be described by the metric
\be\label{lormet} \d s^2=-\d \tau^2+\cos^2 \tau \,\d\phi^2, \ee
where $\phi\cong\phi+b$, with an arbitrary $b>0$.   Thus $\phi$ parametrizes a circle $S_\phi$.    We will call this spacetime $U_b$.
$U_b$ has a big bang singularity
at $\tau=-\pi/2$, and a big crunch at $\tau=\pi/2$.   The interpretation of those singularities in quantum
theory is obscure, to say the least, but they will not be too troublesome for the issues addressed in the present article.

We would like to describe a Hilbert space of quantum states for JT gravity possibly coupled to matter
in such a closed universe.   This is straightforward.  The only modulus of the closed universe is $b$.
Quantizing the matter system   on $U_b$ gives a Hilbert space $\H^\matt_{\cl,b}$.   
The isometry group of the closed universe is just the group $U(1)_\phi$ of constant shifts of $\phi$.  
We have to
impose $U(1)_\phi$ as a group of constraints.   
Let $P$ be the generator of $U(1)_\phi$ (the operator that measures the momentum around $S_\phi$).
Since the group $U(1)_\phi$ is compact, imposing the constraint
means simply restricting to the subspace of $\H^\matt_{\cl,b}$ with $P=0$.    We will call this
subspace $\H^\matt_{\cl_0,b}$.   

In addition, we have to take into account the gravitational sector.
The only dynamical variables of JT gravity in this closed universe are $b$ and its canonical momentum.
Therefore, in addition to its dependence on the matter variables, a quantum state is a function of $b$.

Thus finally we can describe the Hilbert space $\H_\cl$ produced by quantizing
JT gravity coupled to matter in a 
 closed universe.
A general state $\Psi\in\H_\cl$ can be represented by a function $\psi(b)$ that is valued in
$\H^\matt_{\cl_0,b}$.    Inner products of such states are defined by integration over $b$ along with the 
natural inner product in the matter sector.   So if $\Psi_1,\Psi_2$ correspond to functions $\psi_1(b),\psi_2(b)$, then 
\be\label{murky}\la\Psi_1,\Psi_2\ra =\int_0^\infty \d b \,\la \psi_1(b),\psi_2(b)\ra. \ee

If the matter theory is conformally invariant, then $\H_\cl$ can be described more simply.   In that case,
 $\H^\matt_{\cl,b}$ is independent of $b$, and we denote its $P=0$
subspace as $\H^\matt_{\cl_0}$.   The wavefunction $\psi(b)$ then takes values in the fixed,
$b$-independent Hilbert space $\H^\matt_{\cl_0}$.   So $\H_\cl=\H^\matt_{\cl_0}\otimes L^2(\R_+)$,
where $\R_+$ is the half-line $b>0$.   

Now let us discuss what should be the bulk Hilbert space of JT gravity in a world with two asymptotic
boundaries  and with wormholes included in the dynamics.    A spacetime with two asymptotic boundaries
must always have precisely one open component, that is, one component that is noncompact in space (open and closed universes
are both noncompact in time).   The Hilbert space obtained by quantizing JT gravity plus matter in a (connected) open universe
was described in section \ref{incmatter}.   Up to this point, we have denoted that Hilbert space simply as
$\H$, but now that we are including closed universes, it will be helpful to be more precise and write
$\H_\o$ for the open universe Hilbert space.

Once we include wormholes, any number of closed universes can be created and annihilated, so
a bulk description of the Hilbert space (in a spacetime with one open component) 
will include one factor of $\H_\o$ and any number of factors
of $\H_\cl$.   We do, however, have to take into account Bose symmetry among the closed universes.
Bose symmetry means that the Hilbert space for a closed universe with  $k$ components is not 
$\H_\cl^{\otimes k}$,
but its symmetric part, often denoted $\Sym^k\H_\cl$.   Thus the full bulk Hilbert space with one
open component and any number of closed components is $\H_\o\otimes (\C\oplus \H_\cl \oplus \Sym^2
\H_\cl \oplus \cdots)$.    A common abbreviation is to write 
$\Sym^*\H_\cl =\C\oplus \H_\cl \oplus \Sym^2\H_\cl \oplus \cdots$, so 
finally the bulk Hilbert space for the case of one open component is
\be\label{omelk} \H_\bulk=\H_\o \otimes \Sym^*\H_\cl. \ee
 
Since we will study wormhole dynamics with the help of Euclidean path integrals, we also
want to consider the Euclidean analog of the closed universe $U_b$.
Setting   $\tau_E=\i\tau$, the big bang/big crunch spacetime $U_b$ is converted to a complete
spacetime of Euclidean signature:
\be\label{ormet}\d s^2=\d\tau_E^2+\cosh^2\tau_E \d\phi^2. \ee
The curve $\tau=0$ in Lorentz signature, or $\tau_E=0$ in Euclidean signature, is a closed geodesic
$\gamma$ of length $b$.  In Lorentz signature, this geodesic has maximal length in its homotopy class, but in Euclidean signature
it has minimal length.

In general, if $M$ is any Euclidean signature spacetime with $R=-2$
and $\gamma\subset M$ is a simple (non-self-intersecting) 
closed geodesic of length $b$, then $\gamma$ is always locally length-minimizing;  moreover,
near $\gamma$, $M$ is precisely
isometric to $U_b$.   We can regard $b$ as a measure of the size of the
wormhole. As remarked in section \ref{overview}, in JT gravity coupled to matter, wormhole contributions
actually diverge for $b\to 0$, though we will proceed formally and not worry about this.

\subsection{The Hilbert Space From A Boundary Point of View}\label{boundary}

From a boundary point of view, we can repeat many statements from section \ref{euclideanstyle},
but now including wormhole corrections.

\begin{figure}
 \begin{center}
   \includegraphics[width=3in]{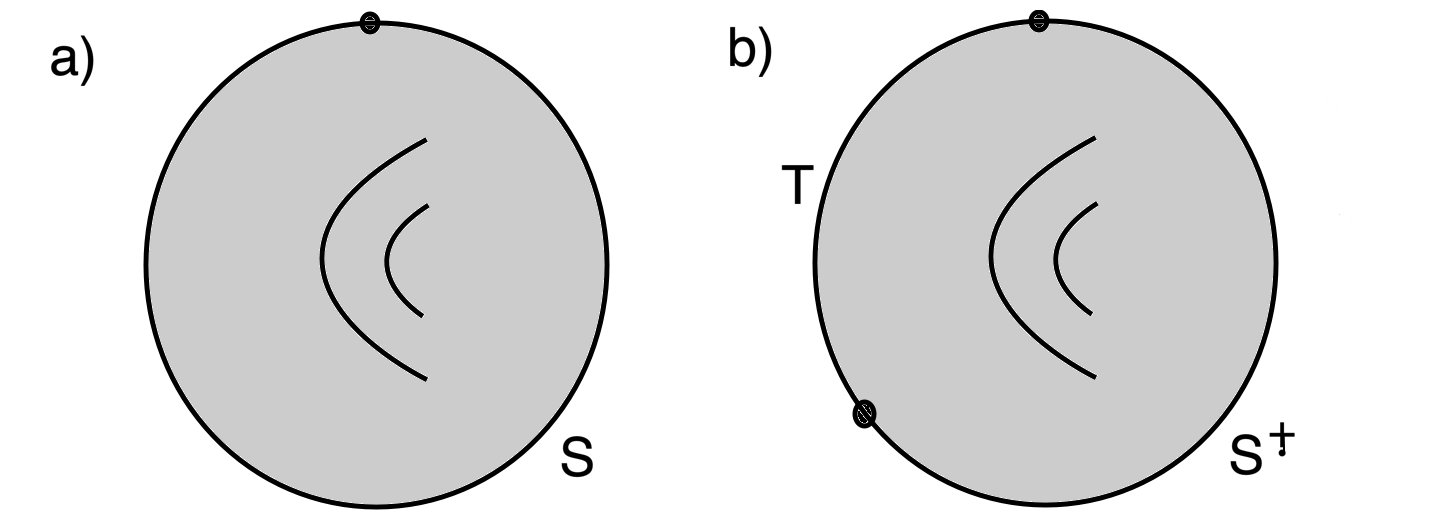}
 \end{center}
\caption{\footnotesize (a) A disc with a handle attached, representing a genus 1 contribution to $\Tr\,\S$ for some string $\S$. The boundary
is labeled by the string $\S$, with its ends glued together to make a circle. (b)  A similar procedure to compute $\la\S,\T\ra=\Tr\,\S^\dagger\T$.\label{EEE}}
\end{figure}

Thus, if $\S$ is a string, we now define $\Tr\,\S$ by a path integral on a two-manifold $M$ that
has a single asymptotic boundary labeled by $\S$ (with its ends joined together), as in fig. \ref{EEE}(a).  This is precisely analogous to
fig. \ref{AAA}, except that
since we want to include wormhole corrections, we no longer
 insist that $M$ should be a disc; rather in defining $\Tr\,\S$, we sum over all isomorphism classes of
 hyperbolic two-manifold of any genus
with a single boundary component.

Similarly, we can define an inner product on the space $\A_0$ spanned by the strings in the familiar
way:
\be\label{innprod}\la \S,\T\ra=\Tr\, \S^\dagger \T. \ee
Concretely, this matrix element is computed on an oriented two-manifold $M$ with a single asymptotic
 boundary that is labeled by $\S^\dagger\T$  and otherwise with any topology (fig. \ref{EEE}(b)).
We will show in section \ref{furthersteps}  that this inner product is positive 
semi-definite.\footnote{This statement is nontrivial only because the inner product 
defined via disc amplitudes 
in section \ref{euclideanstyle} is positive semi-definite rather than positive-definite.   
A strictly positive inner product in the absence of wormholes would automatically
remain positive order by order in the wormhole expansion.  The nontrivial question is whether vectors
that are null vectors in leading order can gain negative norm due to wormhole corrections.  We will see that this does not occur.}

The procedure to construct a Hilbert space is the same as before.   We
 define formally for every string $\S$ a state $\Psi_\S$, and we define the inner products of these
states by  $\la \Psi_{\S_1},\Psi_{\S_2}\ra =\la \S_1,\S_2\ra$.
Dividing by null vectors and taking a Hilbert space completion, we get a Hilbert space that now we call
the boundary Hilbert space $\H_{\bdry}$.   Its relation to the bulk Hilbert space $\H_\bulk$ is more subtle
than was the case in section \ref{euclideanstyle} and will be the subject of section \ref{comparison}.

As before, strings can act on $\H_{\bdry}$ by $\S\Psi_\T=\Psi_{\S\T}$.   This gives an action of $\A_0$
on the Hilbert space $\H_\bdry$.    We will explain in section \ref{furthersteps} that if $\Psi_\S$ or $\Psi_\T$ is
null, then $\S\Psi_\T=\Psi_{\S\T}$ is also null.   
Hence it is possible to take the quotient of $\A_0$ by null vectors to get an algebra $\A_1$.
  Taking a completion of $\A_1$, we get a
von Neumann algebra $\A$ of boundary observables that act on  $\H_\bdry$, acting on a string on the left.     Its commutant
$\A'$ the opposite algebra $\A^\op$, acting on strings on the right.

The trace and therefore also the inner products that were used in this construction receive wormhole
corrections, so they are not the same as they were in the absence of wormholes in section 
\ref{euclideanstyle}.   However, in the presence of appropriate 
matter, the algebra $\A$ is a factor of Type II$_\infty$ just as before. 
Wormhole corrections cannot bring a center into being, so the center is trivial if it is trivial in the absence of wormholes.
Having  a trace that is not defined for all elements of the algebra, $\A$ must then be of Type II$_\infty$ or
Type I$_\infty$.   It is not of Type I$_\infty$, since order by order in the wormhole expansion we are not solving the black hole
information problem.

\subsection{Relating the Boundary and the Bulk}\label{comparison}

In section \ref{bulk}, we constructed a bulk Hilbert space $\H_\bulk$ that includes closed universes.
In section \ref{boundary}, we defined a boundary Hilbert space $\H_\bdry$ and for every string $\S$, a corresponding
vector $\Psi_\S\in \H_\bdry$.   An inner product $\la \Psi_{\S},\Psi_{\T}\ra$ is computed  
by a path integral on an oriented two-manifold
$M$ of any topology with an asymptotic  boundary circle  labeled by $\S^\dagger\T$ and any ``filling'' of $M$ in the interior.
We want to find a map 
\be\label{zombo}\wW:\H_\bdry\to \H_\bulk \ee
that preserves inner products, in the sense that
\be\label{tombo} \la \Psi_\S,\Psi_\T\ra= \la \wW(\Psi_\S), \wW(\Psi_\T)\ra,\ee
where the inner product on the left is in $\H_\bdry$ and the one on the right is in $\H_\bulk$.  
The adjoint of $\wW$ is a bulk to boundary map
\be\label{zombox}\vV:\H_\bulk \to \H_\bdry. \ee
Since the inner product on $\H_\bulk$ is manifestly positive, the
 existence of the map $\wW$ implies that,
as was asserted in section \ref{boundary}, the inner product on states $\Psi_\S$ is positive semi-definite, so that
after dividing by null vectors, the inner product on $\H_\bdry$ is positive-definite.

To find the map $\wW$, we will generalize the procedure of section \ref{euclideanstyle} to allow for the presence of
wormholes.   We start with the path integral  that defines  the inner product $\la\Psi_\S,\Psi_\T\ra$
of states associated with strings.   This is a path integral on a spacetime $M$ whose boundary is a circle made up of segments
labeled by $\S^\dagger$ and by $\T$.   The segments meet at boundary points $p,q$.  
Previously (fig. \ref{BBB}(b)), $M$ was assumed to be a hyperbolic disc,
and therefore there was a unique geodesic $\gamma\subset M$ joining $p$ and $q$.   This divided
$M$ into portions $M_-$ ``below'' $\gamma$ and $M_+$ ``above'' $\gamma$.  (In what follows, we include $\gamma$ itself  in both
$M_-$ and $M_+$ and thus we define $M_-$ and $M_+$ to 
be closed.)   The path integral on $M_-$ gives a description of a ket, the path integral on $M_+$ gives a description of a bra,
and the sum over fields on $\gamma$ computes the  inner product of these two states.
This is how, in the absence of wormholes, we defined a map from boundary states to bulk states that preserved inner products.
In the absence of wormholes, this map was an isomorphism so we did not distinguish the boundary and bulk Hilbert spaces.

   \begin{figure}
 \begin{center}
   \includegraphics[width=3in]{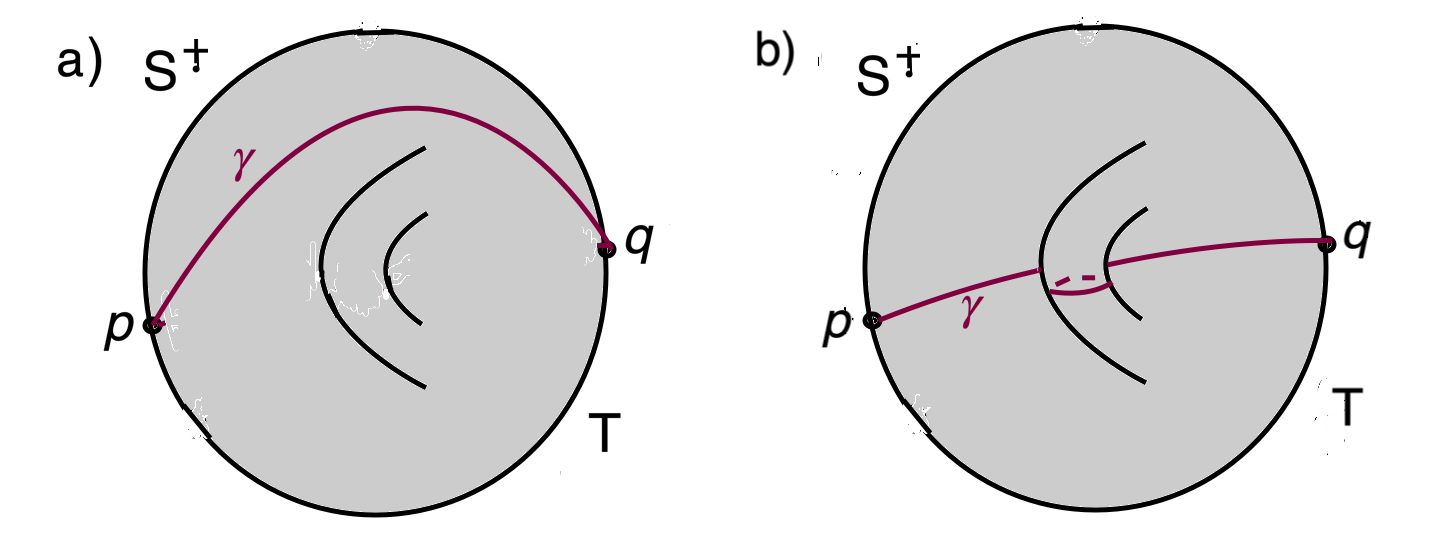}
 \end{center}
\caption{\footnotesize A  contribution to $\Tr\,\S^\dagger \T$ from a genus one spacetime $M$.   
The boundary segments labeled by $\S^\dagger$ and by $\T$
are separated by points $p$, $q$.   Two examples are sketched of a separating geodesic cut $\gamma$ from $p$ to $q$.
In (a), $\gamma$ is connected and consists of a geodesic from $p$ to $q$.   In (b), $\gamma$ is not connected and is the union  
 of a geodesic from $p$ to $q$ and a closed geodesic  ``inside the wormhole.''  In each case, 
 $\gamma$ is separating in the sense that removing it divides $M$ into disconnected components ``above'' and ``below''
$\gamma$.   In (b), this would not be so if we omit from $\gamma$ the wormhole component.    \label{FFF}}
\end{figure} 
This construction needs some modification when wormholes are included, because although $\gamma$ is unique when
$M$ is a disc, it is otherwise far from unique.   In general, when $M$ has higher genus, there are infinitely many geodesics
in $M$ connecting the boundary points $p$ and $q$.   If we simply sum over all possible $\gamma$'s, we will get
an infinite overcounting.   Instead of such a simple sum,  we will stipulate that we pick
$\gamma$
to have  minimal renormalized length among all geodesic cuts from $p$ to $q$.   By a geodesic cut from $p$ to $q$, we mean a one-dimensional
submanifold $\gamma\subset M$ that satisfies the geodesic equation, has asymptotic ends at the points $p$, $q$, and has the property
that if we ``cut'' $M$ along $\gamma$, it divides into an ``upper'' piece $M_+$ (containing in fig. \ref{FFF}  the part
of $\partial M$ labeled by $\S^\dagger)$ and  a ``lower'' piece $M_-$ (containing the part labeled by $\T$).    As discussed in more
detail shortly, we do {\it not} require that $\gamma$ be connected.     The minimal geodesic cut is unique except on a set
of measure zero in the moduli space of hyperbolic metrics on $M$; such a set of measure zero is not important in the
analysis of  inner products between states in Hilbert space.

The choice of the minimal geodesic cut requires some discussion.  First of all, this choice is computationally difficult in the
sense that in general it is difficult to find the minimal geodesic cut.   That is one reason, but probably far from being the
main reason, that the boundary to
bulk map $\wW$ and its adjoint, the bulk to boundary map $\vV=\wW^\dagger$, are computationally difficult.  These
maps are relatively simple to describe (modulo the difficulty in finding minimal geodesic cuts) for states that have
a simple Euclidean description, but for other states, $\vV$ and $\wW$ are probably very difficult to describe explicitly.
For example, acting on a state with a simple Euclidean construction,
real time evolution by  the boundary Hamiltonians $H_L$ and/or $H_R$ probably produces states on which an explicit
description of the maps $\wW$ and $\vV$ is very complicated.    This will become apparent when we describe how to 
define $H_L$ and $H_R$ as operators on $\H_\bulk$ (see the end of section \ref{remarks}).

A second point is that as the moduli of $M$ are changed, the minimal geodesic cut $\gamma$ generically evolves
smoothly but will sometimes jump discontinuously.   We are not sure what to say about this.   
Such jumps are possibly inevitable if one aims to give a Hamiltonian description, with continuous time
evolution, of a theory in which spacetime is modeled as a smooth manifold, so that the distinction between different
topologies is sharp.  
We use the minimal geodesic cut as a sort of gauge choice for the bulk state. 
Although relying on the minimal geodesic cut  will probably seem unnatural to many readers, with its aid we will obtain
 some nice results that
appear hard to obtain otherwise.  
With the help of the minimal geodesic cut, we can describe explicitly the map $\wW$ from states defined by boundary
data to bulk states and prove that it preserves inner products. This also makes it possible to complete
the definition of the boundary Hilbert space $\H_\bdry$.  And the minimal geodesic cut
  will be  a key tool in proving that any pure or mixed state on
$\H_\bulk$ is equivalent, for a boundary observer, to some pure state in $\H_\bdry$.   So the minimal geodesic cut is useful, but perhaps there is another
route to the same results.

Once we decide to base the definition of the boundary to bulk map on minimal geodesic cuts, there is still a choice to make, as 
illustrated in fig. \ref{FFF}:

(1)   We could stipulate that $\gamma$ should be connected and thus should be simply a geodesic from $p$ to $q$.
We will call this the restricted version of the proposal.   In this case, the cut reveals a single open universe and no closed
ones.   Therefore, with this proposal, the boundary to bulk map $\wW$ really maps $\H_\bdry$ to the original open
universe Hilbert space $\H_\o \subset \H_\bulk$.    However, we will see that this version of the proposal does not work.

(2) In the alternative that works, there is no condition for the geodesic cut $\gamma$ to be connected. We do require
that $\gamma$ is embedded in $M$.
  Then $\gamma$
consists of a simple (non-self-intersecting) geodesic $\gamma_0$ from $p$ to $q$ along with  disjoint
simple closed geodesics $\gamma_\alpha$, $\alpha=1,\cdots, n$.   In this case, when we ``cut'' along $\gamma$, we 
reveal an open universe and $n$ closed universes.    So with this definition $\wW$ really maps $\H_\bdry$ to
$\H_\bulk$, not just to the open universe subspace $\H_\o$.       We will call this the natural version of the proposal,
since once wormholes are included, it seems unnatural to exclude closed universe states.

   \begin{figure}
 \begin{center}
   \includegraphics[width=4.8in]{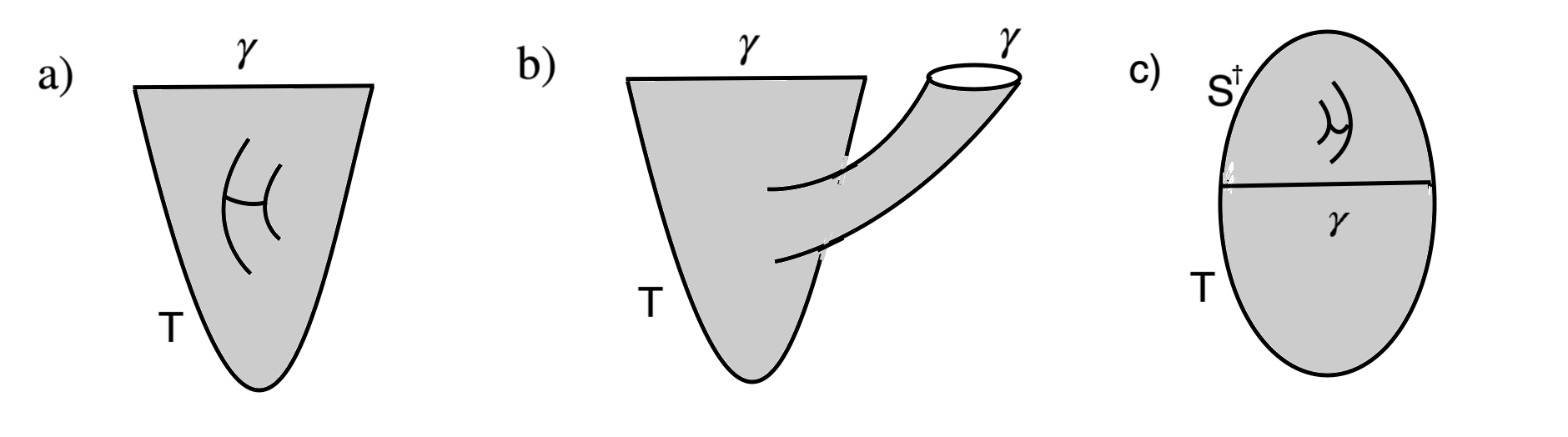}
 \end{center}
\caption{\footnotesize Illustrated here is the procedure to calculate the bulk state $\wW(\Psi_\T)$ for a string $\T$.  The spacetime $M_-$ has
an asymptotic boundary labeled by the string $\T$ as well as a minimal geodesic boundary $\gamma$.   There are two possible cases.  In the restricted proposal,   $M_-$ may
have wormholes but $\gamma$ is connected, as sketched in (a), and $\wW(\Psi_\T)$ is an element of the open universe Hilbert space $\H_\op$.
 In the natural proposal, 
 $\gamma$ is allowed to be disconnected, as in (b),  and $\wW(\Psi_\T)$ is valued in  $\H_\bulk$ but not in $\H_\op$.  As shown in (c), to compute an
 inner product $\la \wW(\Psi_\S)|\wW(\Psi_\T)\ra$, we glue together a bra and ket $|\wW(\Psi_T)\ra$ and $\la \wW(\Psi_\S)|$ defined by this procedure and perform
 a path integral.  Sketched is an example with one wormhole and a connected minimal geodesic cut $\gamma$.
  In case $\S=\T$, the resulting path integral is nonnegative, and vanishes if and only if $\wW(\Psi_\T)=0$, because for any values of the fields
 on $\gamma$, the path integral on the region above $\gamma$ is the complex conjugate of the path integral on the region below $\gamma$.  \label{GGG}}
\end{figure}

In either version of the proposal, one has to explain the rule for describing $\wW(\Psi_\T)$ as a state on $\gamma$.
We expect to compute $|\wW(\Psi_\T)\ra$ by a path integral over two-manifolds $M_-$ that have an asymptotic boundary
segment labeled by $\T$ and a geodesic boundary $\gamma$.  The path integral on $M_-$ as a function
of the fields on $\gamma$ will compute the desired state.     In the restricted proposal, $\gamma$ is required
to be connected (fig. \ref{GGG}(a)), and in the natural
version, $\gamma$ can have disconnected components (fig. \ref{GGG}(b)). The bra $\la \wW(\Psi_\S)|$ will be computed
similarly by a path integral over a two-manifold $M_+$ also with $\gamma$ as a geodesic boundary, and then to compute
the inner product $\la \wW(\Psi_\S)|\wW(\Psi_\T)\ra$, we glue $M_-$ and $M_+$ together along $\gamma$ to make
a two-manifold $M$, as in fig. \ref{GGG}. The inner product $\la \wW(\Psi_\S)|\wW(\Psi_\T)\ra$ defined by this procedure (fig. \ref{GGG}(c)) will hopefully coincide
with the path integral that we would compute on $M$, with asymptotic boundary conditions set by $\S$ and $\T$.

In order for this to be true, in either version of the proposal,  we need a further condition on $\gamma$ to ensure
that once we glue $M_-$ and $M_+$ together along $\gamma$ to make $M$, 
$\gamma$ will be uniquely determined (at least generically) just from the geometry of $M$.  If and only if this is so, pairs $M,\gamma$
will be classified (generically) by the same data that would classify $M$ alone, and 
hence the path integral evaluated with the cutting procedure will coincide
with the path integral that we would have defined on $M$ if we had never introduced $\gamma$ or the decomposition of $M$ 
as $M_+\cup M_-$.   Our strategy to ensure that $\gamma$ is uniquely determined (generically) will be to arrange so that 
$\gamma$ is  a minimal geodesic cut 
from $p$ to $q$ in 
$M$.  For this to have a chance of being true, we have to at least require that $\gamma$ is minimal in $M_-$, meaning
that there is no cut from $p$ to $q$ in $M_-$ (or no connected cut in the restricted version of the proposal) that has a renormalized length less than $\gamma$.     Here in the case of a manifold
$M_-$ with boundary, we allow a geodesic cut to be contained partly or entirely in the boundary (thus the boundary of $M_-$ is regarded
as an example of a cut, even though in this case the part of $M_-$ ``above'' the cut is empty).   In asking that $\gamma$ should be minimal, it does not matter if we ask for $\gamma$ to be minimal among all cuts or only among geodesic cuts;
 if there is a non-geodesic cut from $p$ to $q$
that is shorter than $\gamma$, then it can always be further shortened to a geodesic cut that is also shorter than $\gamma$.
 Similarly
we require that $\gamma$ is minimal in $M_+$.   In either the revised or the natural version of the proposal, stipulating that
we only integrate over metrics on $M_-$ with the property that the boundary $\gamma$ is minimal (among all cuts in the natural
version of the proposal, and among all connected cuts in the restricted version)   completes the definition of what
we mean by the path integral  on $M_-$ that, as a function of fields on $\gamma$,
 is supposed to compute $|\wW(\Psi_\T)\ra$.      A similar restricted path integral on $M_+$ computes a bra of the form $\la \wW(\Psi_\S)|$.   

   \begin{figure}
 \begin{center}
   \includegraphics[width=4in]{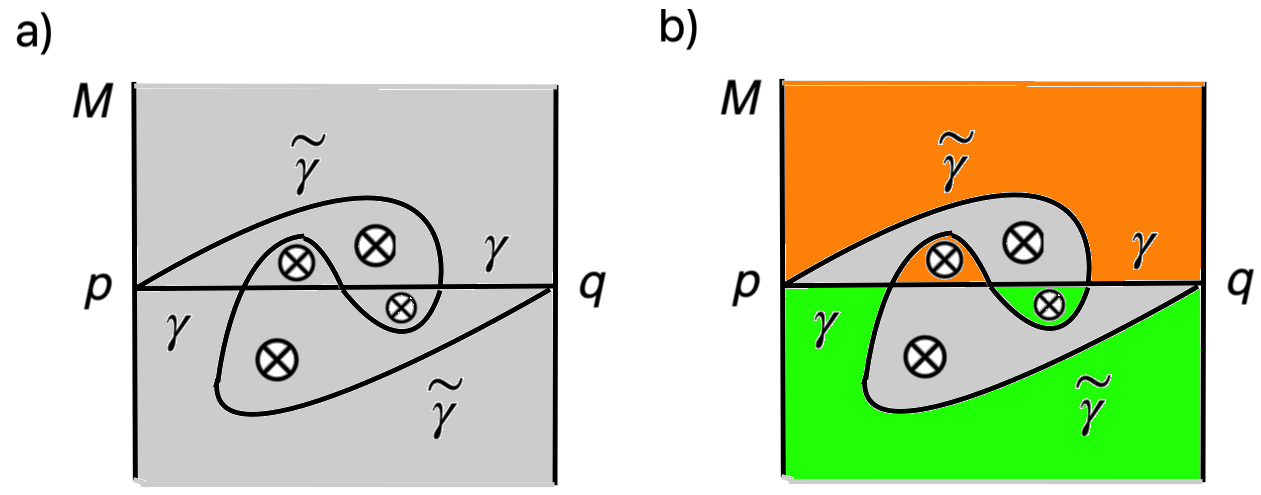}
 \end{center}
\caption{\footnotesize (a)  Sketched is a portion of a two-manifold $M$ with two geodesic cuts $\gamma$ and $\t\gamma$ between boundary points
$p$ and $q$; $\t\gamma$ is partly ``above''  and partly ``below'' $\gamma$.
  The symbols $\otimes$ represent unspecified topological complications (such as an attached genus $g$ surface).   The figure is drawn so that $\gamma$ is a horizontal straight line on the
page that looks like a geodesic.  In the presence of the indicated
wormholes, $\t\gamma$ might be a geodesic as well.   The relation between $\gamma$ and $\t\gamma$ is symmetric; by a diffeomorphism of $M$, one could
make $\t\gamma$ look like a straight line on the page and make $\gamma$ look like a wiggly curve. So either one could have smaller renormalized length. (b) $\gamma_-$ is the
boundary of the green region; $\gamma_+$ is the boundary of the orange region.   They are not connected.  \label{HHH}}
\end{figure}

But when we glue together $M_-$ and $M_+$ to make $M$, is $\gamma$ minimal in $M$, or can it be replaced in 
$M$ by a shorter geodesic cut $\t\gamma$ from $p$ to $q$?    Since $\gamma$ was minimal in $M_-$, there is no
such $\t\gamma $ that is contained entirely in $M_-$, and since $\gamma$ was minimal in $M_+$, there is no such
$\t\gamma$ that is contained entirely in $M_+$.   But could there be a $\t\gamma$ that is partly in $M_+$ and partly in $M_-$
(fig. \ref{HHH}(a))?   

In the natural version of the proposal,  a simple cut and paste argument shows that if $\gamma$ is minimal
in $M_-$ and in $M_+$, then it is minimal in $M$.
This argument does not work for the restricted version of the proposal.   
   That is why only the natural version of the proposal is successful.

To explain the cut and paste procedure, let $\t\gamma$ be any cut from $p$ to $q$.   Just as
$\gamma$ divides $M$ into a lower piece $M_-$ and an upper piece $M_+$, likewise $\t\gamma$ divides $M$ into a lower piece $\t M_-$ and an upper piece $\t M_+$.   Now we can define two
new cuts, $\gamma_-= \gamma\cap \t M_- \cup \t\gamma \cap M_-$, and $\gamma_+=\gamma\cap \t M_+ \cup
\t\gamma\cap M_+$.   
In other words, $\gamma_-$ consists of points in $\gamma$ that are ``below'' (or on) $\t\gamma$ together with points in $\t\gamma$ that are ``below'' $\gamma$, while
$\gamma_+$ consists of points in $\gamma$ that are ``above'' (or on) $\t\gamma$ and points in $\t\gamma$ that are ``above'' $\gamma$.
  Equivalently, $\gamma_-$ is the boundary
of $M_-\cap \t M_-$ and $\gamma_+$ is the boundary of $M_+\cap \t M_+$. 
This last description makes clear that $\gamma_-$ and $\gamma_+$ are cuts.  Note in particular that $\gamma_-\subset M_-$ and $\gamma_+\subset M_+$.
These definitions imply that  $\gamma_-\cup \gamma_+=\gamma\cup \t\gamma$ and\footnote{Unless $\gamma$ and
$\t\gamma$ have one or more components in common (which is possible in the natural version of the proposal if $\gamma$
and $\t\gamma$ are not connected),
$\gamma\cap\t\gamma$ is a set of measure 0, possibly a
finite set.  In the example of fig. \ref{HHH}, $\gamma\cap \t\gamma$ consists of three points.  Common components of 
$\gamma$ and $\t\gamma$, if there are any, are also present in $\gamma_+$ and $\gamma_-$ and cancel out of all relations
in the text.}  $\gamma_-\cap\gamma_+
=\gamma\cap\t\gamma$.  Accordingly, the renormalized lengths of the four cuts satisfy
\be\label{benlen} \ell(\gamma_+) +\ell(\gamma_-)= \ell(\gamma)+\ell(\t\gamma). \ee
In  the natural version of the proposal, minimality of $\gamma$ in $M_-$ means that the renormalized length of $\gamma_-$ is 
no less  than that of $\gamma$:
\be\label{renlen}\ell(\gamma_-)\geq  \ell(\gamma). \ee
Similarly, minimality of $\gamma$ in $M_+$ implies in the natural version
that
\be\label{zenlen} \ell(\gamma_+)\geq \ell(\gamma). \ee
A linear combination of these relations gives
\be\label{xenlen}\ell(\t\gamma)\geq \ell(\gamma),\ee
showing, in the natural version of the proposal, that $\gamma$ is minimal in $M$.

Why does this argument fail in the restricted version of the proposal?   It fails because even if $\gamma $ and $\t\gamma$
are connected, $\gamma_-$ and $\gamma_+$ may not be (fig. \ref{HHH}(b)).
If $\gamma_-$ or $\gamma_+$ is not connected, then in the restricted version of the proposal, we are not entitled to
assume eqn. (\ref{renlen}) or eqn. (\ref{zenlen}), so we cannot deduce eqn. (\ref{xenlen}).    On the contrary,  fig. \ref{HHH}
is essentially symmetrical in $\gamma$ and $\t\gamma$ up to a diffeomorphism of $M$, so in the restricted version of
the proposal, it is entirely possible for $\gamma$
to be non-minimal.

Although the boundary-to-bulk map $\wW:\H_\bdry\to \H_\bulk$ is isometric and well-defined for any boundary state, there is no reason to think that its image is dense is $\H_\bulk$, and hence no reason to think that the adjoint map $\vV:\H_\bulk \to \H_\bdry$ is also isometric. In particular, in the limit $e^{-S} \to 0$ where the wormhole contributions vanish, the boundary path integral defined by a string $\S$ can be used to prepare arbitrary states in the Hilbert space $\H_\o$ of an open geodesic, but does not enable us to create the states that
contain closed universes.   Intuitively, the  ``size'' of $\wW(\H_\bdry)$ is independent of $e^{-S}$, so we expect $\wW(\H_\bdry)$ to be much ``smaller'' than $\H_\bulk$ for all
values of $e^{-S}$.  

   \begin{figure}
 \begin{center}
   \includegraphics[width=3in]{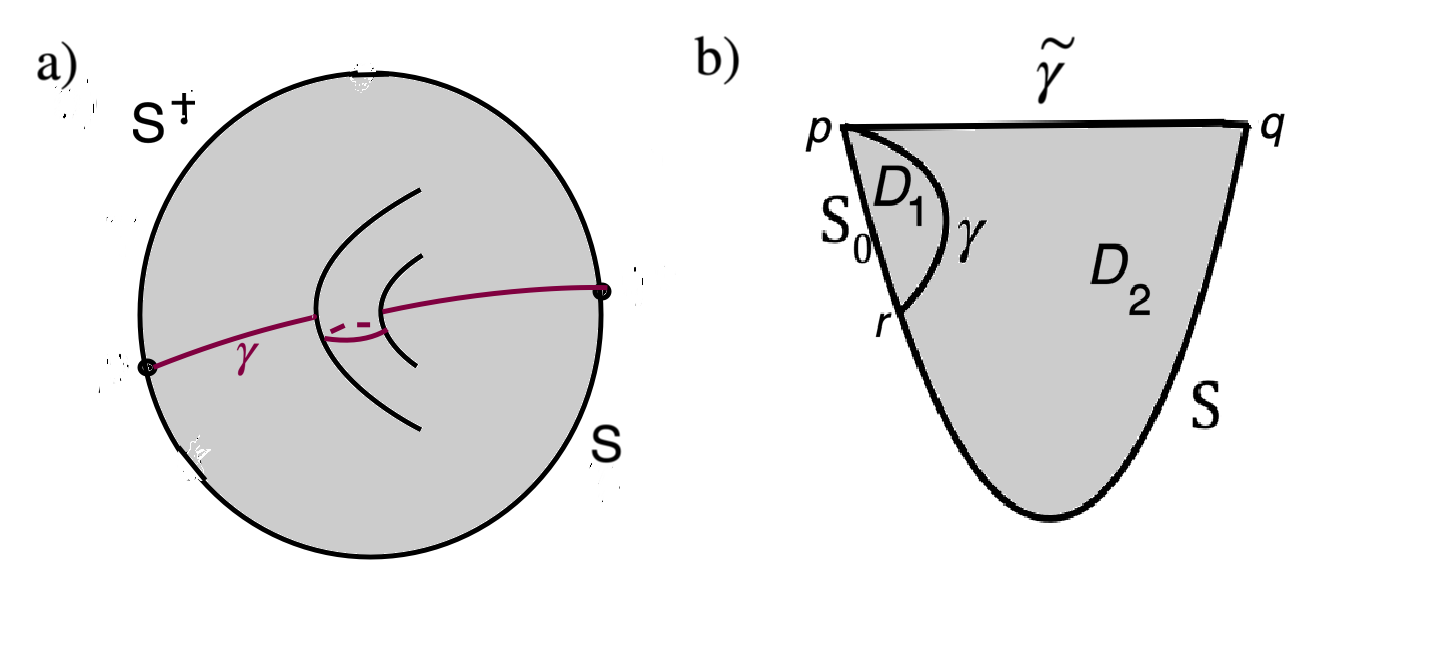}\caption{\footnotesize
    (a) A one wormhole contribution to $\la\Psi_\S,\Psi_\S\ra$.  $\gamma$ is a minimal
   geodesic cut with two components.   For fixed values of the fields along $\gamma$, after summing over all topologies and
   integrating over all moduli,
   the path integrals above and below $\gamma$ are complex
   conjugates, implying that $\la\Psi_\S,\Psi_\S\ra\geq 0$.   (b) This is a repeat of fig. \ref{CCC} except that wormholes may be present (not drawn) and $\gamma$ and
   $\t\gamma$ are now minimal geodesic cuts.   If $\Psi_{\S_0}$ is null, then the path integral in the smaller region $D_1$
   vanishes for any values of the fields along $\gamma$.   This implies vanishing of the path integral in $D_0=D_1\cup D_2$,
   implying that $\Psi_{\S_0\S}$ is null. \label{ZZZ}  }
 \end{center}
\end{figure}

\subsection{Further Steps}\label{furthersteps}

At this point, restricting  to the natural version of the proposal, it is  fairly straightforward to imitate 
other arguments in section \ref{euclideanstyle}, with a few new twists because
 the boundary to bulk map $\wW:\H_\bdry\to \H_\bulk$ is now
not an isomorphism but an embedding in a larger Hilbert space.   

First of all, as promised in section \ref{boundary}, we can now show that the inner products on states $\Psi_\S$,
with $\S\in\A_0$, are positive semi-definite.    In the path integral that computes $\la\Psi_\S,\Psi_\S\ra$, which is sketched in 
fig. \ref{ZZZ}(a), for any values of the fields on the minimal geodesic cut $\gamma$, after integrating over all moduli,
the path integral on the region above
the cut is equal to the complex conjugate of the path integral below the cut.
This is true essentially by reflection positivity of the bulk path integral. (More precisely, it is true because orientation reversal has the effect
of complex conjugating the integrand of the bulk path integral; this is the fact that underlies reflection positivity.)
Hence $\la\Psi_\S,\Psi_\S\ra\geq 0$, with
vanishing only if  the bulk state $\wW(\Psi_\S)$ vanishes identically as a function of the fields on $\gamma$.   
This enables us to define a boundary Hilbert space $\H_{\bdry}$ together with an embedding $\wW:\H_\bdry\to \H_\bulk$.

As before, we declare $\S\in\A_0$ to be null if $\la\Psi_\S,\Psi_\S\ra=0$ and let $\A_1$ be the quotient of $\A_0$ by such null
vectors.  To know that $\A_1$ is an algebra and acts on $\H_\bdry$, we need to know that if $\S_0$ is null, then 
$\S_0\S$ and $\S\S_0$ are also null.   This follows by the same argument as before, with geodesics replaced by minimal
geodesic cuts  (fig. \ref{ZZZ}(b)).  So now we can take the completion of $\A_1$ as an algebra acting on $\H_\bdry$.
This completion is the algebra $\A=\A_L$ of observables on the left boundary.   Acting on $\H_\bdry$, $\A_L$ has a commutant
that consists of a similar algebra $\A'=\A_R$ of observables on the right boundary.

\begin{figure}
 \begin{center}
   \includegraphics[width=4in]{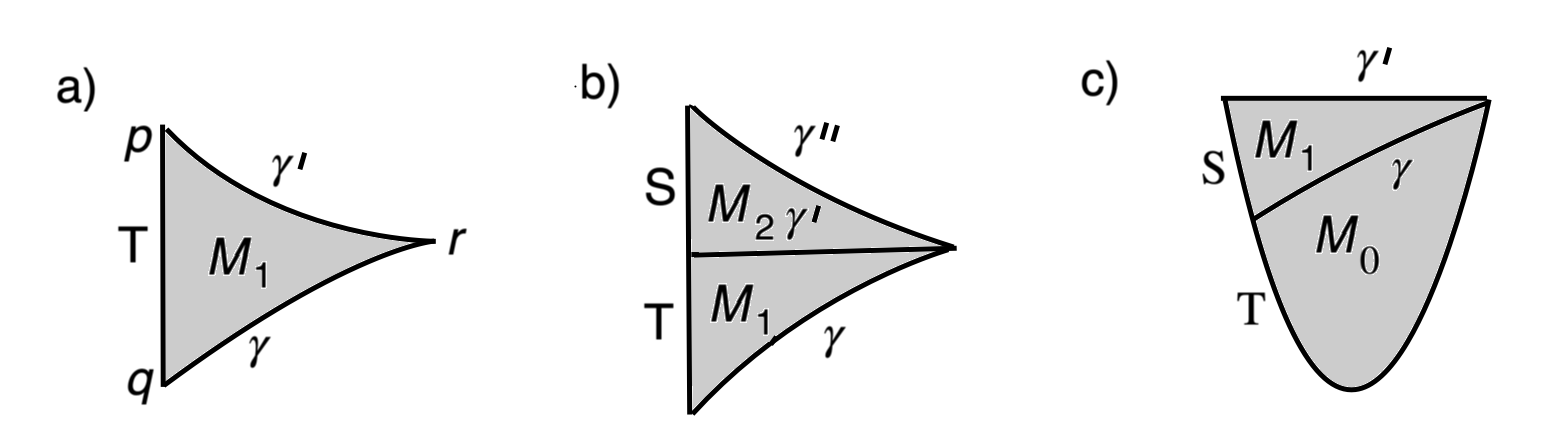}
 \end{center}
\caption{\footnotesize (a) A path integral in region $M_1$ can be used to compute matrix elements $\la\Psi'|\T|\Psi\ra$, where
$\Psi$ and $\Psi'$ are bulk states inserted on $\gamma$ and $\gamma'$, respectively. (Wormholes and initial and final closed
universes may be present and are not drawn.)   (b) The path integral in $M_{12}=M_1\cup M_2$ is used
to prove that the definition in (a) gives an action of the
 algebra $\A$ of boundary observables acts on the bulk Hilbert space. (c) The path integral on
 $M_{01}=M_0\cup M_1$ is used to show that the boundary to bulk map $\wW$ commutes with the action of
 boundary observables.  Note that this picture also implies that if $\Psi_\T=0$ then $\Psi_{\S\T}=0$.    \label{III}}
\end{figure}

We now want to define an action of $\A$ on $\H_\bulk$.  In section \ref{euclideanstyle}, this step was vacuous since $\H_\bdry$
and $\H_\bulk $ coincided.  
   First of all, for a string $\T$ and states $\Psi,\Psi'\in \H_\bulk$, we define the matrix elements
$\la \Psi'|\T|\Psi\ra$ by a path integral in a spacetime region $M_1$ schematically depicted in fig. \ref{III}(a).  $M_1$
has an asymptotic boundary segment labeled by $\T$, and it has past and future boundaries given by geodesic 1-manifolds
$\gamma$ and $\gamma'$, which are not necessarily connected.  Initial and final states $\Psi$ and $\Psi'$ are inserted on $\gamma$ and $\gamma'$. Though not drawn in the figure, $M_1$ may have wormholes and 
$\gamma$ and $\gamma'$ may have any number of disconnected  components, corresponding to the possible presence of closed universes in the initial and final states.
  The path integral over $M_1$ is carried
out only over hyperbolic metrics such that $\gamma$ and $\gamma'$ are minimal.

To show that this definition does give an action of $\A_1$ on $\H_\bulk$, we need to show that for strings $\S,\T$, we have
\be\label{longsum}\la\Psi''|\S\T|\Psi\ra=\sum_{\Psi'} \la\Psi''|\S|\Psi'\ra \la\Psi'|\T|\Psi\ra.\ee
Here the three matrix elements $\la\Psi''|\S\T|\Psi\ra$, $\la\Psi''|\S|\Psi'\ra$, and $ \la\Psi'|\T|\Psi\ra$ are all supposed
to be computed by the recipe just stated, and the sum over $\Psi'$ runs over an orthonormal basis of $\H_\bulk$.
The picture that corresponds to this identity is shown in fig. \ref{III}(b).   
In this picture, the spacetime $M_{12}$  has an asymptotic boundary labeled by $\S$ and $\T$, geodesic boundaries
$\gamma$ and $\gamma''$, on which    initial and final states $\Psi$ and $\Psi''$ are inserted, and an internal geodesic cut $\gamma'$. 
$\T$, $\gamma$, and $\gamma'$ bound a ``lower'' piece $M_1$ of $M_{12}$, while $\S$, $\gamma'$, and $\gamma''$
bound an ``upper'' piece $M_2$.   $M_{12}$ is built by gluing together $M_1$ and $M_2$ along their common boundary $\gamma'$.
 If  $\gamma$ and $\gamma'$ are minimal
in $M_1$, and $\gamma'$ and $\gamma''$ are minimal in $M_2$, then the path integral on $M_{12}$ computes the right hand side of eqn. (\ref{longsum}), with the sum over intermediate states $\Psi'$
coming from the sum over fields on $\gamma'$.  
 On the other hand, if $\gamma$, $\gamma'$, and $\gamma''$ are all minimal in $M_{12}$, then the same path integral computes the
left hand side of eqn. (\ref{longsum}).   Here minimality of
$\gamma'$ means that generically it is uniquely determined by the geometry of $M_{12}$, so including it in the definition of the path
integral has no effect and it can be forgotten; minimality of $\gamma$ and $\gamma''$ in $M_{12}$ is the condition that defines
the path integral on $M_{12}$ that computes $\la\Psi''|\S\T|\Psi\ra$.
    So to complete the proof of eqn.(\ref{longsum}), we just need to know 
that if $\gamma$ and $\gamma'$ are minimal in $M_1$, and $\gamma'$ and $\gamma''$ are minimal in $M_2$, then
all three of them are minimal in $M_{12}$.   This follows by the same cut and paste argument as in section \ref{comparison}.

At this point, it is natural to ask if the boundary to bulk map $\wW:\H_\bdry\to \H_\bulk$ is compatible with the action of the
algebra $\A_1$ on $\H_\bdry$ and on $\H_\bulk$ in the sense that for a string $\S$ and for $\Psi\in\H_\bdry$, one has
$\S\wW(\Psi)=\wW(\S\Psi)$.  It suffices to check this for the case that $\Psi=\Psi_\T$ for some string $\T$, since
states of that form are dense.    Thus we need to verify that 
\be\label{zarko} \S\wW(\Psi_\T)=\wW(\Psi_{\S\T}).\ee   
The relevant picture is fig. \ref{III}(c), where if the relevant cuts are minimal,  then (i) 
the path integral in region $M_0$ computes $\wW(\Psi_\T)$, (ii) the path integral in region $M_0$ computes the action of $\S$
on this state, and (iii) the path integral in $M_{01}=M_0\cup M_1$ computes $\wW(\Psi_{\S\T})$.
For statement (i), $\gamma$ must be minimal in $M_0$, for statement (ii), $\gamma$ and $\gamma'$ must be minimal in $M_1$,
and for statement (iii), $\gamma$ and $\gamma'$ must be minimal in $M_{01}$.    
  (For statement (iii), we reason as in the last paragraph: $\gamma$ being minimal in $M_{01}$ means
that  it is uniquely determined generically and plays no role, and $\gamma'$ being minimal
and the hyperbolic metric of $M_{01}$ being otherwise
arbitrary ensures that the path integral in $M_{01}$ computes $\wW(\Psi_{\S\T})$.)    
A cut and paste argument as before shows that
if $\gamma$ is minimal in  $M_0$ and $\gamma$ and $\gamma'$ are minimal in $M_1$, then $\gamma$ and $\gamma'$
are minimal in $M_{01}$.   So eqn. (\ref{zarko}) is valid.

This argument shows that the algebra $\A_1$ acts on $\H_\bdry$ and $\H_\bulk$, and that $\wW:\H_\bdry\to\H_\bulk$
maps the action on $\H_\bdry$ to the action on $\H_\bulk$.   We can complete  $\A_1$ acting on $\H_\bulk$
to a von Neumann algebra, and we want to know that this is the same von Neumann algebra $\A=\A_L$ that we get
if we complete $\A_1$ acting on $\H_\bdry$.   This statement means  for any sequence $\S_1,\S_2,\cdots
\in \A_1$, the sequence 
$\lim_{n\to\infty}\la\Psi|\S_n|\Psi\ra$ converges for all $\Psi\in \H_\bulk$ if and only if it converges for all $\Psi\in\H_\bdry$.
The ``only if'' statement is trivial since $\H_\bdry$ is isomorphic via $\wW$ to a subspace of $\H_\bulk$.   The ``if'' statement
follows from something superficially stronger but actually equivalent that we will prove presently: for every $\Lambda\in\H_\bulk$, there is $\chi\in\H_\bdry$ such that
$\la\Lambda|\S|\Lambda\ra=\la\chi|\S|\chi\ra$ for all $\S\in \A_1$.   Physically, this statement means that every $\Lambda\in\H_\bulk$
is indistinguishable, from the point of view of a boundary observer, from some $\chi\in\H_\bdry$.

    \begin{figure}
 \begin{center}
   \includegraphics[width=4in]{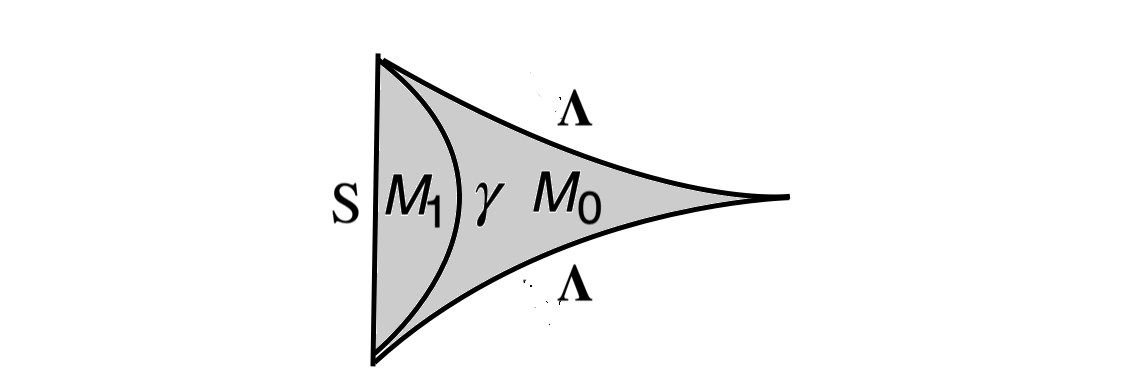}
 \end{center}
\caption{\footnotesize A picture that is used to show that, from the point of view of a boundary observer, any bulk state is equivalent
to a possibly mixed state on $\H_\bdry$.    \label{JJJ}}
\end{figure} 

A tempting but insufficient argument would go as follows.
If $\Lambda\in\H_\bulk$ is any bulk state, the function $\S\to \la\Lambda|\S|\Lambda\ra$ is a  linear functional on the algebra $\A$ that is non-negative (meaning that it is non-negative if $\S=\T^\dagger \T$ for some $\T$), and therefore,
as $\A$ is of Type II$_\infty$, it is $\Tr\,\S\rho$ for some $\rho$.  Here $\rho$ may be  an element of
$\A$, but more generally is  ``affiliated'' to $\A$ (meaning that bounded functions of $\rho$ belong to $\A$ and in particular
that $\rho$ can be arbitrarily well approximated for many purposes by elements of $\A$).  $\rho$ can then also be replaced
by a pure state $|\chi\ra$, as we explain later.
The trouble with this argument is that the function $\la\Lambda|\S|\Lambda\ra$, for $\Lambda\in \H_\bulk$, is initially defined for $\S$ in the algebra $\A_1$ of linear combinations of strings modulo null
vectors.   To know that this linear function 
 is $\Tr\,\S\rho$ for some $\rho$ in 
 (or affilliated to) $\A$, we need to know that it
extends continuously over the completion $\A=\A_L$ of $\A_1$, or equivalently, that the von Neumann algebra completion
of $\A_1$ acting on $\H_\bulk$ is the same as\footnote{If the completion of $\A_1$ acting on $\H_\bulk$ were some other von Neumann algebra $\t\A\not=\A$,
then for a bulk state $\Lambda$, the density matrix $\rho$ satisfying $\la\Lambda|\S|\Lambda\ra=\Tr\,\S\rho$ would be affiliated to $\t\A$, not to $\A$.}  $\A$, which was defined as the completion of $\A_1$ acting on $\H_\bdry$.   But this is  what
we are trying to prove.

In fact, the existence of a suitable $\rho$ affiliated to $\A$ (and thus the equivalence of the two completions of $\A_1$)  can be deduced as follows.  In  fig. \ref{JJJ},  $M_{01}=M_0\cup M_1$ has an asymptotic boundary labeled by $\S$
as well as past and future geodesic boundaries labeled by $\Lambda$  that are assumed to be minimal.
   The path integral on $M_{01}$, with $\Lambda$ inserted as an initial and final state on the geodesic boundaries,
    will compute $\la\Lambda|\S|\Lambda\ra$.
  But
we can also choose a minimal geodesic cut $\gamma$ connecting the two ends of $\S$, as shown.   Then the path integral
on $M_0$ computes a state $\chi\in \H_\bulk$, and the path integral on  $M_1$ computes the inner product $\la\chi|\wW(\Psi_\S)\ra$.
So $\la\Lambda|\S|\Lambda\ra =\la \chi|\wW(\Psi_\S)\ra$.   
Here {\it a priori} $\chi$ is a general bulk state, not in the image of  $\wW:\H_\bdry\to \H_\bulk$.  However, since $\wW(\Psi_\S)$ is in the image of $\wW$,
without changing the inner product  $\la \chi|\wW(\Psi_\S)\ra$
we can replace $\chi$ by its orthogonal projection to the image of $\wW$.   Then, since $\wW$ is invertible when restricted to this image,
there is a unique $\zeta\in\H_\bdry$ with $\wW(\zeta)=\chi$.   
In fact, $\zeta=\vV(\chi)$, where $\vV:\H_\bulk\to \H_\bdry$ is the adjoint of $\wW$.   
So $\la\Lambda|\S|\Lambda\ra=\la\zeta,\Psi_\S\ra$, where now the inner product is between two states in $\H_\bdry$.
  Finally, we recall that $\H_\bdry$ has a dense set of states $\Psi_\rho$,
$\rho\in \A$, so $\zeta=\Psi_\rho$ where $\rho$ is either an element of $\A$ or in general an operator affiliated to $\A$.
Hence  $\la\zeta ,\Psi_\S\ra=\la\Psi_\rho,\S\ra=\Tr\,\rho\S$. Putting all this together, we learn that
\be\label{densmat}\la\Lambda|\S|\Lambda\ra =\Tr\,\rho\S.\ee
  $\rho$ is a non-negative self-adjoint operator of trace 1  or in other words a density matrix, since the functional $\la\Lambda|\S|\Lambda\ra$
is nonnegative and (if $\Lambda$ is a unit vector) equals 1 if $\S=1$.

A general mixed state on $\H_\bulk$ is $\rho_\bulk = \sum_i p_i|\Lambda_i\ra \la \Lambda_i|$, where $\Lambda_i$ are orthonormal pure states
in $\H_\bulk$ and $p_i>0$, $\sum_i p_i=1$.  As just explained, there are density matrices $\rho_i$ affiliated to $\A$ such that
  $\la\Lambda_i |\S|\Lambda_i\ra=\Tr\,\S\rho_i$ for all $\S$.
So if $\rho_\bdry =\sum_i p_i\rho_i$ then 
\be\label{linko} \Tr\,\S\rho_\bulk= \Tr\,\S \rho_\bdry. \ee
On the left, $\S$ is an operator on $\H_\bulk$ and the trace is the natural trace of an operator that acts on  $\H_\bulk$.
On the right, $\S$ and $\rho$ are elements of the Type II$_\infty$ algebra  $\A$ and $\Tr $  is  the trace of  this algebra.
Eqn. (\ref{linko})  expresses the fact that every pure or mixed state on $\H_\bulk$ can be described, from the point of view of a boundary observer, by a density matrix
$\rho_\bdry$ associated to $\A$.   But actually, any 
such density matrix  can be purified
by a pure state in $\H_\bdry$.    For this, let $\sigma=\rho_\bdry^{1/2}$.     Since $\Tr\,\sigma^\dagger\sigma=\Tr\,\rho_\bdry=1$,
the condition of eqn. (\ref{algcond}) is satisfied, and there is a state $|\sigma\ra \in \H_\bdry$ satisfying 
\be\label{purifying}\la\sigma|\S|\sigma\ra =\Tr\,\S\sigma\sigma^\dagger =\Tr\,\S\rho_\bdry\ee
for all $\S$.   Thus, in fact, to a boundary observer, every pure or mixed state $\rho_\bulk$ on the bulk Hilbert space $\H_\bulk$
is indistinguishable
 from some pure state $|\sigma\ra$ in the much smaller Hilbert space $\H_\bdry$.    This pure state is unique only up to the action of a unitary operator
in the commutant $\A'=\A_R$ of $\A=\A_L$.

\subsection{Some Final Remarks}\label{remarks}

We defined $\H_\bdry$  starting with open universe observables that we called ``strings.''   A string corresponds to a piece of the asymptotic boundary of spacetime, topologically
a closed interval, labeled by operator insertions that correspond to boundary observables.
But $\H_\bulk$ is much bigger than $\H_\bdry$. How can states that are in  $\H_\bulk$ but not in
$\H_\bdry$ be accessed?      One answer, in the spirit of \cite{MM}, is that we could generalize the construction of section \ref{euclideanstyle} to include ``closed strings'' as well
as the ``open strings'' that we have considered so far.   A closed string here just means an asymptotic boundary of spacetime that is topologically a circle.   For the closed string,
one can take the same boundary conditions,     labeled by a pair $b,\Lambda$,
 that we assumed in section \ref{operators},\footnote{One can contemplate more general boundary conditions on asymptotic closed boundaries, but we do not expect that this would add anything, since the boundary conditions
considered in section \ref{operators} suffice to create an arbitrary closed universe state.} where we tried to use closed strings to define operators $\O_{b,\Lambda}$ on $\H_{\bdry}$.  The construction would be similar to that of section \ref{operators}, except that now (as in \cite{MM}), we would specify whether a given closed asymptotic 
boundary is creating part of the
initial state or part of the final state.    In section \ref{operators}, there was no reason to make this distinction.

In more detail, we would proceed as follows,  Let $\h \S$ be a not necessarily connected string consisting of a single ``open string'' and any number of closed strings.
For each $\h \S$, formally define a state $\Psi_{\h\S}$, with inner products $\la \Psi_{\h\T},\Psi_{\h\S}\ra$ defined as in section \ref{euclideanstyle} by a path integral
on a spacetime whose asymptotic boundary is built by gluing $\h\S$ onto the adjoint of $\h\T$.   (The adjoint operation is the same as before for open strings, and
is $\sf{CPT}$ for closed strings.)   If the inner products $\la \Psi_{\h\T},\Psi_{\h\S}\ra$ are positive semi-definite, then upon dividing 
 by null vectors and taking a Hilbert space completion, one arrives at what we will call the Marolf-Maxfield Hilbert space $\H_{\MM}$,
since this construction for the closed strings was described in \cite{MM}.   We expect that the construction that we have described with the minimal geodesic cuts, extended
to this more general situation in a natural way, will show that the inner products  $\la \Psi_{\h\T},\Psi_{\h\S}\ra$ are positive semi-definite and
establish an isomorphism between $\H_{\MM}$ and what we have called $\H_\bulk$.

Of course, this reformulation of what one wants to do with asymptotic closed boundaries does not, by itself, eliminate the problem we had in section \ref{operators} with
the divergence that results from the negative Casimir energy.   Exactly what happens in a better theory that resolves this divergence remains to be understood.

Though every state in $\H_\bulk$ is equivalent from the viewpoint of a boundary observer to some pure state in $\H_\bdry$,
there is no natural way to exhibit this equivalence by a linear map from $\H_\bulk$ to $\H_\bdry$.  
 The only natural map that we have found between these spaces is
$\vV:\H_\bulk\to \H_\bdry$, the adjoint of $\wW$.  However, $\vV$ is far from being an isometry:   it is an isomorphism on
$\wW(\H_\bulk)$ and annihilates the orthocomplement of this space.  Still, in the spirit of \cite{AEHPV}, one may wonder
whether for some purposes, after restricting  to a suitable subspace of $\H_\bulk$, such as a subspace obtained by long enough real time evolution starting from
a space of macroscopically similar black hole states,  some multiple of $\vV$  may be very hard 
to distinguish from an isomorphism.

    \begin{figure}
 \begin{center}
   \includegraphics[width=4in]{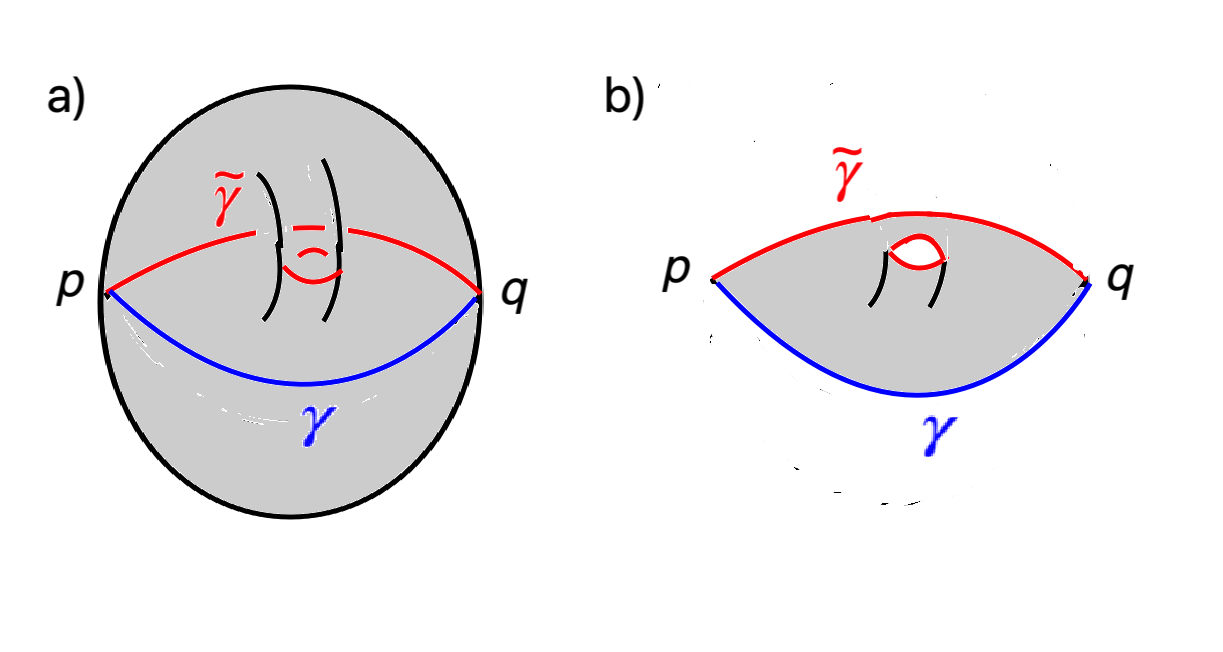}
 \end{center}
\caption{\footnotesize (a)  A disc with a handle attached; shown are geodesic cuts $\gamma$ (blue) and $\t\gamma$ (red) that connect boundary points $p$, $q$.  In this example, $\gamma$
is connected and $\t\gamma$ is not.  As $p$ is moved
``upwards'' along the boundary, the minimal geodesic cut can jump from $\gamma$ to $\t\gamma$. (b)  To reproduce this jumping, the boundary Hamiltonian $H_L$ 
has a matrix element that glues the indicated surface onto an initial state  defined on $\gamma$, producing a final state defined on $\t\gamma$.   In this example, that matrix element involves
creation of a baby universe.   \label{KKK}}
\end{figure} 

We conclude with a discussion of real time evolution.
Since the algebra $\A=\A_L$ of observables on the left boundary acts on the bulk Hilbert space $\H_\bulk$, in particular
this gives an action of its generator $e^{-\beta H_L}$ on $\H_\bulk$.   Taking logarithms, the boundary Hamiltonian $H_L$
is an operator on $\H_\bulk$, and exponentiating again, we can describe the real time evolution of a bulk state by the action of
$e^{-\i t H_L}$.    By the same logic, we can define the evolution of a bulk state under real time evolution of the right boundary
by $e^{-\i t H_R}$.     Apart from topology-changing processes,
 $H_L$ and $H_R$ act very simply; they act as  described in eqn. (\ref{asops}) on the open universe Hilbert space $\H_\o$, and they
annihilate the closed
universe Hilbert space $\H_\cl$, since the total energy of a closed universe is 0.   However, this is far from the whole story;
$e^{-\beta H_L}$ and similarly
$e^{-\beta H_R}$ have matrix elements that describe topology-changing processes in the bulk, and therefore the same is true
of $H_L$ and $H_R$.    For an example, see fig. \ref{KKK}.     
Because the Hamiltonian has topology-changing matrix elements, real 
time evolution over any substantial time interval is likely to be quite complicated, even if the starting
point is a state with a simple Euclidean description.

\section{Multiple Open Universes}\label{multiple}

In section \ref{wormhole}, we studied universes with a single open component and any number of closed components.   From the standpoint
of General Relativity, it is certainly  possible to contemplate universes with multiple open components.   This generalization is nontrivial in the presence
of wormholes, since different open universe components can interact by exchanging wormholes.
The analysis presented so far in this article extends naturally to the case of multiple open universes, as we will now discuss.   

The most significant conclusion that we will reach is that an observer with access to only one asymptotic boundary has no way to
determine by any measurement  how many other such boundaries there are.   The reasoning that leads to this conclusion will be similar
to arguments that we have seen already.

As in footnote \ref{reflection}, we do not assume time-reversal or reflection symmetry, so we distinguish left and right asymptotic
boundaries.  In the absence of time-reversal symmetry, spacetime is oriented, and its boundary is also oriented.   In all
pictures in this section,
the orientation comes from the counterclockwise orientation of the plane.

In the absence of reflection symmetry, left and right boundaries are inequivalent; there is no Bose symmetry between them.
But we also do not impose Bose symmetry between boundaries of the same type.   The different left or right boundaries
are considered inequivalent, since we want to analyze
 the operators available to an observer who has access to  one specified left or right
boundary.

In generalizing our previous results  to a universe with multiple open components, we will not be as detailed
as we have been up to this point.   We just briefly describe the analogs of the main steps in section \ref{wormhole}.
In doing so, for brevity we consider the case of a universe with two open components. The generalization to any number of open components is immediate.

\vskip.5cm
\noindent {\bf 1) The Algebra} \\  The algebra of observables on a particular left (or right) boundary is taken to be precisely the
same algebra as in section \ref{wormhole} (with wormhole corrections included).   Thus the algebra is defined as before by starting with strings $\S,\T$,
computing inner products $\la\Psi_\S,\Psi_\T\ra$ from a spacetime with one asymptotic boundary and any topology,
and taking a completion to get a Hilbert space $\H_\bdry$ and an algebra $\A=\A_L$ that acts on it.

     \begin{figure}
 \begin{center}
   \includegraphics[width=4.5in]{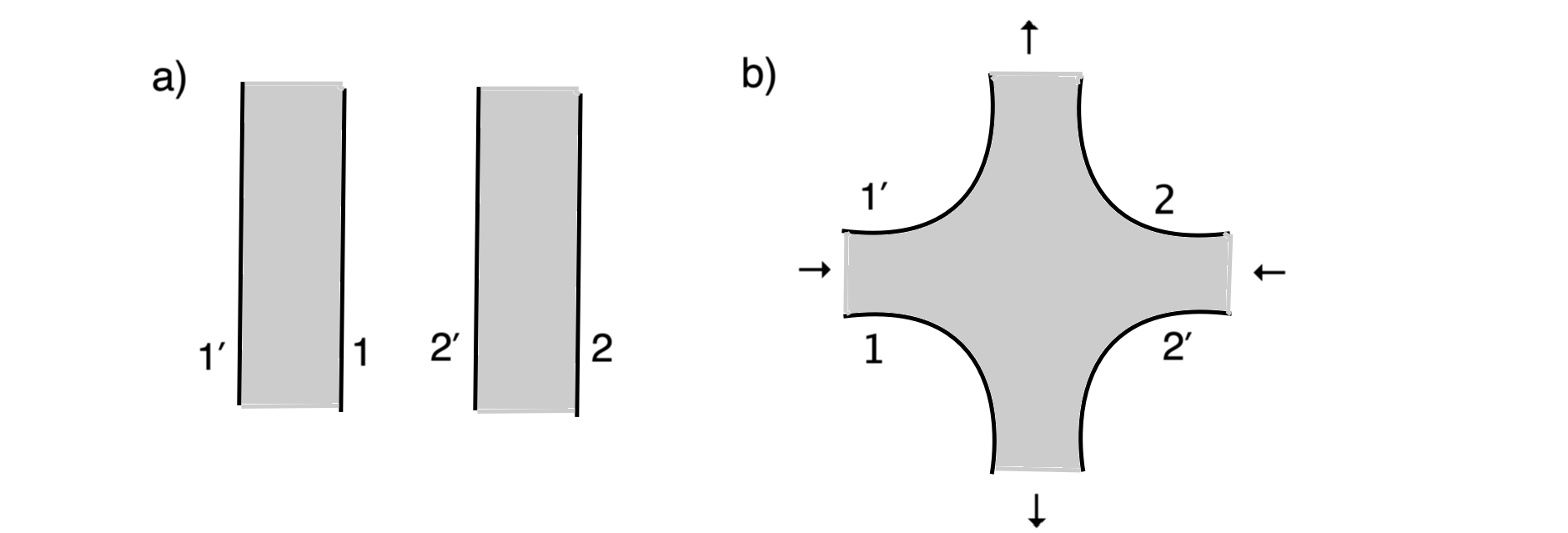}
 \end{center}
\caption{\footnotesize (a) Free propagation of an open  universe of type $1'1$ and one of type $2'2$. (b) A transition from a universe with components 
$1'1+2'2$ to one with components $1'2+2'1$.   Incoming arrows mark open universe components in the initial state, and outgoing arrows mark open universe components
in the final state.   Topologically, the spacetime in (a) is a disjoint union of two discs, with total Euler characteristic 2, and the spacetime in (b) is a single disc,
with Euler characteristic 1.  So the process in (b) is suppressed by a single power of $e^{-S}$.    This is the lowest order ``interaction'' between distinct
open universe components.  \label{LLL}}
\end{figure} 

\vskip.5cm\noindent
{\bf 2) The Bulk Hilbert Space}  \\ For the case of two open components, let us label the left boundaries as $1'$ and $2'$ and the
right boundaries as $1$ and $2$.  To make a world with these two asymptotic boundaries, we pair up the boundaries
as $1'1+2'2$ or as $1'2+2'1$.   Correspondingly, the Hilbert space with two open universe components (and no closed universes)
is, in an obvious notation, $\H_{\o,[2]}=\H_{1'1}\otimes \H_{2'2}\oplus \H_{1'2}\otimes \H_{2'1}$.     Including closed universes
as before, the bulk Hilbert space with two open components is
\be\label{gentwo}\H_{\bulk,[2]}=\H_{\o,[2]}\otimes \Sym^*\H_\cl =( \H_{1'1}\otimes \H_{2'2}\oplus \H_{1'2}\otimes\H_{2'1})\otimes
\Sym^*\H_\cl. \ee
The dynamics leads to transitions between configurations of type $1'1+2'2$ and those of type $1'2+2'1$.
Such transitions are suppressed by one factor of $e^{-S}$ (where $S$ is the entropy); see fig. \ref{LLL}.

     \begin{figure}
 \begin{center}
   \includegraphics[width=5in]{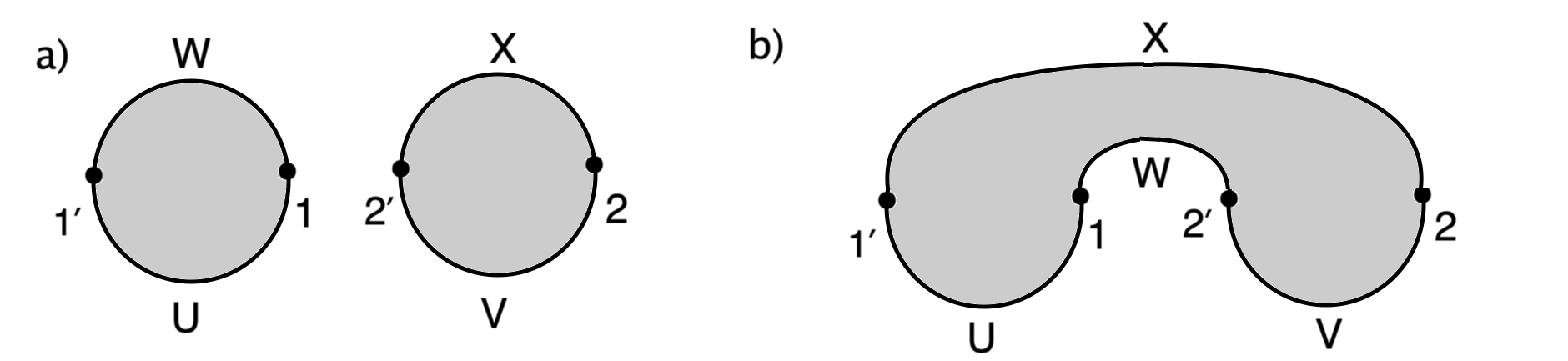}
 \end{center}
\caption{\footnotesize (a) The lowest order contribution to the inner product
$\la\Psi_{\W_{1'1}\times \X_{2'2}} |\Psi_{\U_{1'1}\times \V_{2'2}}\ra$, between states created by string pairs both
of type $1'1+2'2$.    (b)  The lowest order contribution to an inner product
$\la \Psi_{ \W_{2'1}\times\X_{1'2}}|\Psi_{\U_{1'1}\times \V_{2'2}}\ra$ between states created
by string pairs of opposite types $1'1+2'2$ and  $1'2+2'1$.    Either picture
can be decorated with wormholes, including wormholes that connect the two components  in (a).  The spacetime
in (b) has Euler characteristic 1, compared to 2 in (a), so the inner product between states
created by string pairs of opposite type is suppressed by one factor of $e^{-S}$. \label{MMM}}
\end{figure} 

\vskip.5cm\noindent
{\bf 3) The Boundary Hilbert Space} \\  To specify a state via boundary data now requires a pair of strings labeled by their endpoints,
for example $\S_{1'1}\times \T_{2'2}$ or $\S_{1'2}\times \T_{2'1}$.  To define the adjoint of a pair, we apply the adjoint operation that was introduced in section
\ref{euclideanstyle} to each string separately, exchanging its left and right endpoints, and formally replacing a string with the adjoint string.  
So   the adjoint of, for example, $\S_{1'2}\times \T_{2'1}$
is $\S^\dagger_{21'}\times \T^\dagger_{12'}$  (the adjoint strings correspond to bras rather than kets and the right boundary is written first).
   To each such pair we formally associate a state $\Psi_{\S_{1'1}\times\T_{2'2}}$ or $\Psi_{\S_{1'2}\times \T_{2'1}}$.
  We refer to such pairs or states as being of type $1'1\times 2'2$ or $1'2\times 2'1$, as the case may be.  
Inner products of these states, for example
$\la\Psi_{\W_{1'1}\times \X_{2'2}} |\Psi_{\U_{1'1}\times \V_{2'2}}\ra$ or
$\la \Psi_{ \W_{2'1}\times\X_{1'2}}|\Psi_{\U_{1'1}\times \V_{2'2}}\ra$,
are defined in the obvious way by gluing one string to the adjoint of the other and summing 
over all possible fillings (fig. \ref{MMM}).  With this rule,  inner products between states created by string pairs of opposite type are nonzero but are suppressed by one factor of
$e^{-S}$, as illustrated in the figure.     That these inner products are positive semi-definite follows from an embedding in the bulk Hilbert space, as discussed
shortly.   Given this, the boundary Hilbert space $\H_{\bdry,[2]}$ for a universe with two open components and any number of closed components is defined in the usual way
by dividing out null vectors and then taking a completion to get a Hilbert space.

     \begin{figure}
 \begin{center}
   \includegraphics[width=4.5in]{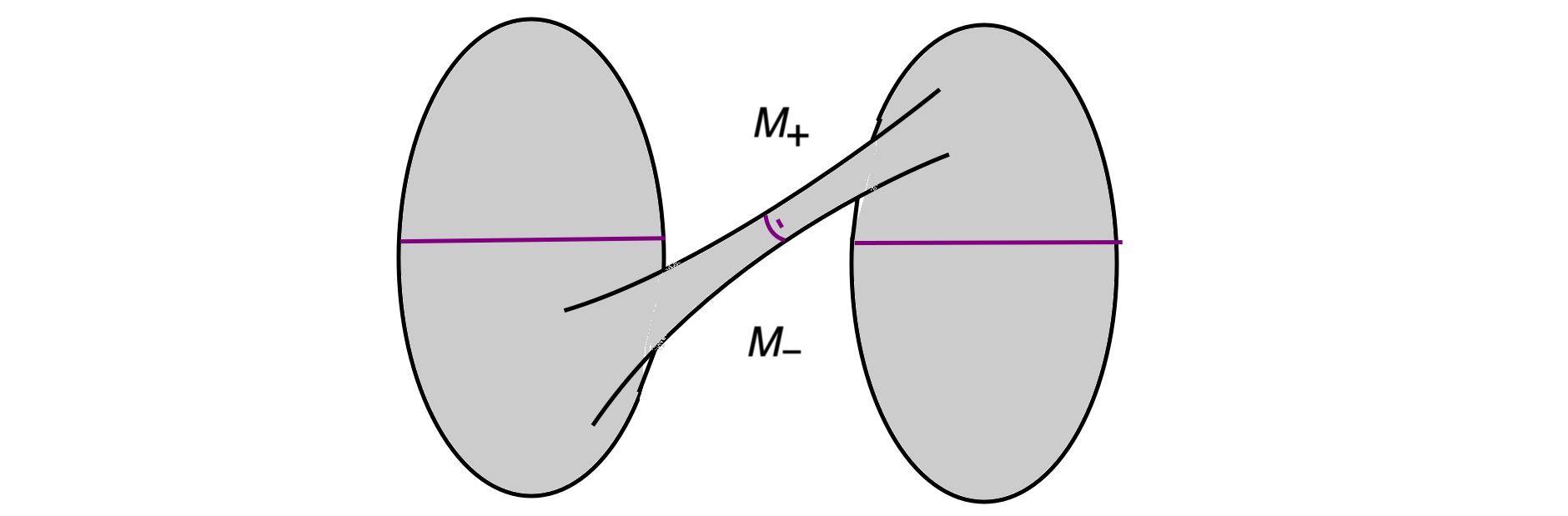}
 \end{center}
\caption{\footnotesize   A minimal geodesic cut that separates the region $M_-$ below
the cut from the region $M_+$ above the cut. In this example, the cut has three components, one of which is ``in the wormhole.''  \label{NNN}}
\end{figure} 
\vskip.5cm\noindent
{\bf 4) Geodesic Cuts}   By a boundary cut of a spacetime $M$ with several left and right asymptotic boundaries,  we mean
simply the choice of a point on each asymptotic boundary (if an asymptotic boundary has an endpoint where it meets a geodesic boundary, that endpoint can be part of the geodesic cut).   
By a geodesic cut $\gamma$   asymptotic to a given boundary cut of $M$,
we mean a collection of oriented disjoint geodesics that include geodesics that pair up the left and right boundary points in the
given boundary cut, together with possible closed geodesics, satisfying the condition that $\gamma$ divides $M$ into disjoint components $M_-$ and $M_+$.  $M_+$ is on the side
of $\gamma$ that is specified by the orientation of the right boundaries of $M$, and $M_-$ is on the opposite side of $\gamma$.  
 Pictures are generally drawn to place $M_+$ ``above'' the cut and $M_-$ ``below'' it (fig. \ref{NNN}).    If $M$ has geodesic boundaries as well as asymptotic
boundaries, then we allow the case that a component of $\gamma$ is a boundary component of $M$; in particular, we allow the case that all components of $\gamma$ are
boundary components of $M$, and $M_-$ or $M_+$ is empty.
  A geodesic cut $\gamma$ is minimal if it has minimal renormalized length among all geodesic
cuts asymptotic to a given boundary cut.

     \begin{figure}
 \begin{center}
   \includegraphics[width=5.9in]{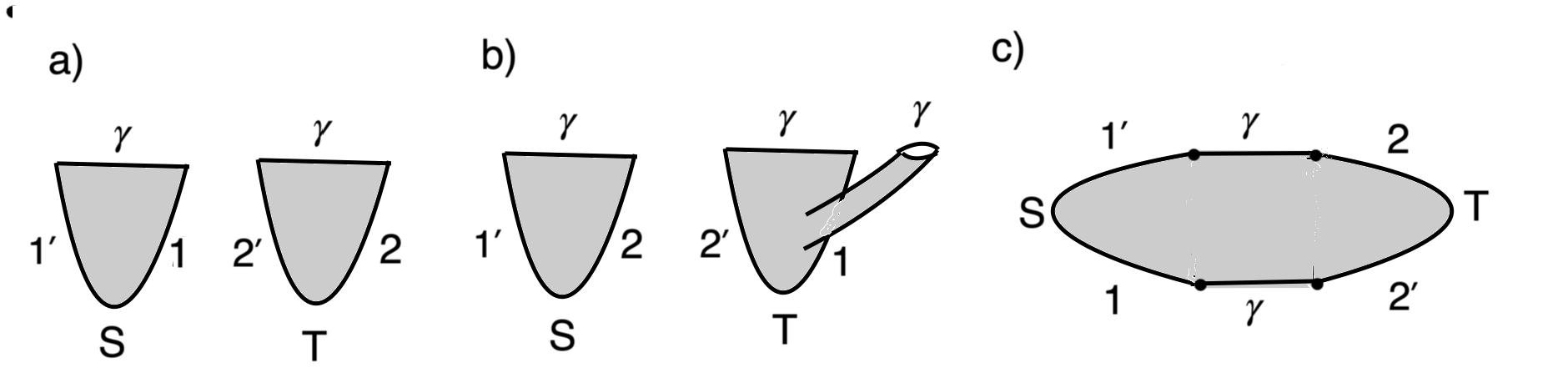}
 \end{center}
\caption{\footnotesize (a) In leading order, the bulk state $\wW(\Psi_{\S_{1'1}\times \T_{2'2}})$ created
by a string pair of type $1'1+2'2$ is of the same type.   In this figure, $\gamma$ is a minimal geodesic cut with multiple
components and the state 
the state $\wW(\Psi_{\S_{1'1}\times \T_{2'2}})$
is a function of fields on $\gamma$.   (b) In order $e^{-S}$,  $\wW(\Psi_{\S_{1'1}\times \T_{2'2}})$ has a component with a closed baby universe.  (c) In the same order, it has a component consisting
 of two open universes 
 of types $1'2+2'1$.  As usual, all figures can be decorated with wormholes, and additional closed universes
 can be added to the final state. 
 \label{OOO}}
\end{figure} 

\vskip.5cm\noindent
{\bf 5) Boundary to Bulk Map }   We want to define an isometric map $\wW$  from states defined by boundary data (such as a pair of strings) to $\H_{\bulk,[2]}$. 
For example, to define  $\wW(\Psi_{\S_{1'1}\times \T_{2'2}})$, we sum over spacetimes $M$ that have an asymptotic boundary
defined by  $\S_{1'1}\times \T_{2'2}$
as well as geodesic boundaries that make up a minimal geodesic cut $\gamma$ (fig. 
\ref{OOO}).
The dependence of the path integral on the fields on $\gamma$ then gives a state in $\H_{\bulk,[2]}$.   Note that by this definition,
$\wW$  maps a string of type $1'1+2'2$ to a bulk state  that for large $S$ is mostly of type $1'1+2'2$, but that in order $e^{-S}$ also has a component of type $1'2+2'1$ (fig. \ref{OOO}(c)).  
As in section \ref{wormhole}, the map $\wW$ is isometric, that is, it preserves inner products.   This is proved by 
 showing that if
a geodesic cut $\gamma$ of $M$, in, say, fig. \ref{NNN}, is minimal in $M_-$ and in $M_+$, then it is minimal in $M$. The proof of this involves the same cut and paste argument as in section \ref{wormhole}. Hence, as in fig. \ref{GGG}(c), for 
string pairs $\S,\T$ and $\U,\V$,
if one glues together the path integral construction of $|\wW(\Psi_{\S\times \T})\ra$ and that of
$\la\wW(\Psi_{\U\times \V})|$, one gets the same path integral that computes the inner product $\la \Psi_{\U\times\V}|\Psi_{\S\times \T}\ra$ between states defined by string
pairs, implying that $\la\Psi_{\U\times \V}|\Psi_{\S\times \T}\ra=
\la \wW(\Psi_{\U\times\V})|\wW(\Psi_{\S\times \T})\ra$.
This embedding implies that the inner products of states
defined by linear combinations of string pairs are positive semi-definite.   Dividing out null vectors and taking a Hilbert space completion, one arrives at
 the definition of the boundary Hilbert space $\H_{\bdry,[2]}$,
which then comes with an embedding $\wW:\H_{\bdry,[2]} \to\H_{\bulk,[2]}$.

\vskip.5cm\noindent
{\bf 6) Action of the Boundary Algebra}  The boundary algebra $\A$ was originally defined as an algebra of operators acting on $\H_\bdry$,
the Hilbert space accessible to a boundary observer in a universe with just one open component.    However, precisely the same algebra acts
on both $\H_{\bdry,[2]} $ and on $\H_{\bulk,[2]}$, and moreover, these actions commute with the map $\wW$ between those two spaces.  
To understand this,  consider an observer with access to the left boundary labeled $1'$.   To define the action on $\H_\bdry$, we start with an obvious action
on pairs of strings.   We take a string $\S$ on the $1'$ boundary to act on a pair of strings in an obvious way, for example
$\T_{1'1}\times \U_{2'2}\to (\S\T)_{1'1}\times \U_{2'2}$.    Starting with this action on strings, we want to define an action of $\S$ on $\H_{\bdry,[2]}$
by (for example) $\S\Psi_{\T_{1'1}\times \U_{2'2}}=\Psi_{(\S\T)_{1'1}\times \U_{2'2}}$.  For this definition to make sense, we need to know that if $\S$
is null (meaning that $\Psi_\S=0$ in $\H_{\bdry}$ and therefore $\S=0$ in $\A$) or $\T_{1'1}\times \U_{2'2}$ is null (meaning that $\Psi_{\T_{1'1}\times \U_{2'2}}=0$ in
$\H_{\bdry,[2]}$), then $\Psi_{{\S\T}_{1'1}\times \U_{2'2}}=0$.  The proof of the first statement precisely follows fig. \ref{CCC} or fig. \ref{ZZZ}(b), and the proof
of the second precisely follows fig. \ref{DDD}(b) or fig. \ref{III}(c).

     \begin{figure}
 \begin{center}
   \includegraphics[width=5in]{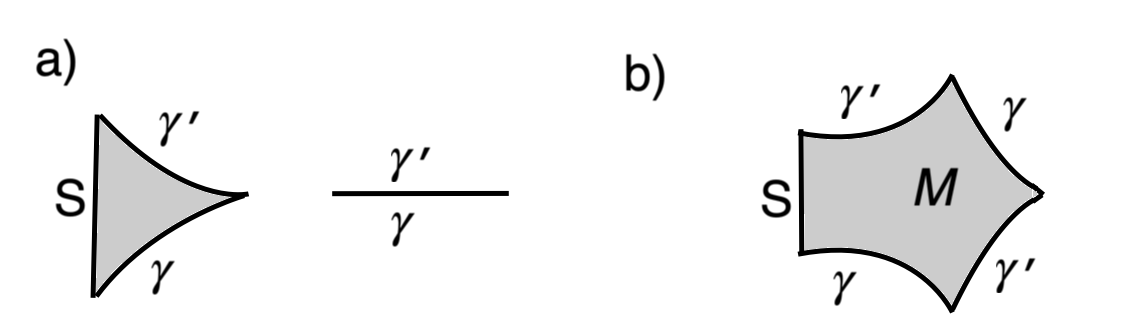}
 \end{center}
\caption{\footnotesize Computation of a matrix element $\la\Psi'|\S|\Psi\ra$, where $\Psi,\Psi'$ are bulk states in
a world with two open universe components, and the string $\S$ acts only on one specified left boundary.
(a) The most obvious possibility is that  $\S$ acts on the state on one open component and does nothing
to the state on the other open component. (b)  There are less obvious possibilities.
In general, the spacetime $M$ has an asymptotic boundary labeled by $\S$ and geodesic boundaries labeled
by minimal geodesic cuts $\gamma$ and $\gamma'$.   $\gamma$ and $\gamma'$ have three of their
four endpoints in common and the fourth at opposite ends of $\S$.   Initial and final states $\Psi$ and $\Psi'$
are functions of boundary data on $\gamma$ and on $\gamma'$, respectively.  $\gamma$ and $\gamma'$
are allowed to have  components in common, as in (a).     \label{PPP}}
\end{figure}

We also want to define an action of $\A$ on $\H_{\bulk,[2]}$.   This again is done by imitating previous definitions, though the presence of more than one asymptotic component makes the resulting pictures  harder to
draw or visualize.  For example, let us define a matrix element $\la\Psi'|\S|\Psi\ra$, where $\S$ is a string
acting on a specified left  boundary and $\Psi,\Psi'\in \H_{\bulk,[2]}$.   
For this, we consider a spacetime $M$ whose boundary consists of an asymptotic boundary labeled by $\S$
and two minimal geodesic cuts $\gamma$, $\gamma'$ on which states $\Psi,\Psi'$ are inserted. 
We assume that  $\gamma$ and $\gamma'$ each have two noncompact
components (along with possible closed components) and therefore four endpoints, and also that    $\gamma$ and $\gamma'$
have three endpoints in common, and that their fourth endpoints are at opposite ends of $\S$.
   Some obvious and less obvious choices of $M$ are sketched in fig. \ref{PPP}. 
   The matrix element  $\la\Psi'|\S|\Psi\ra$  is defined 
by a sum over all such spacetimes $M$.  The proof that this does give an action of
$\A$ on $\H_{\bulk,[2]}$ follows fig. \ref{III}(b).  The presence of an additional open universe component
makes the drawing of representative pictures more complicated but does not affect the logic of the argument.
Similarly, the proof that the action of $\A$ on $\H_{\bulk,[2]}$ and $\H_{\bdry,[2]}$ commutes with the map $\wW$,
in the sense that $\wW(\S\Psi_\T)= \S \wW(\Psi_\T)$, follows fig. \ref{III}(c).

\vskip.5cm\noindent
{\bf 7) Any Bulk State Is Equivalent To A Boundary State} Finally, we come to showing that from the perspective of a boundary observer, with access say to a specified
left boundary, any pure or mixed state on the bulk Hilbert space $\H_{\bulk,[2]}$ is indistinguishable from some pure state in $\H_{\bdry}$.   The key picture is
fig. \ref{JJJ}, generalized now to the case of more than one open universe component (and any number of closed components).  By the same logic as before,
this picture can be used to show that any pure or mixed state on $\H_{\bulk,[2]}$ is equivalent, to a boundary observer, to some density matrix
$\rho$ affiliated to $\A$.  Setting $\sigma=\rho^{1/2}$, it then follows as before that any pure or mixed state on $\H_{\bulk,[2]}$ is actually equivalent for
a boundary observer to the pure state $|\sigma\ra\in\H_{\bdry}$.

\vskip1cm
 \noindent {\it {Acknowledgements}}  
   GP is supported by the UC Berkeley physics department, the Simons Foundation through the ``It from Qubit" program, the Department of 
  Energy via the GeoFlow consortium (QuantISED Award DE-SC0019380) and an early career award, and AFOSR (FA9550-22-1-0098); he also acknowledges support from an 
  IBM Einstein Fellowship at the Institute for Advanced Study.
  Research of EW supported in part by NSF Grant PHY-2207584.   We thank Feng Xu for very helpful explanations about von Neumann algebras, Don Marolf for valuable discussions,
  and David Kolchmeyer for pointing out an error in section \ref{operators} in the original version of this article.
  
 \bibliographystyle{unsrt}

\end{document}